\documentclass[a4paper,11pt]{article}
\pdfoutput=1 
\usepackage{jheppub}
\usepackage[T1]{fontenc} 
\usepackage{graphicx}
\usepackage{soul}
\usepackage{amsmath}
\usepackage{amssymb}
\usepackage{subcaption}
\def\ba{\begin{eqnarray}}
\def\ea{\end{eqnarray}}
\def\be{\begin{equation}}
\def\ee{\end{equation}}
\def\als{\alpha_{\rm s}}

\title{Bottomonium suppression in an open quantum system using the quantum trajectories method}
\preprint{TUM-EFT 140/20; HU-EP-20/36-RTG}

\author[a,b]{Nora Brambilla,}
\author[c]{Miguel \'{A}ngel Escobedo,}
\author[d]{Michael Strickland,}
\author[a]{Antonio Vairo,}
\author[a]{Peter Vander Griend,}
\author[e,f]{Johannes Heinrich Weber}

\affiliation[a]{Physik-Department, Technische Universit\"{a}t M\"{u}nchen, James-Franck-Str. 1, 85748 Garching,
Germany}
\affiliation[b]{Institute for Advanced Study, Technische Universit\"{a}t M\"{u}nchen, Lichtenbergstrasse 2 a, 85748
Garching, Germany}
\affiliation[c]{Instituto Galego de F\'{i}sica de Altas Enerx\'{i}as (IGFAE), Universidade de Santiago de Compostela. E-15782, Galicia, Spain}
\affiliation[d]{Department of Physics, Kent State University, Kent, OH 44242, United States}
\affiliation[e]{Department of Computational Mathematics, Science and Engineering, and Department of Physics and Astronomy,
Michigan State University, East Lansing, MI 48824, USA}
\affiliation[f]{Institut f\"ur Physik, Humboldt-Universit\"at zu Berlin \& IRIS Adlershof, D-12489 Berlin, Germany}

\emailAdd{nora.brambilla@ph.tum.de}
\emailAdd{miguelangel.escobedo@usc.es}
\emailAdd{mstrick6@kent.edu}
\emailAdd{vandergriend@tum.de}
\emailAdd{antonio.vairo@tum.de}
\emailAdd{johannes.weber@physik.hu-berlin.de}

\abstract{We solve the Lindblad equation describing the Brownian motion of a Coulombic heavy quark-antiquark pair 
in a strongly coupled quark-gluon plasma using the highly efficient Monte Carlo wave-function method.
The Lindblad equation has been derived in the framework of pNRQCD and fully accounts for the quantum and non-Abelian nature of the system.
The hydrodynamics of the plasma is realistically implemented through a 3+1D dissipative hydrodynamics code.
We compute the bottomonium nuclear modification factor and compare with the most recent LHC data.  
The computation does not rely on any free parameter, as it depends on two transport coefficients that have been evaluated independently in lattice QCD.
Our final results, which include late-time feed down of excited states, agree well with the available data from LHC 5.02 TeV PbPb collisions.}

\begin{document}

\maketitle
\flushbottom

\section{Introduction}
\label{sect:intro}
The aim of relativistic heavy-ion collisions is to recreate and study the quark-gluon  plasma (QGP),
a primordial state of matter that existed microseconds after the Big-Bang.
In such a state quarks and gluons are not confined within hadrons and, in the limit of high temperatures, they behave almost as free particles.
The study of the QGP is very challenging due to its short life time: 
we can only infer its properties  from the way it affects particles that are detected at the end after freeze-out.

Among the hard probes of the QGP there is quarkonium suppression. 
The original idea was put forward by Matsui and Satz in~\cite{Matsui:1986dk} who assumed the quarkonium interaction to be screened in the
hot QGP leading to a suppression in the number of quarkonia.
Quarkonia would eventually be detected through their decays into muon-antimuon pairs.
In this scenario, the observation of quarkonium suppression is a signal of QGP formation and, by measuring its  strength, it is possible to learn information about the medium.
Such suppression is quantified in the quarkonium nuclear modification factor  $R_{AA}$.
It is this quantity, among others, that is measured in heavy ion collision experiments at LHC and at RHIC.

The screening mechanism underlying the idea of Matsui and Satz originates from the chromoelectric screening of the medium.
It sets in at the momentum scale given by the Debye mass, $m_D$.
Hence, the typical screening distance for a heavy quark-antiquark pair ($Q\bar{Q}$) is of order $r\sim 1/m_D$ or larger.
Sequential screening as a function of the radius of the quarkonium state is a consequence of this mechanism.
This paradigm was challenged when the quark-antiquark potential in the medium was first calculated
in weak coupling in the screening regime $r\sim 1/m_D$~\cite{Laine:2006ns}.
The calculation showed that, in addition to the screening of the real part, the potential also posseses an imaginary part.
The imaginary part also leads to quarkonium dissociation, which turns out to 
happen at a temperature lower than the screening one~\cite{Brambilla:2008cx,Escobedo:2008sy,Beraudo:2007ky,Margotta:2011ta},
i.e., at least in weak coupling, quarkonium has already dissociated when reaching the screening temperature. 
Since then, many  phenomenological studies solving the Schr\"{o}dinger equation with a complex potential
have appeared~\cite{Strickland:2011mw,Strickland:2011aa,Krouppa:2015yoa,Krouppa:2016jcl,Krouppa:2017jlg,Bhaduri:2018iwr,Boyd:2019arx,Bhaduri:2020lur,Islam:2020gdv,Islam:2020bnp}.

Nonrelativistic  Effective  Field  Theories (NREFTs) allow one to appropriately define the heavy quark-antiquark  potential
and supply a scheme for the systematic calculation of quarkonium properties.
They exploit the separation of  energy scales characteristic of nonrelativistic bound states.
At zero temperature, the energy scales are the heavy-quark mass, $m$, the inverse of the Bohr radius of the bound state
$1/a_0 \sim mv$, and the binding energy $E \sim mv^2$, where  $v\ll  1$ is the relative quark velocity in the bound state.
The EFT that is obtained by integrating out degrees of freedom associated with the scale $m$ is Non Relativistic QCD
(NRQCD)~\cite{Caswell:1985ui,Bodwin:1994jh} and the  EFT obtained by integrating out gluons with momentum or energy scaling like the
inverse of the Bohr radius is potential NRQCD (pNRQCD)~\cite{Pineda:1997bj,Brambilla:1999xf,Brambilla:2004jw}.
At leading order in $v$, the equation of motion of pNRQCD is the quantum mechanical Schr\"odinger equation for a
nonrelativistic bound state. Differently from a pure quantum mechanical treatment of the bound state, 
however, pNRQCD provides an unambiguous field theoretical definition of the potential.  
The potential encodes contributions coming from modes with energy and momentum above the scale of the binding energy.
Moreover, pNRQCD adds systematically to the leading order Schr\"odinger equation higher order corrections.
The first one in a weak coupling regime is carried by chromoelectric dipole terms.

In the last decade, pNRQCD has been  applied also to study quarkonium
at finite temperature~\cite{Brambilla:2008cx,Escobedo:2008sy,Brambilla:2010vq}.
At finite temperature more scales are relevant, for instance the temperature $T$ and, at weak coupling, the Debye mass $m_D$.
Nevertheless, at leading order in $v$ the equation of motion of pNRQCD is still a Schr\"odinger equation, which 
describes the real time evolution of the $Q\bar{Q}$ pair in the medium.
The potential encodes now also thermal contributions if there are thermal modes associated with energy scales larger than $mv^2$.
The thermal part of the potential has a real part (well described in the weak coupling regime by the singlet free energy in Coulomb gauge~\cite{Berwein:2017thy}) and an imaginary part.
The real part of the potential is screened only if $m_D$ is of the order of the inverse of the Bohr radius.
If it is smaller, then the potential gets at most thermal corrections that are power like in $T$.
It is in this regime, and not in the screened one, that dissociation happens at weak coupling due to the imaginary part
of the potential being as large as the real one~\cite{Laine:2006ns}.
At one loop, the imaginary part is a consequence of two distinct phenomena:
Landau damping~\cite{Laine:2006ns,Escobedo:2008sy,Brambilla:2008cx}, an effect that exists also in QED,
and singlet-octet transitions, which are specific of QCD~\cite{Brambilla:2008cx}.
The Landau damping originates from the inelastic scattering of the heavy quark or antiquark
with the partons in the medium~\cite{Brambilla:2013dpa}, while the singlet to octet transition
originates from the gluodissociation of quarkonium~\cite{Brambilla:2011sg}. 
Which phenomenon dominates depends on the ratio  between the scales $m_D$ and $E$.
These findings in the EFT, mostly in the weak coupling regime, have inspired several subsequent nonperturbative
calculations of the  static potential at finite  $T$. In  particular, the  Wilson loop  at finite  $T$ has  been computed
on the  lattice~\cite{Rothkopf:2011db,Rothkopf:2019ipj} finding  hints of a large  imaginary part.
These calculations  are rather  challenging  and  refinements   of  the  extraction  methods  are
currently in development~\cite{Petreczky:2018xuh,Bala:2019cqu}.

Quarkonium scattering in the medium, quarkonium dissociation into an unbound color octet $Q\bar{Q}$ pair,
and the inverse processes of $Q\bar{Q}$ pair generation call for an appropriate framework to describe the quarkonium non-equilibrium evolution in the QGP:
the open quantum system framework (OQS) (see~\cite{Akamatsu:2020ypb} for a review and~\cite{Akamatsu:2014qsa} for a seminal paper).
The system is in non-equilibrium because through interaction with the environment
color singlet and octet $Q\bar{Q}$ states continuously transform into each other although the total number of heavy quarks is conserved.
In~\cite{Brambilla:2016wgg,Brambilla:2017zei,Brambilla:2019tpt}, an OQS framework rooted in pNRQCD
has been developed that is fully quantum, conserves the number of heavy quarks and takes into account
both the color singlet and the color octet $Q\bar{Q}$ degrees of freedom.
In this framework, the QGP plays the role of the environment characterized by a scale $\pi T$
and the quarkonium is the system characterized by the scale $E$.
The inverse of the energy can  be identified  with the intrinsic time scale of the system, $\tau_S\sim 1/E$, 
and the inverse of $\pi T$ with the correlation time of the environment, $\tau_E\sim 1/(\pi T)$.
If the medium is in thermal equilibrium, or locally in thermal equilibrium, we may understand $T$ as a temperature,
otherwise it is just the inverse of the correlation time of the environment. The medium can be strongly coupled.
The evolution of the system is characterized by a relaxation time $\tau_R$ that is proportional to the inverse
of the $Q\bar{Q}$ self-energy in pNRQCD, i.e. $\tau_R\sim 1/[a_0^2(\pi T)^3]$.
Under the condition that the quarkonium has a small radius (i.e. that it is a Coulombic bound state) such that $1/a_0 \gg \pi T, \Lambda_{\rm QCD}$, and that $\pi T \gg E$,
a set of master equations governing the time evolution of the heavy $Q\bar{Q}$ pairs in the medium has been derived in~\cite{Brambilla:2016wgg,Brambilla:2017zei}.
The equations express the time evolution of the density matrices  of the  color singlet and color octet $Q\bar{Q}$ states.
They account for the mass shift of the heavy $Q\bar{Q}$ pair induced by the medium,
the decay widths induced by the medium, the generation of $Q\bar{Q}$ color singlet states from $Q\bar{Q}$ color octet states
interacting with the medium and the generation of $Q\bar{Q}$ color  octet states from $Q\bar{Q}$
(color singlet or octet) states interacting with the medium.
At leading order the interaction between a heavy $Q\bar{Q}$ field and the  medium is encoded in pNRQCD in a chromoelectric dipole interaction.

The master equations are, in general, non Markovian.
They become Markovian if $\tau_R \gg \tau_E$, while  the condition $\tau_S\gg \tau_E$ qualifies the regime as a quantum Brownian motion.
Under these conditions, the master equations assume a Lindblad form~\cite{Lindblad:1975ef,Gorini:1975nb}.
If we further consider the medium isotropic and the quarkonium at rest with respect to the medium,
in the large time limit the Lindblad equation for a strongly coupled medium turns out to depend, remarkably, on only two transport coefficients:
the heavy quark momentum diffusion coefficient, $\kappa$, and its  dispersive counterpart $\gamma$.
Both coefficients have a field theoretical definition:
they are given by time integrals of gauge invariant correlators of chromoelectric
fields~\cite{Brambilla:2016wgg,Brambilla:2017zei,Brambilla:2019tpt}. 
These coefficients can be evaluated nonperturbatively within QCD at finite $T$ regularized on the lattice.\footnote{
  It is remarkable that the OQS/pNRQCD framework allows to use input coming from a lattice QCD calculation
  of a quantity in thermal equilibrium to describe the out of equilibrium evolution of quarkonium in the medium.}
Recently, a quenched lattice QCD evaluation of $\kappa$ in an unprecedented large window of temperatures
has been completed displaying a significant temperature dependence~\cite{Brambilla:2020siz}.
Similar studies are ongoing for $\gamma$. Unquenched values for both parameters have been obtained in~\cite{Brambilla:2019tpt}.
Once the Lindblad equation has been solved and evolved up to freeze out time, the result can be used to compute
observables like $R_{AA}$ and $v_2$ by projecting on the quarkonium states of interest.
These observables can be compared with data from the LHC~\cite{Brambilla:2016wgg,Brambilla:2017zei}. 
We are, therefore, in a position to determine much needed information about the properties of the QGP by solving 
evolution equations for the out of equilibrium  dynamics of the heavy quarkonium in the QGP 
that have been derived in a controlled and systematic fashion from QCD.
However, for this purpose we need to develop an efficient algorithm to solve the resulting dynamical equations
and we need to couple them properly to the hydrodynamical evolution of the medium.

In this work, we revisit the solution of the evolution equations found in~\cite{Brambilla:2016wgg,Brambilla:2017zei}
with the aim of determining $R_{AA}$ and comparing with experimental data.
We extend this previous work by relaxing several approximations that were implemented in the numerical solution of the Lindblad equation
due to the high computational cost associated with it.
The main goal of this paper is to present a new method for solving the Lindblad equation that substantially increases numerical efficiency through massive parallelization.
In addition, the new framework allows for realistic hydrodynamical evolution of the medium.

Solving the Lindblad equation is challenging, the reason being that, for a Hilbert space of dimension $N$, one needs to compute the evolution of a density matrix with $N^2$ entries.
In the case of quarkonium, the Hilbert space has infinite  dimensions.
For numerical  studies, quarkonium is simulated using a finite size lattice.
The main approximations that were used in~\cite{Brambilla:2016wgg,Brambilla:2017zei} were the following.
{\it (i)} A lattice of size 40 times the Bohr radius of the $\Upsilon(1S)$, $a_0$, with a spacing of $0.1a_0$ was used.
Note that doubling the size of the lattice makes solving the Lindblad equation four times more expensive.
{\it (ii)} An expansion in spherical harmonics was used and only $S$-wave and $P$-wave states were considered. 
{\it (iii)} For each centrality window only a simulation with the average temperature of the window was performed.
Moreover, a temperature profile given by boost-invariant ideal Bjorken evolution was used for its simplicity and its analytic closed form.

In the present study, we relax these approximations by using the Monte Carlo wave-function method~\cite{Dalibard:1992zz}.
Other methods exist and have been used to simplify the computation of the Lindblad equation in the context of quarkonium suppression~\cite{Miura:2019ssi,Sharma:2019xum}.
However, for reasons  that we will explain later, we believe that the Monte Carlo wave-function
method is more suitable for quarkonium studies, especially when color degrees of freedom are taken  into account.
The method reduces to simulating the evolution of an ensemble of vector states that, on average, behave like the density matrix.
The vector states evolve either according to an  effective non-Hermitian Hamiltonian or by quantum jumps at  random times.
In our case, the Hamiltonian does not mix states with different color or orbital angular momentum.
Therefore, the  most numerically demanding task of the algorithm is to solve a Schr\"{o}dinger equation in one  dimension
and this even when all possible orbital angular momenta are taken into account.
In  summary,  the Monte  Carlo wave-function  method allows to decrease the lattice spacing and increase the volume without
increasing the numerical cost as much as it would require directly solving the Lindblad equation and, additionally,
without truncating the spherical harmonics as was the case in~\cite{Brambilla:2016wgg,Brambilla:2017zei}.

We believe that our method may be useful also for similar phenomenological applications.
The semiclassical limit of the Lindblad equation has been studied in~\cite{Blaizot:2017ypk}
and the relevance of correlated versus non correlated noise in~\cite{Sharma:2019xum}.
In~\cite{Yao:2018nmy, Yao:2020xzw,Yao:2020eqy}, using the same pNRQCD and OQS framework that we use here and a specific
scale hierarchy, transport equations and, in particular, a semiclassical Boltzmann equation, 
have been obtained for the evolution of quarkonium in medium.
For the differential  reaction  rate, the  information about the QGP is contained
in a chromoelectric gluon correlator that involves also staple-shaped Wilson lines.
Similar correlators show up at $T=0$ in the gluon parton distribution functions,
the gluon transverse momentum dependent parton distribution functions 
and in the quarkonium production cross section expressed in pNRQCD~\cite{Brambilla:2020ojz}.
Other applications  of   the  OQS framework that do not use the EFT approach can be found
in~\cite{Akamatsu:2011se,Blaizot:2015hya,Katz:2015qja,Blaizot:2017ypk,Blaizot:2018oev}.

The outline of the paper is as follows.
In section \ref{sect:theory}, we review the Monte Carlo wave-function method and how it is implemented in order to compute the quarkonium evolution.
In section \ref{sect:initstate}, we discuss  the initial  conditions.
In section \ref{sect:codetests}, we present our results and compare with what was previously obtained by directly solving the Lindblad equation.
In section \ref{sect:hydro}, we outline  the hydrodynamical evolution used for the QGP background.
In section \ref{sect:results}, we compare with the data collected at LHC and finally,
in section  \ref{sec:conclusions}, we give our conclusions.

\section{The quantum trajectories algorithm}
\label{sect:theory}

\subsection{The Monte Carlo wave-function method}
\label{ssec:MCWF}
In this paper, we focus on the evolution of a heavy quark-antiquark pair that follows the GKSL or Lindblad equation \cite{Gorini:1975nb,Lindblad:1975ef}
derived in \cite{Brambilla:2016wgg,Brambilla:2017zei} in the regime in which all the thermally induced energy scales are much smaller than the inverse Bohr radius and much larger than the binding energy $E$.
The general form of the Lindblad equation is 
\begin{equation}  \frac{d\rho}{dt}=-i[H,\rho]+\sum_n \left(  C_n\,\rho\,C_n^\dagger-\frac{1}{2}\{C_n^\dagger C_n,\rho\} \right)\,,
\label{eq:Lindblad}
\end{equation}
where $\rho$ is the reduced density matrix, $H$ is  a Hermitian  operator and  the operators $C_n$'s are called collapse operators.
The specific values of $H$ and $C_n$ in our case will be discussed later.

In \cite{Brambilla:2016wgg,Brambilla:2017zei}, eq.~\eqref{eq:Lindblad} was solved numerically by expanding in spherical harmonics,
keeping only $S$-wave and $P$-wave states and discretizing the radial component on a lattice.
This is a very computationally demanding  procedure that limits phenomenological applications.
This is a common problem when solving the Lindblad  equation, which is due to the fact that the size of the density matrix scales with  the square of the number of degrees of freedom.
In our case, for example, this means that  doubling the lattice size implies multiplying by four the computational  cost.
In the literature, several techniques have  been developed to tackle this problem, and they are named \textit{master equation unraveling} (see \cite{Breuer:2002pc} and references therein).
One such unraveling, the Quantum  State  Diffusion method \cite{Gisin:1992xc}, has been already applied to the study of quarkonium,
see for example the recent papers \cite{Miura:2019ssi,Sharma:2019xum}.
However, we believe that a more robust method is provided by the Monte Carlo wave-function (MCWF) method  \cite{Dalibard:1992zz}.
The reason is that it makes a more efficient use  of the symmetries of the evolution of quarkonium, particularly when color degrees of freedom are taken into account.
For the case of the equations derived in \cite{Brambilla:2016wgg,Brambilla:2017zei}, it allows to take into account all possible orbital angular momenta
using only a one dimensional lattice in the numerical computation.

Let us now review the basis of the MCWF method. First, we define a partial decay width operator as
\begin{equation}
\Gamma_n=C_n^\dagger C_n\,.
\label{eq:width}
\end{equation}
Then, the total decay width is simply given by $\Gamma=\sum_n \Gamma_n$.
We can also define an effective Hamiltonian (which is non Hermitian whenever $\Gamma\neq 0$)
\begin{equation}
H_\text{eff}=H-\frac{i}{2}\Gamma\,.
\label{eq:Heff}
\end{equation}
We can now rewrite the Lindblad equation as
\begin{equation}
\frac{d\rho}{dt}=-iH_\text{eff}\rho+i\rho H^\dagger_\text{eff}+\sum_n C_n\,\rho \,C_n^\dagger\,.
\end{equation}
 Note that in QCD $H_\text{eff}$ does not mix states with different color or angular momentum. Now, we consider the evolution of $\rho(t)$ during a small time step $dt$
\begin{equation}
\rho(t+dt) = \rho(t)-iH_\text{eff}\rho(t)dt+i\rho(t)H_\text{eff}^\dagger dt+\sum_n C_n\,\rho(t) \,C_n^\dagger dt\,;
\end{equation}
the density matrix $\rho$ is a Hermitian semi-positive definite matrix with trace equal to one.
Then $\rho(t)=\sum_n p_n|\Psi_n(t)\rangle\langle\Psi_n(t)|$, where $p_n\geq0$ and $\sum_n p_n=1$.
Because the Lindblad equation is linear in the density matrix, if we are able to solve the evolution for an initial condition
that is a pure state then we can solve the evolution for any initial condition.
Therefore, we focus in the present discussion on the case $\rho(t)=|\Psi(t)\rangle\langle\Psi(t)|$.
\begin{eqnarray}
  \rho(t+dt) & = & (1-iH_\text{eff}dt)|\Psi(t)\rangle\langle\Psi(t)|+i|\Psi(t)\rangle\langle\Psi(t)|H_\text{eff}^\dagger dt
                   +\sum_n C_n\,|\Psi(t)\rangle\langle\Psi(t)| \,C_n^\dagger dt\nonumber \\
& = & (1-\langle\Psi(t)|\Gamma|\Psi(t)\rangle dt) \, \frac{(1-iH_\text{eff}dt)|\Psi(t)\rangle\langle\Psi(t)|(1+iH_\text{eff}^\dagger dt)}{1-\langle\Psi(t)|\Gamma|\Psi(t)\rangle dt} \nonumber \\
& + & \sum_n\langle\Psi(t)|\Gamma_n|\Psi(t)\rangle dt \, \frac{C_n\,|\Psi(t)\rangle\langle\Psi(t)| \,C_n^\dagger dt}{\langle\Psi(t)|\Gamma_n|\Psi(t)\rangle dt}+\mathcal{O}(dt^2)\,.
\end{eqnarray}
We can understand the previous equation in the following way.
\begin{itemize}
\item[(a)]
  With probability $1-\langle\Psi(t)|\Gamma|\Psi(t)\rangle dt$ the wave-function follows the evolution
  $$ \frac{(1-iH_\text{eff}dt)|\Psi(t)\rangle}{\sqrt{1-\langle\Psi(t)|\Gamma|\Psi(t)\rangle dt}}.$$
  This evolution can be computed by performing the operation $(1-iH_\text{eff}dt)|\Psi(t)\rangle$ and then normalizing the result.
  \item[(b)]
    With probability $\langle\Psi(t)|\Gamma_n|\Psi(t)\rangle dt$ the wave-function makes what is called a quantum jump, and transforms into
    $$\frac{C_n\,|\Psi(t)\rangle}{\sqrt{\langle\Psi(t)|\Gamma_n|\Psi(t)\rangle}}.$$
    Similarly to the case of no jump, the evolution can be computed by performing the operation $C_n\,|\Psi(t)\rangle$ and then normalizing the result.
\end{itemize}
We see that the evolution of a density matrix that fulfils the Lindblad equation is equivalent to making the previously discussed operations
to a wave-function with the corresponding probability.
This can potentially reduce the numerical cost of the simulation of the evolution because,
although the algorithm involves averaging over many trajectories,
the cost of computing each trajectory scales like $N$ (the number of entries in the wave-function), while directly solving the Lindblad equation scales like $N^2$.
There is an additional overhead due to the requirement of sampling a large enough number of trajectories, which however does not vary strongly with $N$, see e.g. appendix~\ref{sec:finite_size}, where the box size $N$ has been varied by a factor two with the same number of trajectories\footnote{Similar observations have been made in other applications of quantum trajectories, see e.g. Ref.~\cite{PhysRevA.97.022116}.}.
How many trajectories are a large enough number may depend on the state in question. 
Generally more trajectories are necessary for reliably sampling final states whose $l$ value is further away from the $l$ value of the initial state.
This property is shared with the Quantum State Diffusion method, however,
the advantages of the MCWF will become clearer when we review the specific form of the Lindblad equation derived in \cite{Brambilla:2016wgg,Brambilla:2017zei}.

\subsection{Quarkonium evolution in the regime $1/a_0 \gg  T,m_D \gg E$}
\label{ssec:hq}
Here we review the evolution equation obtained from pNRQCD in \cite{Brambilla:2016wgg,Brambilla:2017zei} in the regime $1/a_0 \gg T,m_D\gg E$ for Coulombic quarkonia.
The evolution equation has the Lindblad form of eq. \eqref{eq:Lindblad}.
In our case, $\rho$ is the reduced density matrix of the heavy quark-antiquark pair.
We assume it to be block diagonal in the color degrees of freedom such that 
\begin{equation}
\rho=\left(\begin{array}{c c}
\rho_s & 0 \\
0 & \rho_o\end{array}\right)\,.
\label{eq:rhoso}
\end{equation}
The symmetries of the evolution equations ensure that if we start with a density matrix with this structure at some given time $t_0$
it will keep the same structure during all the evolution; $\rho_s$ and $\rho_o$ are formally operators in a continuous Hilbert space
labeled by the relative coordinate ${\bf r}$.
The probability for the quarkonium to be in a specific color state is $p_{s,o}= {\rm Tr}(\rho_{s,o})$. 

The Hamiltonian is given by 
\begin{equation}
H = \left(\begin{array}{c c}
h_s & 0\\
0 & h_o
\end{array}\right)
+ \frac{r^2}{2}\,\gamma\, 
\left(\begin{array}{c c}
1 & 0\\
0 & \frac{N_c^2-2}{2(N_c^2-1)}
\end{array}\right)\,,
\label{eq:hamiltonian}
\end{equation}
where $h_{s,o}$ is the in vacuum Hamiltonian of a singlet or an octet in pNRQCD \cite{Pineda:1997bj,Brambilla:1999xf}:
\begin{equation}
h_s = \frac{{\bf p}^2}{m} -C_F\frac{\als}{r},\qquad h_o = \frac{{\bf p}^2}{m} + \frac{\als}{2N_c\,r};
\label{hsoCoul}
\end{equation}
${\bf p}$ is the relative momentum of the $Q\bar{Q}$ pair, $C_F = (N_c^2-1)/(2N_c)$ and $N_c$ is the number of colors.
We assume the quarkonium under examination to be a Coulombic bound state,
hence the potential in \eqref{hsoCoul} is the Coulomb potential in the color singlet (attractive) and color octet (repulsive) representation.
The coefficient $\gamma$ is obtained by matching pNRQCD with QCD; it can be expressed as the imaginary part of the integral of a chromoelectric correlator: 
\begin{equation}
\gamma  \equiv  \frac{g^2}{6\,N_c} \, {\rm Im}\int_{-\infty}^{+\infty}dt \, \langle T\,E^{a,i}(t,{\bf 0}) E^{a,i}(0,{\bf 0})\rangle
= -i \frac{g^2}{6\,N_c} \, \int_0^{\infty} dt \, \langle [E^{a,i}(t,{\bf 0}), E^{a,i}(0,{\bf 0})]\rangle\,,
\label{gamma}
\end{equation} 
where $\langle \cdots \rangle$ stands for the in medium average and $T$ for time ordering. 
The field ${\bf E}(t,{\bf 0})$ should be understood as in~\cite{Brambilla:2016wgg,Brambilla:2017zei}, i.e., 
as $\Omega^\dagger(t) \, {\bf E}(t,{\bf 0}) \, \Omega(t)$, where now ${\bf E}(t,{\bf 0})$ stands for the usual chromoelectric field in QCD
and $\Omega(t)$ is a Wilson line going from $-\infty$ to $t$: $\displaystyle \Omega(t) = \exp\left[ -ig \int_{-\infty}^t d\lambda \, A_0(\lambda,{\bf 0}) \right]$.
The Wilson lines ensure that $\gamma$ is gauge invariant.

Regarding the collapse operators, there are six of them: $C^0_i$ and $C^1_i$ (the subindex $i$ corresponds to the spatial directions and may assume the values 1,2,3).
They read
\begin{equation}
C^0_i=\sqrt{\frac{\kappa}{N_c^2-1}}\,r^i\left(\begin{array}{c c}
0 & 1\\
\sqrt{N_c^2-1} & 0
\end{array}\right)\,,
\label{eq:c0}
\end{equation}
\begin{equation}
C^1_i=\sqrt{\frac{(N_c^2-4)\kappa}{2(N_c^2-1)}}\,r^i\left(\begin{array}{c c}
0 & 0\\
0 & 1
\end{array}\right)\,.
\label{eq:c1}
\end{equation}
The coefficient $\kappa$ is the heavy quark momentum diffusion coefficient ~\cite{CasalderreySolana:2006rq,CaronHuot:2007gq}.
Like $\gamma$ it is obtained by matching pNRQCD with QCD; it can be expressed as the real part of the integral of a chromoelectric correlator:
\begin{equation}
\kappa  \equiv  \frac{g^2}{6\,N_c} \,  {\rm Re} \int_{-\infty}^{+\infty}dt \, \langle T\,E^{a,i}(t,{\bf 0}) E^{a,i}(0,{\bf 0})\rangle
= \frac{g^2}{6\,N_c} \, \int_0^{\infty} dt \, \langle \{E^{a,i}(t,{\bf 0}), E^{a,i}(0,{\bf 0})\}\rangle\,.
\label{kappa}
\end{equation} 
The chromoelectric fields are defined as in the case of $\gamma$, hence also the above expression of $\kappa$ is gauge invariant.
Both $\gamma$ and $\kappa$ depend on the medium and, because the medium is evolving, on time.
In the case of a medium that is in local thermal equilibrium, we can define a temperature and encode in it the time dependence of $\gamma$ and $\kappa$.
How the temperature depends on time relies on the hydrodynamical description of the medium.
Under the assumption of local thermal equilibrium, lattice determinations of $\gamma$ and $\kappa$ at different temperatures coupled to the
hydrodynamical evolution of the medium can, therefore, describe how these coefficients evolve in time.

The collapse operators define two partial decay widths (see eq. \eqref{eq:width})
\begin{equation}
\Gamma^0=\kappa r^i\left(\begin{array}{c c}
1 & 0\\
0 &\frac{1}{N_c^2-1}
\end{array}\right)r^i\,,
\end{equation}
and
\begin{equation}
\Gamma^1=\frac{\kappa(N_c^2-4)}{2(N_c^2-1)}r^i\left(\begin{array}{c c}
0 & 0\\
0 & 1
\end{array}\right)r^i\,.
\end{equation}
Their sum gives the total decay width,
\begin{equation}
\Gamma=\kappa r^i\left(\begin{array}{c c}
1 & 0\\
0 & \frac{N_c^2-2}{2(N_c^2-1)}\end{array}\right)r^i\,.
\end{equation}

Let us now discuss the general structure of the density matrix regarding the orbital angular momentum.
We can always expand the density matrix in terms of spherical harmonics: 
\begin{equation}
\rho^{lm;l'm'}=\int\,d\Omega(\hat{r})\,d\Omega(\hat{r}') \, Y^{lm}(\hat{r}) \, \rho \, {Y^{l'm'}}^*(\hat{r}')\,.
\end{equation}
As stated above, we assume the density matrix to be block diagonal with respect to color (see eq. \eqref{eq:rhoso}).
We further assume that the same happens regarding the orbital angular momentum quantum numbers $l$ and $m$.
Hence, we only need to consider the case $l'=l$ and $m'=m$. Moreover, spherical symmetry enforces that all polarizations are equally possible.
If we define 
\begin{equation}
\rho^l \equiv \sum_m\rho^{lm;lm}\,,
\end{equation} 
the reduced density matrix can be written as $\rho=\sum_l \rho^l$ and we can understand ${\rm Tr}(\rho^l)$ as the probability that the value of the angular momentum squared is $l(l+1)$.
The main difference with the studies in \cite{Brambilla:2016wgg,Brambilla:2017zei} is that in this work, thanks to the application of the MCWF method,
we do not need to truncate the sum of the quantum number $l$, while in previous works only $S$-wave and $P$-wave states were considered. 

In summary, the MCWF algorithm applied to the case of quarkonium evolution in the $1/a_0\gg T,m_D\gg E$ regime consists of the following steps:
\begin{enumerate}
\item Write the density matrix at the initial time as a sum of pure states
  $$ \rho(t_0)= \sum_{n} \, p_n |\Psi_n(t_0)\rangle\langle\Psi_n(t_0)|.$$
  The symmetries of the problem ensure that each of these pure states will have a well defined color and angular momentum. 
\item With probability $p_n$ take $|\Psi_n(t_0)\rangle$ as the initial condition.
\item \label{i:step} For the first time step, follow the recipe given in section \ref{ssec:MCWF} to determine whether there is no jump or,
  if there is a jump, which collapse operator has to be applied.
\begin{enumerate}
\item If there is no jump, the evolution is given by a Schr\"{o}dinger equation. Since neither $|\Psi_n(t_0)\rangle$ nor $H_{\rm eff}$ mix different colors or angular momenta,
  this is a problem that can be solved numerically using a 1D lattice.
\item If there is a jump, the collapse operators can change color or angular momenta.
  However, in the case of the Lindblad equation considered here this is done in such a way that the resulting state
  also has a well defined color and angular momentum (although different from the previous one).
  The probability of transition between different angular momentum states are discussed in appendix \ref{Ap:om}.
  Regarding color, a color singlet $Q\bar{Q}$ state always jumps to a color octet $Q\bar{Q}$ state.
  In the case that the state that jumps is an octet, it has a $2/(N_c^2-2)$ chance to jump to a singlet.\footnote{
    If $|\Psi_o\rangle$ is the wave-function of an octet state, then $\frac{\langle\Psi_o|\Gamma^0|\Psi_o\rangle}{\langle\Psi_o|\Gamma|\Psi_o\rangle}=\frac{2}{N_c^2-2}$.\label{footoctet}}  
\end{enumerate}
\item Repeat step \ref{i:step} for each time step until the end of the evolution.
\end{enumerate}

\subsection{The waiting time approach}
In the numerical implementation of the algorithm, we use an approach that we call the waiting time approach
(see section IIID of \cite{Daley:2014fha} and references therein).
Let us consider the case in which the initial wave-function is $|\Psi(t_0)\rangle$.
The jump rate at time $t$ is
\begin{align}
p_{\rm jump}(t)&=\langle \Psi(t_0)|e^{i\int_{t_0}^t\,dt'H_\text{eff}^\dagger(t')}\Gamma(t) e^{-i\int_{t_0}^t\,dt'H_\text{eff}(t')}|\Psi(t_0)\rangle\nonumber\\
&=-\frac{d}{dt}\langle \Psi(t_0)|e^{i\int_{t_0}^t\,dt'H_\text{eff}^\dagger(t')} e^{-i\int_{t_0}^t\,dt'H_\text{eff}(t')}|\Psi(t_0)\rangle,
\end{align}
which means that at the time $t$ the norm of the state evolving according to $H_{\text{eff}}$ is equal to the probability of no jumps up to the time $t$.
With this in mind we can make a more efficient algorithm to compute one trajectory.
\begin{enumerate}
\item Given an initial wave-function $|\Psi(t_0)\rangle$, we generate a random number between $0$ and~$1$. 
\item We evolve the wave-function with the effective Hamiltonian $H_{\text{eff}}$ until the norm is equal to the random number or smaller. 
\item We compute the effect of the jump. 
\item We repeat the process with the wave-function resulting from the jump. 
\end{enumerate}
In this way, we only need to generate a few random numbers per trajectory.

\subsection{Algorithm for updating the wave-function between quantum jumps}
For the evolution of the wave-function between jumps we use a split-step pseudospectral method \cite{Fornberg:1978,TAHA1984203,Boyd:2019arx}.
Due to the spherical symmetry of the underlying potentials we can compute the evolution of a state with a given angular momentum quantum number $l$ in a color representation $c$ using 
\begin{equation}
u(r,t + \Delta t) = \exp(- i H_{l,c} \Delta t) u(r,t) \, ,
\label{eq:uUpdate}
\end{equation}
where $u(r,t) = rR(r,t)$ with $R(r,t)$ being the radial part of the wave-function.

The Hamiltonian operator $H_{l,c}$ contains the angular momentum term and a potential that depends on whether the pair is in a color-singlet or a color-octet configuration.
For the normalization of the various states, we take
\begin{equation}
\int_0^L dr \, u^*(r,t) u(r,t) = 1 \, ,
\label{eq:normcond}
\end{equation}
with $L$ being the upper bound on the radius in the simulation.
\paragraph*{}

To enforce the boundary condition on $u(r,t)$ at the origin, we use real-valued Fourier sine series in a domain $r \in (0,L]$ to describe both the real and imaginary parts of the wave-function.
To perform the update specified in eq.~\eqref{eq:uUpdate} we split the Hamiltonian into kinetic and potential contributions $H_{l,c} = T +  V_{l,c}$
and use the Baker--Campbell--Hausdorff theorem to approximate, using a split-step decomposition, the time evolution operator as
\be
\exp(- i H_{l,c} \Delta t) = \exp(- i V_{l,c} \Delta t/2) \exp(- i T \Delta t)  \exp(- i V_{l,c} \Delta t/2) + {\cal O}((\Delta t)^2) \, .
\ee
The resulting time evolution steps are 
\begin{enumerate}
\item Update in configuration space using a half-step: $ \psi_1 = \exp(- i V \Delta t/2) \psi_0$.
\item Perform Fourier sine transformations ($\mathbb{F}_s$) on real and imaginary parts separately: $\tilde\psi_1 = \mathbb{F}_s[{\rm Re} \,\psi_1] + i \mathbb{F}_s[{\rm Im} \,\psi_1]$.
\item Update in momentum space using: $\tilde\psi_2 =  \exp\!\left(-i T \Delta t\right) \tilde\psi_1$.
\item Perform inverse Fourier sine transformations ($\mathbb{F}_s^{-1}$) on real and imaginary parts separately: $\psi_2 = \mathbb{F}_s^{-1}[{\rm Re} \,\tilde\psi_2] + i \mathbb{F}_s^{-1}[{\rm Im} \,\tilde\psi_2]$.
\item Update in configuration space using a half-step: $ \psi_3 = \exp(- i V \Delta t/2) \psi_2$.
\item Repeat until the next jump is triggered.
\end{enumerate}
The discrete sine transforms (DST) above can be implemented using standard routines for Fast Fourier Transforms (FFTs).
By repeating this procedure, we can evolve the wave-function forward in time in a manner that is manifestly unitary for real-valued potentials.
For the final results reported herein, we take the momentum-space operator to be $T = p^2/m$ where $m$ is the mass of the heavy quark;
however, for testing we also consider a kinetic energy operator that uses centered discrete differences to compute the second derivative.
In this second case, the momentum-space representation of the kinetic energy operator is of the form $T_{\rm discrete} = 2 [1-\cos(p\Delta r)]/(m\Delta r^2)$.  

We note that besides the improved performance observed, one major benefit of the DST algorithm is that, when dealing with the derivative operator, $T$,
the derivatives are effectively computed using all points in the lattice, not just a fixed number of them.
As a result, the evolution obtained using the DST algorithm is more accurate than the one one would obtain using, for example, a Crank--Nicolson (CN) scheme with a three-point second-derivative \cite{Boyd:2019arx}.
Using the same sized derivative stencil as the default DST algorithm, the CN scheme scales as ${\cal O}(N^2)$ where $N$ is the number of lattice points in the wave-function,
whereas the split-step DST evolution scales as ${\cal O}(N \log N)$~\cite{Boyd:2019arx}.
In order to maximize the calculation speed, we use the CUDA Fast Fourier Transform (CUFFT) library which leverages massively parallel graphics processing units (GPU) to compute the necessary FFTs more efficiently~\cite{cuda}.
As an additional optimization, we have implemented batched FFTs in order to simulate hundreds of quantum trajectories simultaneously, which maximizes GPU utilization.
When run in computing environments containing multiple high-end GPUs, e.g. NVIDIA Tesla P100, the resulting quantum trajectory code allows us to efficiently simulate millions of quantum trajectories in parallel.
The resulting code is called QTraj.

\section{Initial production}
\label{sect:initstate}
In this section, we discuss the initial conditions that we use in order to solve the Lindblad equation.
The authors of \cite{Brambilla:2016wgg,Brambilla:2017zei} assumed that the initial density matrix is diagonal in color and consistent of only $S$-wave states.
We are going to assume the same here, except when studying the evolution with initial $P$-wave states.
In \cite{Brambilla:2016wgg,Brambilla:2017zei}, it was additionally assumed that the initial wave-function was a delta function in coordinate space.
The physical reason is the following. The production of a heavy quark-antiquark pair is a process that involves energies of the order of $m$ or higher.
In coordinate space this means that the process is localized in a region with a radius of the order of $1/m$ or smaller.
This is tiny compared to the size of the typical wave-function of a bound state, which is related with the Bohr radius $a_0\sim 1/(mv)$.
This is the same reason why in NRQCD studies of production and decay the leading order result for $S$-waves are only sensitive to the wave-function of the bound state at the origin.

However, using a delta function for the initial state in the MCWF method is problematic.
The reason is that a wave-function proportional to a delta function cannot be properly normalized.
A solution consists in regularizing the delta function on a finite lattice~\cite{Laine:2007gj}.
However, this solution is still problematic when using the MCWM.
The reason is that the delta function regularized on a finite lattice contains momenta much higher that the ones that are relevant for the bound state physics.
The wave-function of high energy modes grows faster than that of low energy ones and, since the decay width increases with distance, higher energy modes undergo more quantum jumps.
Starting with a delta function regularized on a finite lattice generates many trajectories of high momentum that jump several times and have no impact on quarkonium physics.
This leads to the fact that the number of trajectories that need to be simulated to obtain precise results becomes extremely large.
The situation gets worse as the lattice spacing is reduced since even higher momentum modes enter in the delta function.

\begin{figure}[ht]
\begin{center}
\includegraphics[width=0.495\linewidth]{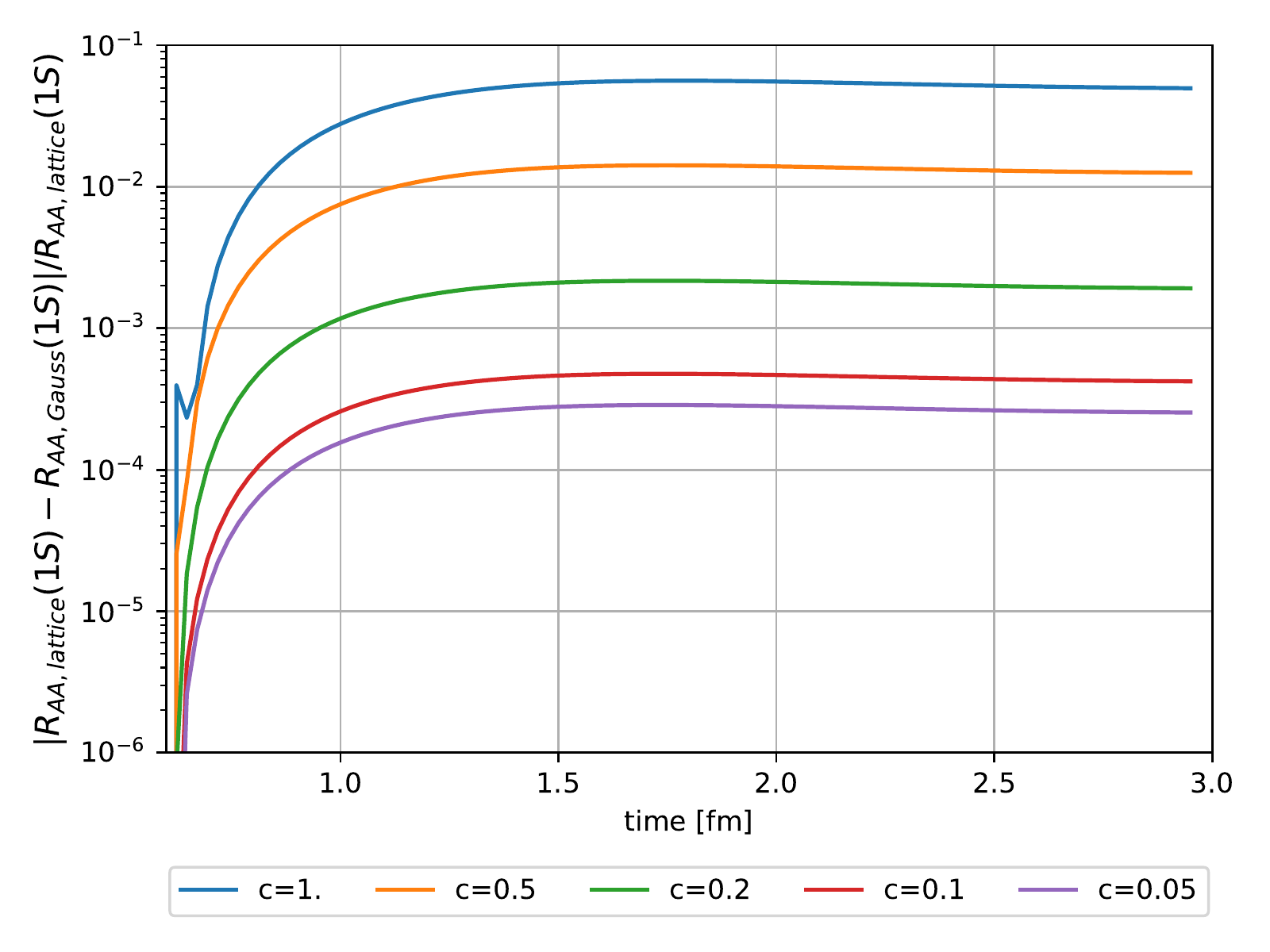}
\includegraphics[width=0.495\linewidth]{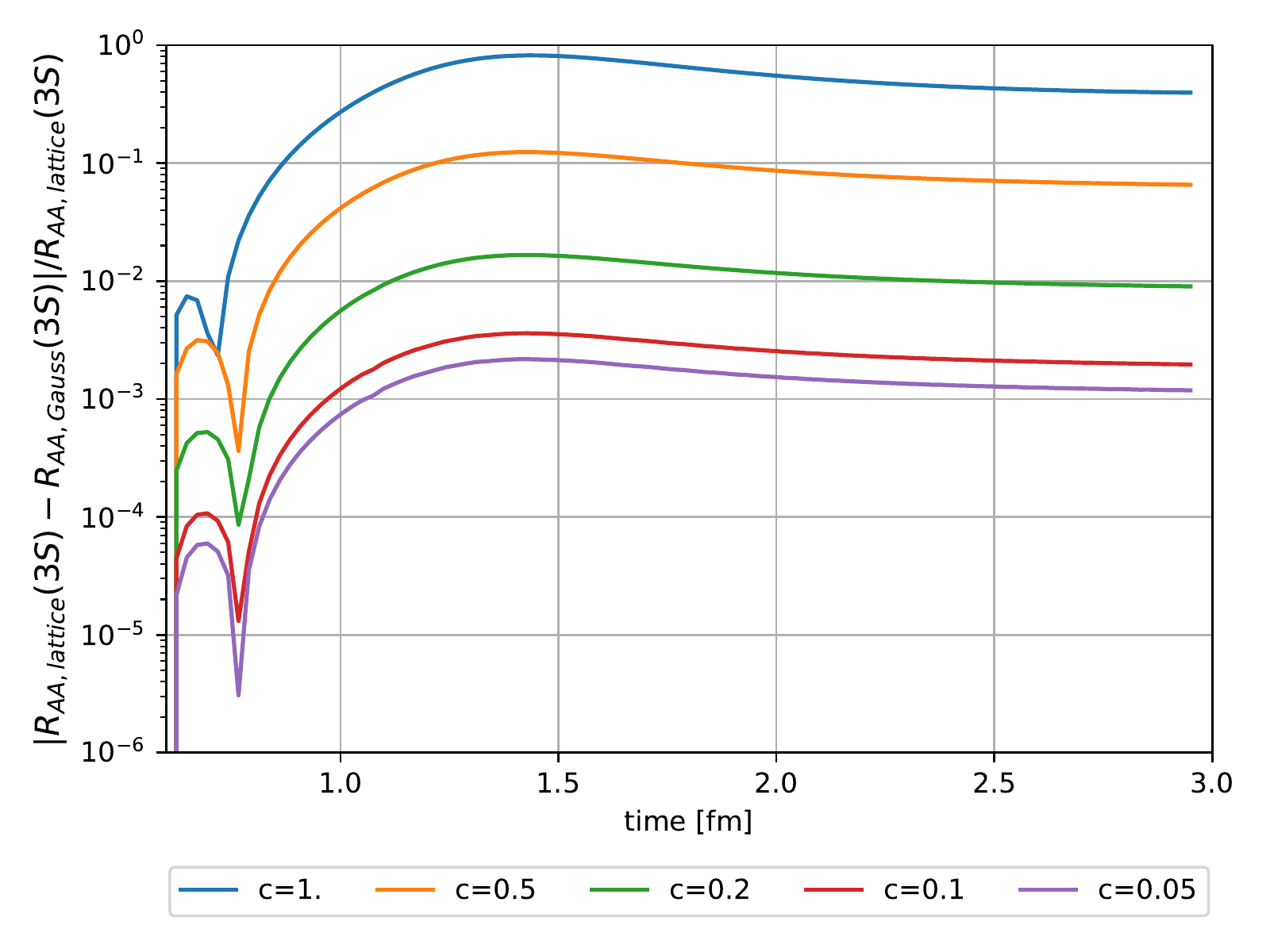}
\end{center}
\caption{Relative difference in $R_{AA}$, computed as in \cite{Brambilla:2016wgg,Brambilla:2017zei}, when using a Gaussian regulated delta function
  and a finite lattice regulated delta function defined according to~\cite{Laine:2007gj}.}
\label{fig:width_study}
\end{figure}

To overcome the above issues, we use an initial wave-function in which very high momentum modes are cut off, but whose size is still much smaller than the Bohr radius $a_0$.
This can be achieved using a Gaussian with a small width instead of a delta function.
To fix the notation, we consider the Gaussian wave-function 
\begin{equation}
\Psi_{Ga}(r)=N\,e^{-r^2/(c a_0)^2}\,,
\label{eq:gaussian}
\end{equation}
where $N$ can be computed by demanding that the state is normalized.
In figure~\ref{fig:width_study} we compare the results of \cite{Brambilla:2016wgg,Brambilla:2017zei} for $R_{AA}$ with the ones that are obtained using the same code and parameters
but with a Gaussian instead of a delta function regularized on a finite lattice as initial condition.\footnote{
For purposes of comparison with \cite{Brambilla:2016wgg,Brambilla:2017zei}, here we ignored late-time feed-down effects.}
Note that the agreement increases as $c$ is reduced, but it also makes it harder to implement in the MCWF method.
We find that $c=0.2$ is a good compromise, and this is the value that we will use in the following.

In this work, we also study the survival probability of $P$-wave states that are produced at collision time. The same arguments that we used in the $S$-wave case apply also here.
Production takes place in a very small region compared to the size of the bound state, but this time the angular distribution is that of a $P$-wave state.
Then, at a resolution comparable with the size of the quarkonium, the wave-function after production behaves as a derivative of a delta function.
This has the same problems as the delta function in the $S$-wave case and the solution is analogous: use instead of the derivative of a delta function the derivative of a Gaussian,
which happens to be proportional to the Gaussian multiplied by $r$.

\section{Comparisons between QuTiP and QTraj}
\label{sect:codetests}
Previous studies~\cite{Brambilla:2016wgg,Brambilla:2017zei} made use of the open-source QuTiP 2 Python package~\cite{qutip1,qutip2}
to solve the Lindblad equation with the Hamiltonian $H$ of eq.~\eqref{eq:hamiltonian} and the collapse operators $C_{i}^{n}$ of eqs.~\eqref{eq:c0}
and \eqref{eq:c1} truncated at $l=1$ in the spherical harmonic expansion discussed in section \ref{ssec:hq}.
The newly developed QTraj code that we present in this work possesses several distinct advantages over the previously utilized QuTiP code.\footnote{
  We note that a Monte Carlo solver is also implemented in the {\tt qutip.mcsolve} function of QuTiP; it still requires, however, the implementation of a cutoff in $l$ in contrast to the QTraj code.}
Foremost among these is, for a system simulated with $N$ discrete points, 
the reduced size in memory from $\mathcal{O}(N^{2})$ to $\mathcal{O}(N)$,
due to calculating with the wave-function rather than with the density matrix.
It should be also added that in QTraj many single calculations of $\mathcal{O}(N)$ must be performed and averaged to arrive at a final result with sufficiently small statistical error;
the overall computational complexity\footnote{
A systematic scaling study of computational complexity and number of trajectories needed will be included in a forthcoming publication by some of the authors~\cite{forthcoming}.} 
may, therefore, exceed $\mathcal{O}(N)$.
However, as each trajectory is independent, this can be counterbalanced by running trajectories 
in an embarrassingly parallel setup.
Other advantages of the QTraj code include working to infinite order in the orbital quantum number $l$ and the use of an all-points derivative rather than forward-backward finite differences.
Compared to the QuTiP based code developed for refs.~\cite{Brambilla:2016wgg,Brambilla:2017zei} QTraj also includes the capability to couple to realistic 3+1D hydrodynamical backgrounds.

We can use the  QuTiP code as a benchmark for the QTraj code.
In order to do so, first we set up the QTraj code as the QuTiP code by implementing a cutoff in $l$, a forward-backward finite difference derivative, $T_\text{discrete}$, and Bjorken evolution.
Finally, we compare results obtained in this way with the QTraj code with the results obtained with the QuTiP code.

\begin{figure}[t]
\begin{center}
\includegraphics[width=0.9\linewidth]{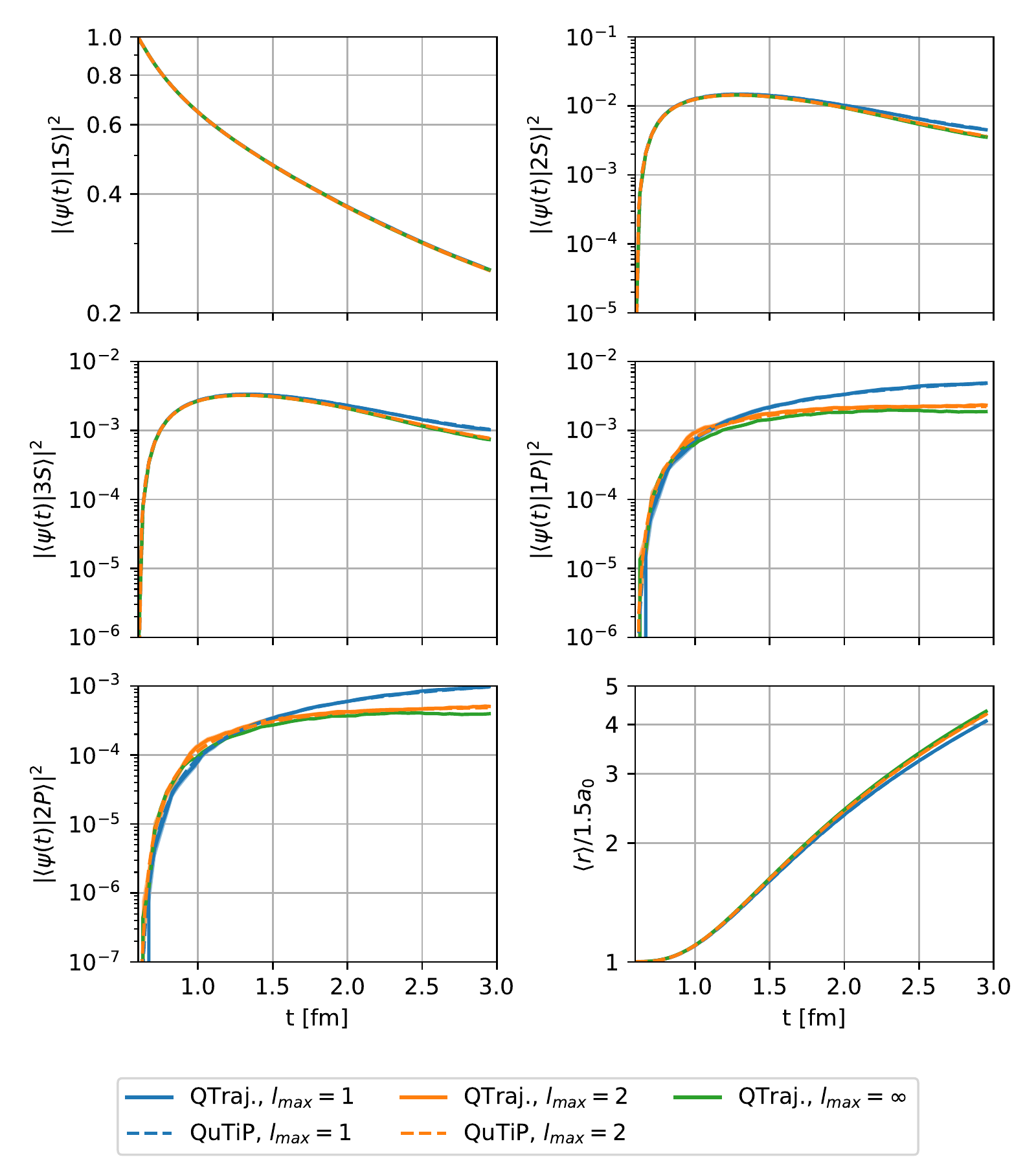}
\end{center}
\caption{A comparison of the overlap of an initial $1S$ state with the lowest lying Coulomb states
  and the expectation value of the radius computed using QuTiP (dashed lines) and 196608 QTraj trajectories (continuous lines)
  for $l_{\text{max}}=1$ (blue lines), 2 (orange lines) and $\infty$ (green line, QTraj only).
  The parameters are taken $\kappa/T^{3} = 2.6$ and $\gamma = 0$.}
\label{fig:oneS_lmax_comp}
\end{figure}

We perform two series of tests to establish the agreement of the results obtained from the new QTraj code with the results obtained from QuTiP.
Our procedure is to simulate the evolution of the state from $t=0$~fm to $t=0.6$~fm in the vacuum
and from $t=0.6$~fm to $t=2.95$~fm in a medium of initial temperature $T_{0}=425$~MeV undergoing Bjorken evolution.
On the QuTiP side, we solve the Lindblad equation and use the time evolved density matrix to calculate the expectation values of the $1S$, $2S$, $3S$, $1P$, and $2P$ Coulombic states and of the radius $r$.
On the QTraj side, we run 196608 trajectories and calculate the corresponding expectation values; finally we compare with the QuTiP results.
We perform these tests on a lattice of spatial extent $L=51.2\,a_{0}$ and lattice spacing $a_{s}=0.1\,a_{0}$.
We take $\kappa / T^{3} = 2.6$ and $\gamma = 0$.
In QuTiP, we work with angular momentum cutoffs $l_{\text{max}}=1$ and $l_{\text{max}}=2$;
in QTraj, we implement an angular momentum cutoff $l_{\text{max}}$ and run simulations with $l_{\text{max}}=1,$ $2$, and $\infty$.
There is a subtlety to consider when working with a cutoff in angular momentum in the quantum trajectories algorithm. 
The implementation of a cutoff in $l$ in QTraj leaves the evolution of states of orbital angular momentum $l < l_{\text{max}}$ unaffected,
i.e., they proceed according to the prescriptions of section \ref{ssec:hq}.
However, a state of angular momentum $l = l_{\text{max}}$ must be evolved with a reduced width
\begin{equation}
\Gamma_{n} = \frac{l_{\text{max}}}{2l_{\text{max}} + 1} C_{n}^{\dagger} C_{n},
\label{eq:Gammanlmax}
\end{equation}
cf. eq.~\eqref{eq:width}.
The reason for this is that a state of angular momentum $l_{\text{max}}$ can only jump down to a state of angular momentum $l_{\text{max}}-1$.
Hence, the total width of the state is reduced by an amount equal to the probability of jumping down by one unit in angular momentum
with respect to the width of the state without cut off on the maximal orbital angular momentum.
This probability is given by $P_{d}^{l_{\text{max}}} = l_{\text{max}}/(2l_{\text{max}}+1)$, see appendix~\ref{Ap:om}.
It is therefore the reduced width, $\Gamma = \sum \Gamma_n$ with $\Gamma_n$ given by eq.~\eqref{eq:Gammanlmax},
that enters the effective Hamiltonian describing the evolution of the state with $l = l_{\text{max}}$.
Once a jump is triggered, this is deterministically a jump down.
The color evolution remains unchanged.
Under the above conditions, we test the two programs using identical initial conditions.

In the first test, we run both programs with a Coulombic $1S$ wave-function as initial condition and plot the results in figure~\ref{fig:oneS_lmax_comp}.
We observe excellent agreement for all measured quantities at identical angular momentum cutoffs $l_{\text{max}}$.

\begin{figure}[ht]
\begin{center}
\includegraphics[width=0.9\linewidth]{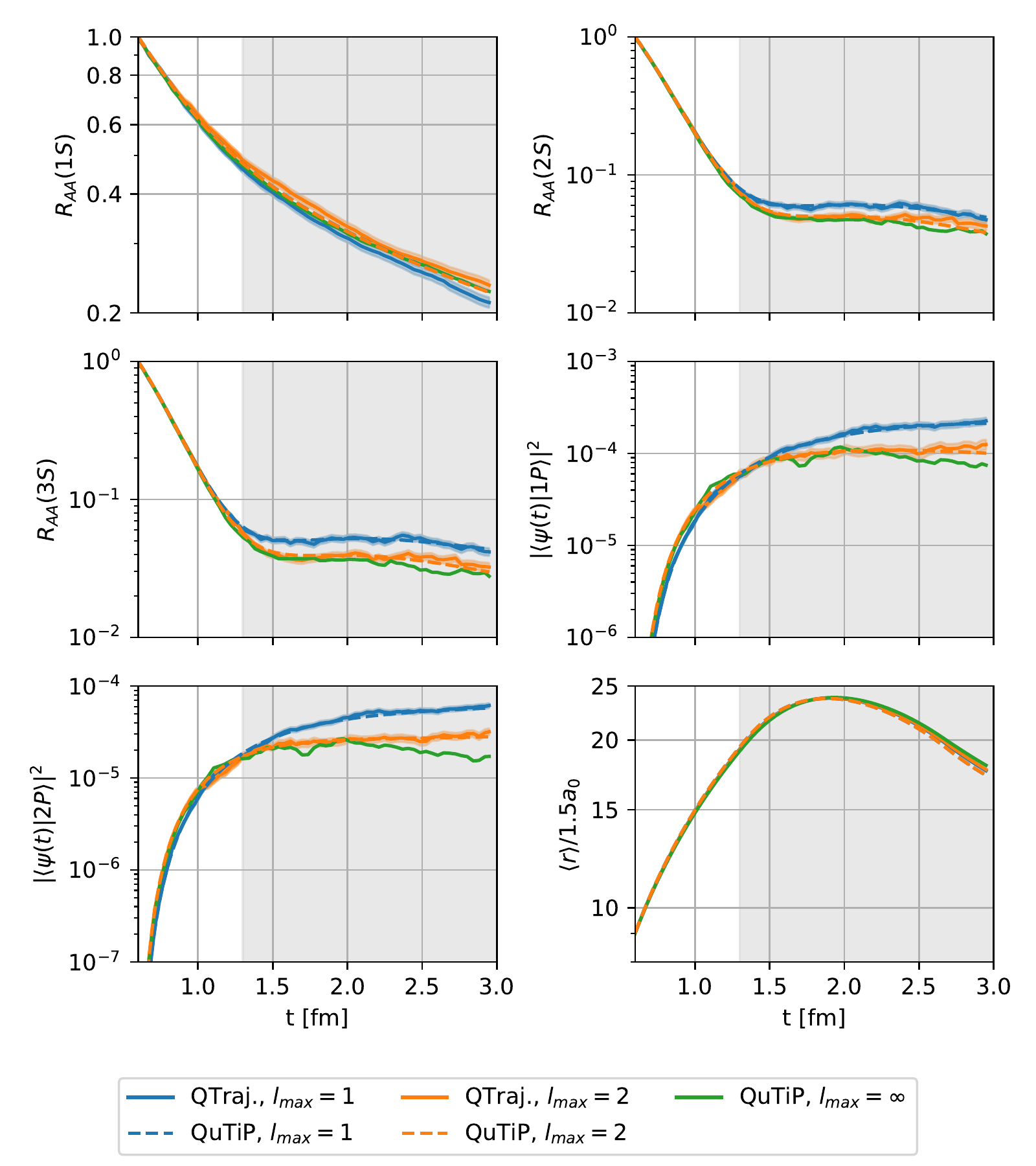}
\end{center}
\caption{A comparison of the overlap of an initial Gaussian state of width $c=0.2$ with the lowest lying Coulomb states
  and the expectation value of the radius computed using QuTiP and the QTraj algorithm.
  Curves and simulation conditions are as in figure~\ref{fig:oneS_lmax_comp}.
  The gray band denotes the area where the wave-function hits the edge of the lattice, i.e., finite size effects become relevant, which occurs at $t=1.3\,\mathrm{fm}$ (see text for discussion).
  We note that for the $S$-wave states we report their overlaps normalized to their initial overlaps and denote this quantity $R_{AA}$.  }
\label{fig:gaussian}
\end{figure}

In the second test, we run both programs with a Gaussian of width $c=0.2$ as initial condition, see section \ref{sect:initstate}.
We plot the results in figure~\ref{fig:gaussian}.
In this case, we do observe some disagreements at late times that cannot be fully accounted for by the statistical errors of QTraj. 
The position space wave-function of a narrowly peaked Gaussian quickly expands, and for our simulation parameters reaches the outer edge of the box before the end of the simulation.  
The reflection behavior exhibited at the spatial boundary of the lattice,
which may be deduced also from the shrinking of the average radius after an initial expansion in the last plot of figure~\ref{fig:gaussian}, differs between QuTiP and QTraj. 
We attribute the late time (small) discrepancies in the expectation values between the two programs to this phenomenon.
Since we find that a Gaussian wave-function evolved in QuTiP hits the edge of the box at approximately $t=1.3$~fm,
we take $t>1.3$~fm as the region where finite size effects become significant.
When running the QTraj code to simulate our final results for the quarkonium nuclear modification factors in section \ref{sect:results},
we will avoid finite size effects by increasing the spatial extent of the lattice.
We note that while increasing the extension of the lattice is rather cheap for QTraj, it is computationally unfeasible for QuTiP. 

In summary, we find that the results of the QuTiP and QTraj programs are in excellent agreement when run in their regions of validity with identical parameters.
They agree on both the extracted overlaps and the expectation value of $r$ including their dependence on $l_{\rm max}$.
This gives us confidence to proceed with the use of QTraj for our calculations.

\section{Hydrodynamic background evolution}
\label{sect:hydro}
In order to compute the in-medium survival probability of a given quantum trajectory one must specify the temperature evolution of the QGP.
For this study we make use of a 3+1D dissipative hydrodynamics code that is based on the quasiparticle anisotropic hydrodynamics (aHydroQP) framework
for dissipative relativistic hydrodynamics \cite{Alqahtani:2015qja,Alqahtani:2016rth,Alqahtani:2017mhy}.
The aHydroQP framework has been shown to well-reproduce a variety of experimental soft-hadronic observables such as the total charged hadron multiplicity,
identified hadron spectra, integrated and identified hadron elliptic flow,
and HBT radii at both RHIC and LHC nucleus-nucleus collision energies~\cite{Alqahtani:2017jwl,Alqahtani:2017tnq,Nopoush:2014pfa,Almaalol:2018gjh,Alqahtani:2020daq,Alqahtani:2020paa}.
The aHydroQP framework is an extension of the originally formulated conformal anisotropic hydrodynamics \cite{Florkowski:2010cf,Martinez:2010sc,Tinti:2013vba}
to include the effects of non-conformal transport coefficients such as the bulk viscosity.
The resulting code uses a realistic equation of state determined from lattice QCD measurements \cite{Bazavov:2013txa}
and self-consistently computed second- and higher-order transport coefficients.
Due to the resummation to all orders in the inverse Reynolds number, the anisotropic hydrodynamics framework can be applied at early times after an $AA$ collision,
which is a period of time when strong non-equilibrium effects
are present~\cite{Chesler:2009cy,Florkowski:2013lza,Florkowski:2013lya,Florkowski:2014sfa,Denicol:2014xca,Denicol:2014tha,Heller:2015dha,Keegan:2015avk,Strickland:2017kux,Strickland:2018ayk,Strickland:2019hff,Almaalol:2020rnu}.
This allows us to more accurately describe the dynamics of the QGP during the entire evolution of the produced bottomonium states since they are created at early times in the QGP's lifetime.

For this paper we use the aHydroQP tuning recently reported in ref.~\cite{Alqahtani:2020paa}.
We assume a smooth optical Glauber initial condition that provides the initial spatial energy density profile of the QGP as a function of the impact parameter.
The study of ref.~\cite{Alqahtani:2020paa} concluded that the best fit to soft hadron observables is achieved
using an initial central temperature of $T_0 = 630$~MeV and a constant specific shear viscosity of $4\pi\eta/s = 2$.\footnote{
The initial longitudinal proper time corresponding to this initial central temperature is 0.25 fm~\cite{Alqahtani:2020paa}.}
To determine the temperature experienced by bottomonium states produced in the QGP,
we assume that the initial transverse spatial distribution for bottomonium production is proportional to the binary overlap profile of the two colliding nuclei, $N_{AA}^\text{bin}(x,y)$,
and then use Monte-Carlo sampling to generate the initial production points.
We take the initial transverse momentum ($p_T$)-distribution to be proportional to $p_T/(p_T^2 + \langle M \rangle^2)^2$ for all states,
where $\langle M \rangle$ is the average mass of all states being considered.  We then Monte-Carlo sample the $p_T$ for each particle generated.
We take into account the approximate boost-invariance of the QGP and assume all bottomonia to have zero momentum rapidity, $y=0$.
Finally, we sample the initial azimuthal angle $\phi$ from a uniform distribution between 0 and $2\pi$.  

\begin{figure}[ht]
\begin{center}
\includegraphics[width=0.6\linewidth]{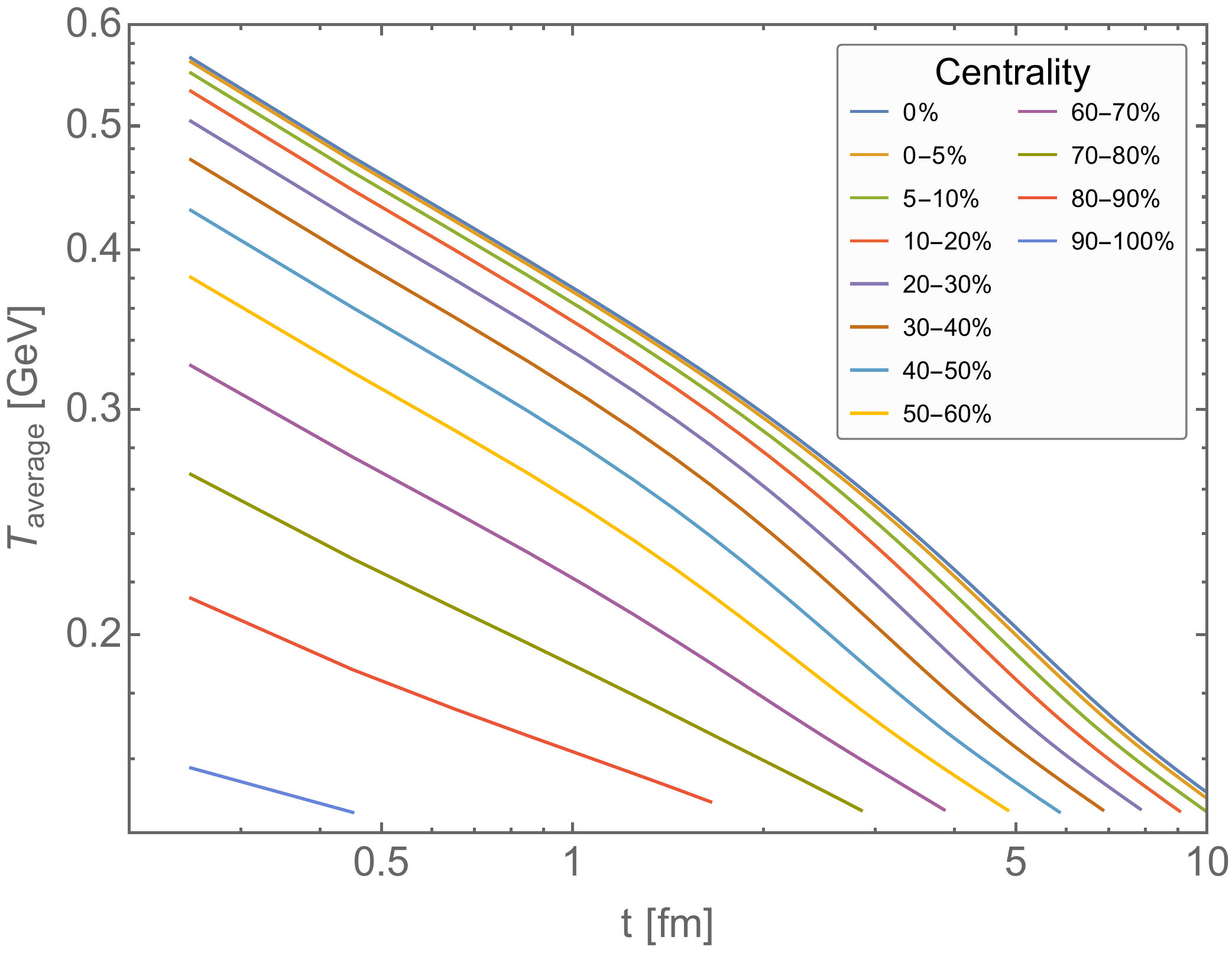}
\end{center}
\caption{Path-averaged temperature evolution used for our calculations.  Lines from top to bottom correspond to the centrality classes listed in table~\ref{tab:collisiondata}.}
\label{fig:temperature-evol}
\end{figure}

\begin{table}[ht]
\begin{center}
\begin{tabular}{|c|c|c|c|c|}
\hline
Centrality & $\langle b \rangle$ [fm] & $\langle N_{\rm part} \rangle$ & $T_0^{\rm central}$ [GeV]  & $T_0^{\rm average}$ [GeV] \\
\hline
0\% 			& 0            & 406.1 & 0.630 & 0.565 \\
0-5\% 		& 2.32       & 374.0  & 0.625 & 0.561 \\
5-10\% 		& 4.25       & 315.9  & 0.614 & 0.550 \\
10-20\% 		& 6.01       & 243.5  & 0.597  & 0.533 \\
20-30\%		& 7.78       & 168.5  & 0.571  & 0.504 \\
30-40\%		& 9.21      &  112.4 & 0.538  & 0.470 \\
40-50\%		& 10.45      & 70.8  & 0.497  & 0.430 \\
50-60\%		& 11.55       & 41.1  & 0.446 & 0.381 \\
60-70\%		& 12.56       & 21.3  & 0.386  & 0.325 \\
70-80\%		& 13.49       & 9.7 & 0.322  & 0.267 \\
80-90\%		& 14.38       & 3.8 & 0.258  & 0.214 \\
90-100\%		& 15.66       & 0.97 & 0.180 & 0.157 \\
\hline                                                          
\end{tabular}
\end{center}
\caption{The average impact parameter, number of participants, initial central temperature, and path-averaged temperatures in centrality classes appropriate for a $\sqrt{s_{NN}} = 5.02$~TeV PbPb collision.} 
\label{tab:collisiondata}
\end{table}

We then averaged the temperature obtained along each Monte-Carlo generated path using approximately 132000 samples per centrality bin.
The resulting path-averaged temperature evolution is plotted in figure~\ref{fig:temperature-evol} for each centrality bin used in this work.
Note that at late times ($t \gtrsim 3$ fm) one can see the onset of 3D expansion for centralities $\lesssim 50\%$, which results in more rapid cooling of the QGP.\footnote{
Since we consider only $y=0$ bottomonium production, the longitudinal proper time $\tau$ is equal to $t$.}
In table~\ref{tab:collisiondata}, for each centrality class considered, we list the average impact parameter, the average number of participating nucleons, the initial central temperature,
and the path-averaged initial temperature.
We note that since many bottomonium states are created away from the center of the collision, the  path-averaged temperature is always lower than the central temperature.
We evolve the quantum wave-packets using the in vacuum potential starting at $t = 0$~fm and turn on the in medium complex potential at $t = 0.6$~fm.
Finally, when the averaged temperature drops below $T_{\rm f} = 250$~MeV, we again use the in vacuum potential for the wave-function evolution. 
Hence the hydrodynamical evolution of the medium does not play any role between $t = 0$~fm and $t = 0.6$~fm and after bottomonium freeze out.
By neglecting medium effects below $T_{\rm f}$ we are ignoring physical effects that may be relevant specially for excited states.
  The inclusion of physical effects in the regime $T\sim E$ is beyond the scope of this paper.
  It would involve solving a more complicated master equation \cite{Brambilla:2017zei} in which the information of the medium cannot be encoded in two transport parameters,
  as it is the case in the regime that we study in this work.
  In Appendix \ref{sec:Tf}, we try to assess the size of the physical effects we are leaving out by varying the value of $T_{\rm f}$.

\section{Results and discussion}
\label{sect:results}
In this section, we present the QTraj results for the survival probability of various bottomonium states and quantify the effect of quantum jumps on this observable.
Using the resulting survival probabilities, we then include the effect of late-time feed down of excited states
and compare the QTraj results with available LHC data for $R_\text{AA}$ and double ratios of various states.

\subsection{Parameters}
\label{sect:potentialmodel}
We look at the quarkonium evolution in the QGP in the regime $1/a_0 \gg  T,m_D,\Lambda_{\rm QCD} \gg E$;
we further assume that the quarkonium is mostly Coulombic.\footnote{
    In this work, we investigate the $\Upsilon(1S)$, $\Upsilon(2S)$, $\Upsilon(3S)$, $\chi_b(1P)$ and $\chi_b(2P)$ under these assumptions.
      In particular, we remark that the $\Upsilon(1S)$ is commonly treated as a Coulombic bound state,
      whereas the $\Upsilon(2S)$, $\chi_b(1P)$ and more critically the $\Upsilon(3S)$ and $\chi_b(2P)$ have been investigated as Coulombic bound states,
      for instance, in~\cite{Brambilla:2001fw,Penin:2005eu,Sumino:2016sxe,Mateu:2017hlz,Peset:2018jkf,Segovia:2018qzb}.}
This has been discussed in section~\ref{ssec:hq}.
From that discussion it follows that the evolution depends on four parameters.
One parameter is the bottom quark mass $m_b$, which has to be understood as the pole mass. 
We take the bottom quark mass to be $m_b = 4.881$~GeV.
Another parameter is the strength of the Coulomb potential.
We may trade this parameter for the Bohr radius $a_0$, which we take to be $a_0= 0.742~\text{GeV}^{-1}  = 0.146~\text{fm}$
to reproduce the value used in~\cite{Brambilla:2016wgg,Brambilla:2017zei}.
This value is the solution of the self consistency equation $a_0 = 2/(C_F \als(1/a_0) m_b)$
when the strong coupling $\als$ in the $\overline{\text{MS}}$ scheme is taken at one loop accuracy.
The strong coupling $\als$ at the scale of the inverse of the Bohr radius is then $\als(1/a_0) = 0.414$.
Finally, the evolution equations also depend on the two coefficients $\gamma$ (see eq.~\eqref{gamma}) and $\kappa$ (see eq.~\eqref{kappa}).
The coefficients $\gamma$ and $\kappa$ have mass dimension three. Hence it is convenient to define
\begin{equation}
\hat{\gamma} = \frac{\gamma}{T^3}, \qquad \hat{\kappa} = \frac{\kappa}{T^3}.
\label{rescaledgammakappa}
\end{equation}  

The coefficient $\gamma$ has been taken equal to zero in~\cite{Brambilla:2016wgg,Brambilla:2017zei}.
More recently, in \cite{Brambilla:2019tpt} $\gamma$ has been extracted using 2+1 flavor lattice QCD data for the quarkonium mass shift in a thermal bath.
Note that $\gamma$ and $\kappa$ are flavor independent.
It was found that the lattice data for the $J/\psi$ and $\Upsilon(1S)$ thermal mass shifts at the temperatures 251~MeV and 407~MeV ($\Upsilon(1S)$ case only)
fall inside the interval $-3.8<\hat{\gamma}<-0.7$. In this work, we will take
\begin{equation}
\hat{\gamma} = -1.75 \pm 1.75
\label{eq:gammahat-values}
\end{equation}
to encompass also the value used in~\cite{Brambilla:2019tpt}.
On general grounds we expect that $\hat{\gamma}$ depends on the temperature.
Indeed, the data in \cite{Brambilla:2019tpt} seem to suggest that $\hat{\gamma}$ is closer to zero at higher temperatures.
Nevertheless, our present knowledge of $\hat{\gamma}$ is clearly insufficient to parameterize $\hat{\gamma}$ in terms of the temperature.
Hence, we will assume $\hat{\gamma}$ constant (in the temperature, and, therefore, also in time) in the present analysis.

\begin{figure}[ht]
\begin{center}
\includegraphics[width=0.5\linewidth]{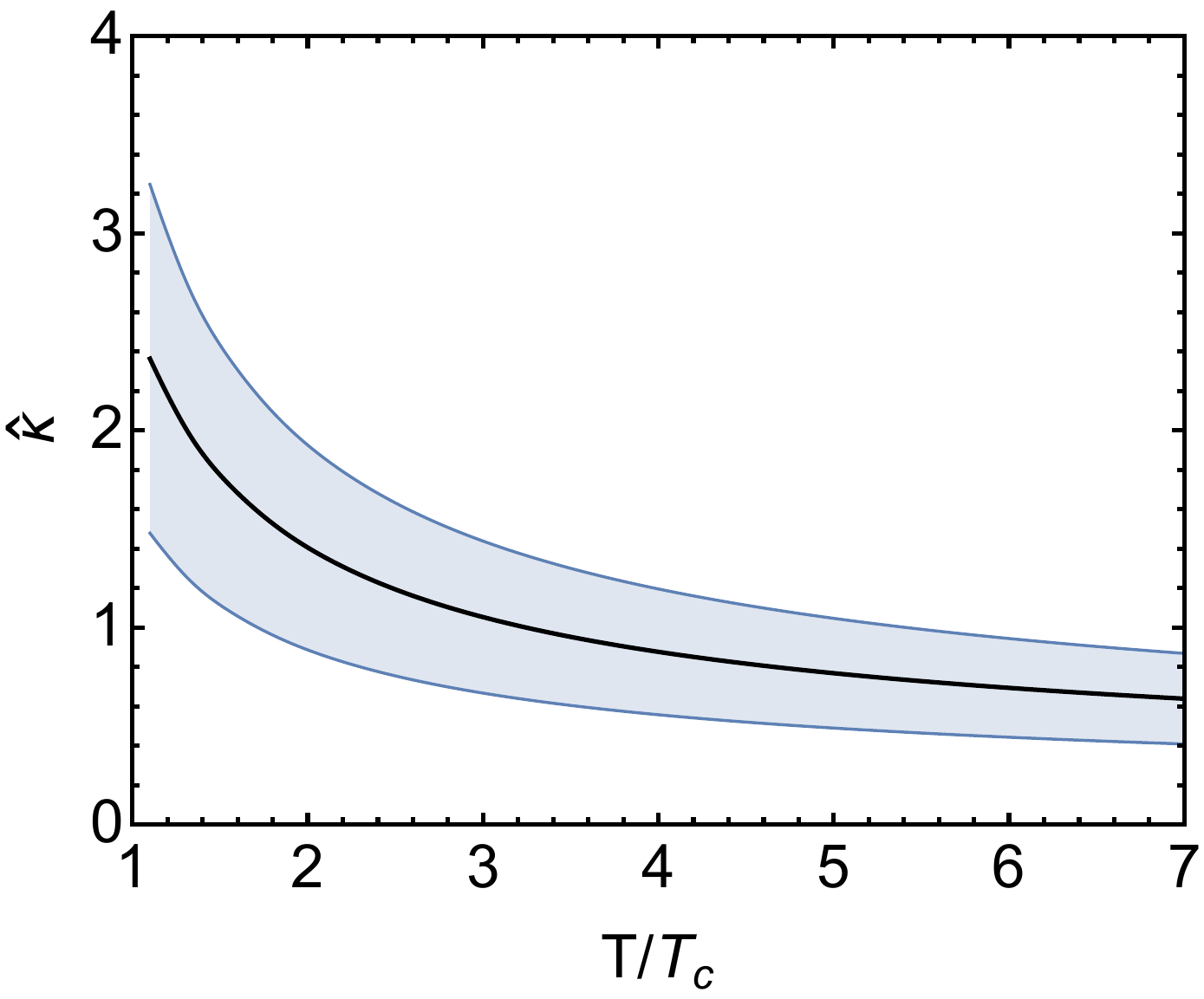}
\end{center}
\caption{Plot of $\hat\kappa$ as a function of the temperature in units of $T_{\rm c}$, where $T_{\rm c} = 155$~MeV.
  The plot reproduces the fit done in~\cite{Brambilla:2020siz} for a set of pure SU(3) lattice data ranging from $T/T_{\rm c}=1.1$ to $T/T_{\rm c} = 10^4$.
  The fit function reproduces the next-to-leading order expression of  $\hat\kappa$ up to the constant multiplying the ratio of the Debye mass over $T$, which is fitted.
  The shaded band includes the errors coming from varying the renormalization scale by a factor of 2 and the statistical error.}
\label{fig:kappaHat}
\end{figure}

The coefficient $\kappa$ has been recently computed at temperatures in the range $1.1 \lesssim T/T_{\rm c} \lesssim 10^4$,
where $T_{\rm c} = 155$~MeV, from pure SU(3) gauge lattice data in~\cite{Brambilla:2020siz}.
Because of the wide range of temperatures, it has been possible to parameterize the temperature dependence of $\hat{\kappa}$.
The change of $\hat{\kappa}$ with the temperature is well parameterized by its next-to-leading order expression
when the coefficient of the term proportional to the ratio of the Debye mass over $T$ is fitted.
The parameterization of $\hat{\kappa}$ as a function of the temperature with the corresponding error band is shown in figure~\ref{fig:kappaHat}.
In this work, we will take $\hat{\kappa}$ and its uncertainties according to figure~\ref{fig:kappaHat}.
We will call $\hat{\kappa}_C(T)$ the central line, $\hat{\kappa}_U(T)$ the upper boundary and  $\hat{\kappa}_L(T)$ the lower one.
Recall that, once coupled with the hydrodynamical evolution of the QGP, the temperature dependence of $\hat{\kappa}$ translates into a time dependence.

We emphasize that in the regime  $1/a_0 \gg  T,m_D,\Lambda_{\rm QCD} \gg E$
all parameters describing the quarkonium evolution in the QGP are determined independently of the quarkonium nuclear modification factors that we eventually compute.
Indeed, $m_b$ and $\als$ are parameters of the QCD Lagrangian, and the coefficients $\gamma$ and $\kappa$ are determined from first principle (lattice) computations in QCD.

\subsection{Numerical setup}
\label{sect:latticesetup}
For all results reported in this section we use a lattice size of $N=4096$ points with $L = 80~\text{GeV}^{-1} \approx 108\,a_0$.
The temporal step size for evolving the wave-function between jumps is $\Delta t = 0.001~\text{GeV}^{-1} \approx 2 \times 10^{-4}~\text{fm}$.  
We initialize the wave-function at $t=0$ and evolve it with the vacuum potential until $t = 0.6$~fm.
From this point forward in time, in between quantum jumps, we evolve the wave-function with the potential appropriate
for the considered $Q\bar{Q}$ state labeled by its integer orbital angular momentum $l \geq 0$ and color state (singlet or octet).
We terminate the evolution when the path-averaged temperature, $T^{\text{average}}$, in a given centrality bin drops below $T = T_{\rm f} = 250$~MeV.
We then compute the final quantum mechanical overlaps of the vacuum eigenstates with the QTraj evolved wave-function to obtain the survival probability of the state.

For all results presented in this section we initialize the wave-function using a smeared Gaussian delta function \eqref{eq:gaussian} of the form
\be
u_{l}(r,t=0) = N r^{l+1} \exp(-r^2/(ca_0)^2) \, ,
\ee
with $c=0.2$ and $N$ being fixed by the normalization condition \eqref{eq:normcond}.

\subsection{Bottomonium survival probabilities}
\label{sect:survival}
We present, first, results for the survival probability of various bottomonium states.
The survival probability tells us the probability to find a given quantum state after the wave-function is evolved in the QGP.
This measure does not yet take into account late-time feed down of excited states, but does allow us to assess the impact of $\hat\kappa$ and $\hat\gamma$.
In addition, we can use the pre feed down results to more cleanly quantify the effect of quantum jumps on the resulting survival probabilities.

\begin{figure}[ht]
\begin{center}
\includegraphics[width=0.475\linewidth]{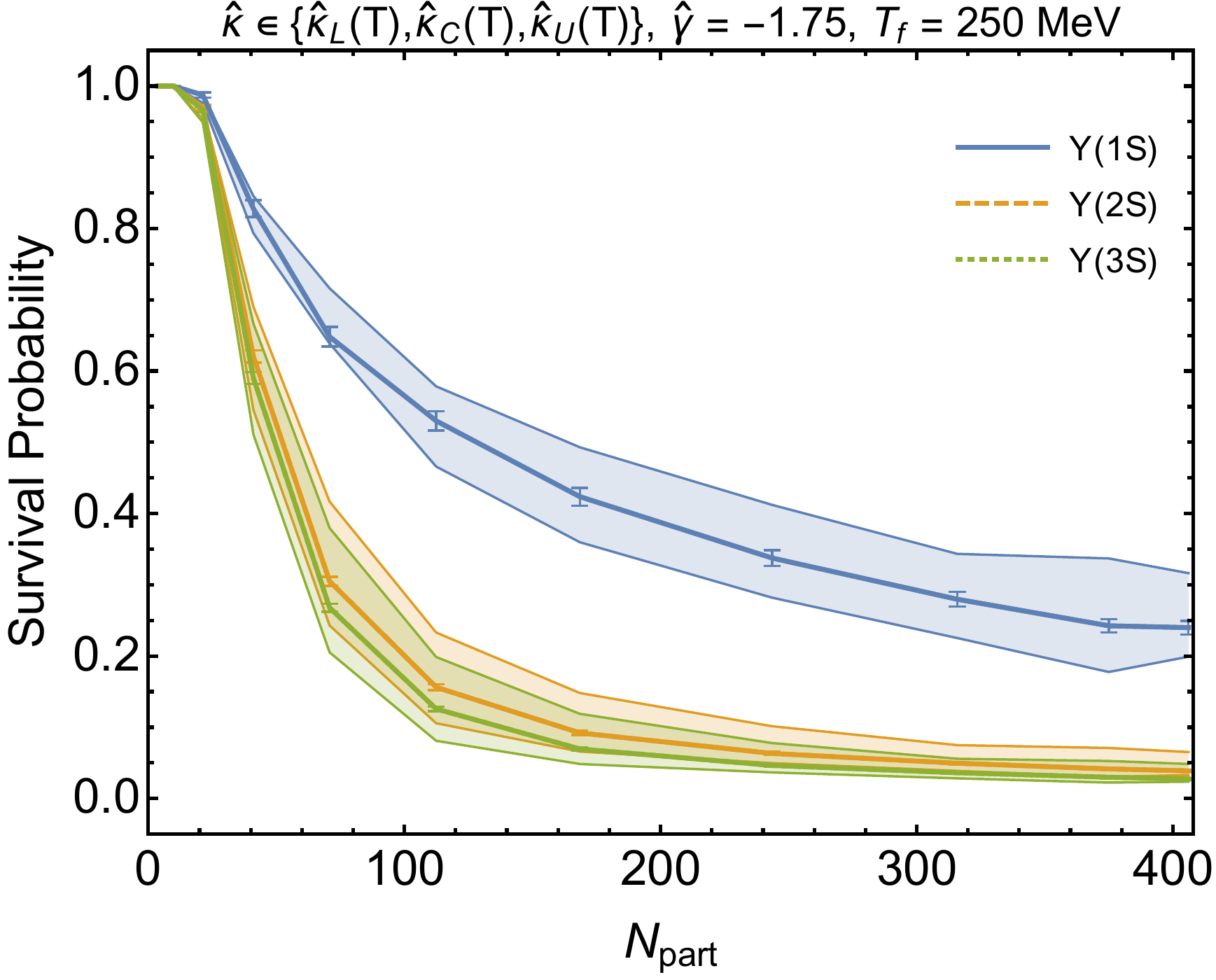} $\;\;\;$
\includegraphics[width=0.475\linewidth]{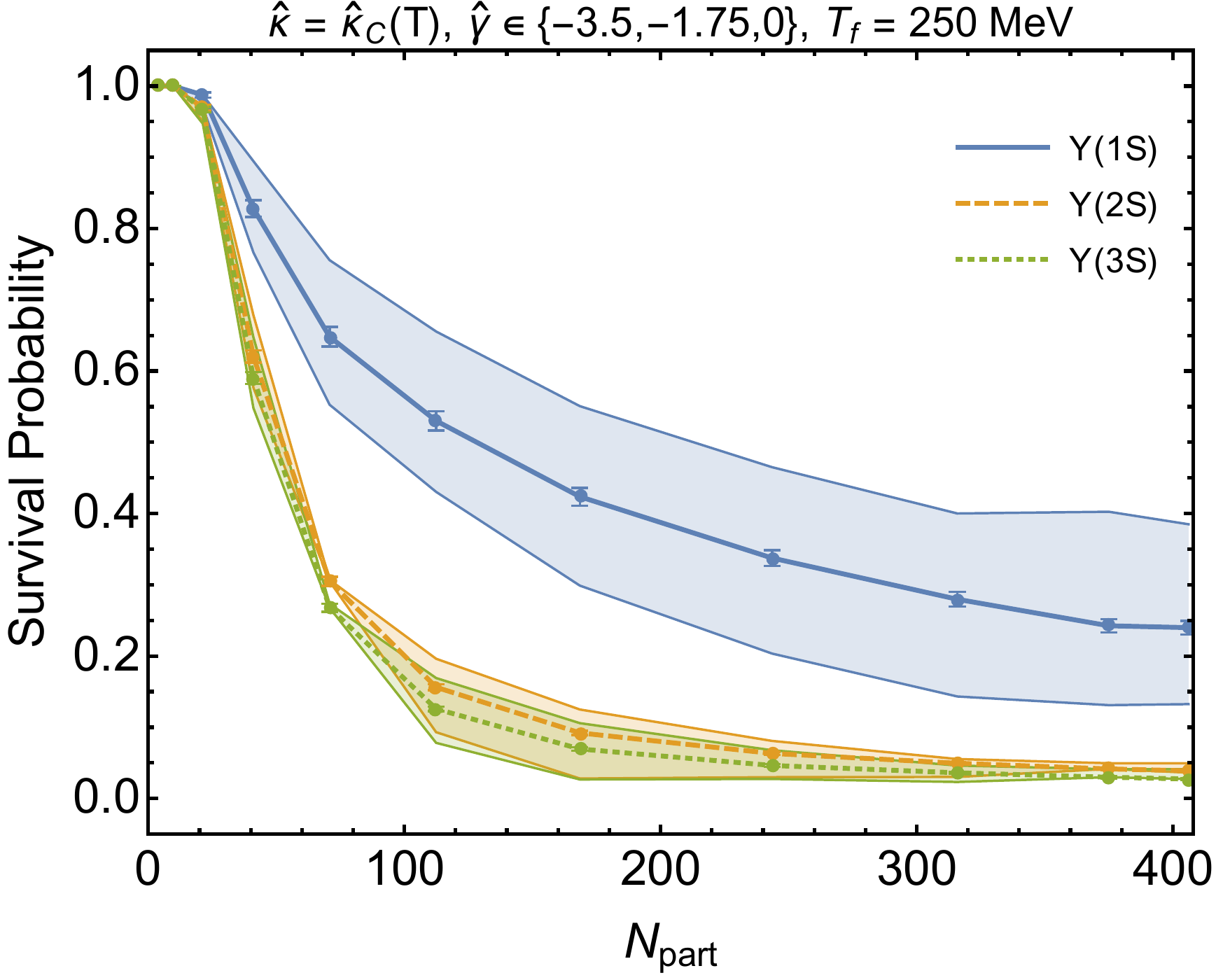} 
\end{center}
\caption{Survival probability of $S$-wave bottomonium states versus $N_{\rm part}$ using Gaussian initial conditions to regularize the delta function.  
  In the left panel, the shaded bands show the change in the prediction for survival probability of $S$-wave states when varying
  $\hat\kappa \in \{ \hat\kappa_L(T), \hat\kappa_C(T),\hat\kappa_U(T) \}$, while assuming $\hat\gamma = -1.75$.
  In the right panel, the shaded bands are obtained by varying $\hat\gamma \in \{-3.5,-1.75,0\}$ while assuming $\hat\kappa = \hat\kappa_C(T)$.
  The error bars on the central lines indicate the statistical uncertainty associated with the average over the quantum trajectories.}
\label{fig:swaveRAA}
\end{figure}

\subsubsection*{Singlet delta $S$-wave initial conditions}
In figure~\ref{fig:swaveRAA}, we present our results for the pre feed down survival probability of color singlet $S$-wave states
as a function of the number of participants in the nuclear collision, $N_\text{part}$.
In the left panel, the shaded bands result from varying $\hat\kappa \in \{ \hat\kappa_L(T), \hat\kappa_C(T),\hat\kappa_U(T) \}$ while assuming $\hat\gamma = -1.75$,
whereas, in the right panel, the shaded bands correspond to varying $\hat\gamma \in \{-3.5,-1.75,0\}$ while assuming $\hat\kappa = \hat\kappa_C(T)$.
Note that, in the left panel, for all three states shown, the upper/lower bounds for $\hat\kappa$ map to the lower/upper bounds in the survival probability.
However, in the right panel, for the $\Upsilon(1S)$, the lower bound for $\hat\gamma$ maps to the lower bound in the survival probability,
while for the $\Upsilon(2S)$, the lower bound for $\hat\gamma$ maps to the upper bound in the survival probability.
The error bars indicated on the central lines are the statistical errors of the averages over the quantum trajectories.
For this figure we have used approximately 98304 quantum trajectories for each point in $N_\text{part}$.
As the figure demonstrates, our resulting statistical errors are smaller than the systematic uncertainties coming from the choice of $\hat\kappa$ and $\hat\gamma$.
We also find that the QTraj predictions for the survival probability are sensitive to the choice of $\hat\kappa$ and $\hat\gamma$,
with the variation of $\hat\gamma$ resulting in the larger variation of the survival probability for the $\Upsilon(1S)$.
The eventual comparison of the quarkonium nuclear modification factors with the experimental data will, therefore, be in the position to validate
or invalidate our independent choice of values for the coefficients $\hat\kappa$ and $\hat\gamma$ given respectively in figure~\ref{fig:kappaHat} and eq.~\eqref{eq:gammahat-values}.

\begin{figure}[t!]
\begin{center}
\includegraphics[width=\linewidth]{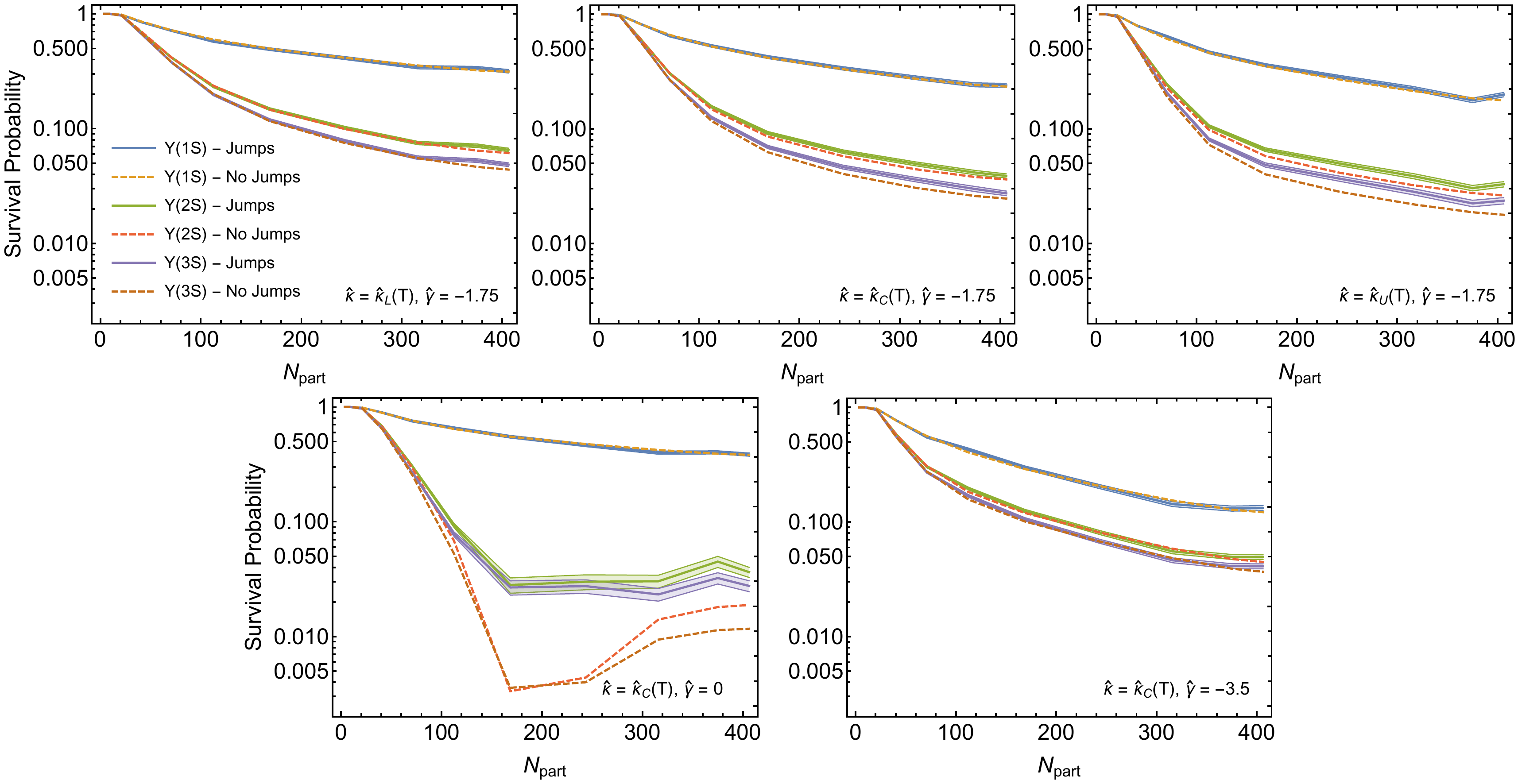}
\end{center}
\caption{Survival probability of $S$-wave bottomonium states versus $N_{\rm part}$ in the presence of quantum jumps
  (continuous lines; these are the same curves as in figure~\ref{fig:swaveRAA}, but now with a logarithmic scale for the probabilities) and without quantum jumps (dashed lines).}
\label{fig:jumpNojump}
\end{figure}

In order to quantify the effect of quantum jumps on the $S$-wave survival probabilities,
in figure~\ref{fig:jumpNojump} we present five panels
where we compare the result of evolving the system with no quantum jumps to the full QTraj result.
In the case that no jumps are allowed, this reduces to evolving the wave-function solely with the complex Hamiltonian $H_\text{eff}$.
For the full QTraj results including quantum jumps, we once again use 98304 quantum trajectories for each point in $N_\text{part}$.
In the top row of figure~\ref{fig:jumpNojump}, we present the results obtained when varying $\hat\kappa$ in the same range used in the left panel of figure~\ref{fig:swaveRAA}.
In the bottom row of figure~\ref{fig:jumpNojump}, we present the results obtained when varying $\hat\gamma$ in the same range used in the right panel of figure~\ref{fig:swaveRAA}.
From these figures we firstly note that the survival probability of the $\Upsilon(1S)$ is well reproduced by the $H_\text{eff}$ (no jump) evolution.
For the excited states, we see a larger effect from the correct implementation of the quantum jumps.
Focusing on the top row of figure~\ref{fig:jumpNojump}, we see that the effect of jumps on the excited states increases with increasing $\hat\kappa$.
We also notice that increasing $\hat\kappa$ generally decreases the survival probability of the states.
The importance of quantum jumps is largest in the case of $\hat\gamma=0$, which is shown in the bottom left panel of figure~\ref{fig:jumpNojump}.
In this case, the $H_{\rm eff}$ evolution (dashed lines) predicts strong suppression of the $2S$ and $3S$ states.
When this occurs, the corrections from the jumps become comparatively more important and result in a flattening of the survival probability as a function of $N_{\rm part}$
at a magnitude that is similar to the survival probabilities of the $2S$ and $3S$ states seen in other panels of figure~\ref{fig:jumpNojump}.

One of the reasons that we see small effects of quantum jumps on the singlet $S$-wave survival probabilities when using $\hat\kappa$ and $\hat\gamma$ at their central values is that,
once such an initial state makes one quantum jump, it necessarily changes its angular momentum quantum number $l$ and its color state to octet.
Later quantum jumps are more likely to cause the state to jump to color octet states with higher angular momentum.
For example, starting from a singlet $1S$ state one can only jump to an octet $1P$ state.
Once the state is in a color octet configuration,
the potential becomes repulsive and the wave-function will spread to larger radii.\footnote{
This does not happen in models that take into account only color singlet configurations, since the color singlet potential is always attractive.}
After evolving the wave-function, a second jump can occur that causes the state to jump back down to the singlet $1S$ state, however, this is a disfavored transition
since the probability of an octet to singlet transition is $2/(N_c^2-2) = 2/7$ (see footnote~\ref{footoctet})
and the probability of an $l = 1$ to $l = 0$ transition is $P_d^1 = 1/3$ (see eq.~\eqref{eq:angprob2}),
resulting in a combined probability of $2/21 \approx 10$\% for transitioning from a $P$-wave color octet to an $S$-wave color singlet.
Moreover, the compact $1S$ color singlet state overlaps only marginally with the wide color octet state and the impact of the jump on the evolution equation is small.
The other 90\% of the time the state will transition to an $l=2$ singlet or octet state or to an $l=0$ octet state.
Since for any finite $l$ transitioning to higher $l$ is favored (see eq.~\eqref{eq:angprob2}), the result is, on average, a directed random walk towards higher $l$.
As the state transitions to higher and higher angular momentum states, the probability to jump up or down in $l$ approaches $P_d^\infty = P_u^\infty = 0.5$
and the state eventually does a balanced random walk in $l$.
The net result of all this is that the survival probability for the singlet $1S$ state is strongly dominated by the case of no quantum jumps
and that the corrections, while important to quantify, are small.
For the $2S$ and $3S$ states, one sees larger effects from the quantum jumps because they have larger average radii than the $1S$ state
and hence will have a larger overlap with color octet states after a series of jumps.
Nevertheless, also in this case transitioning to higher $l$ is favored for the same argument exposed above.

\begin{figure}[t]
\begin{center}
\includegraphics[width=0.6\linewidth]{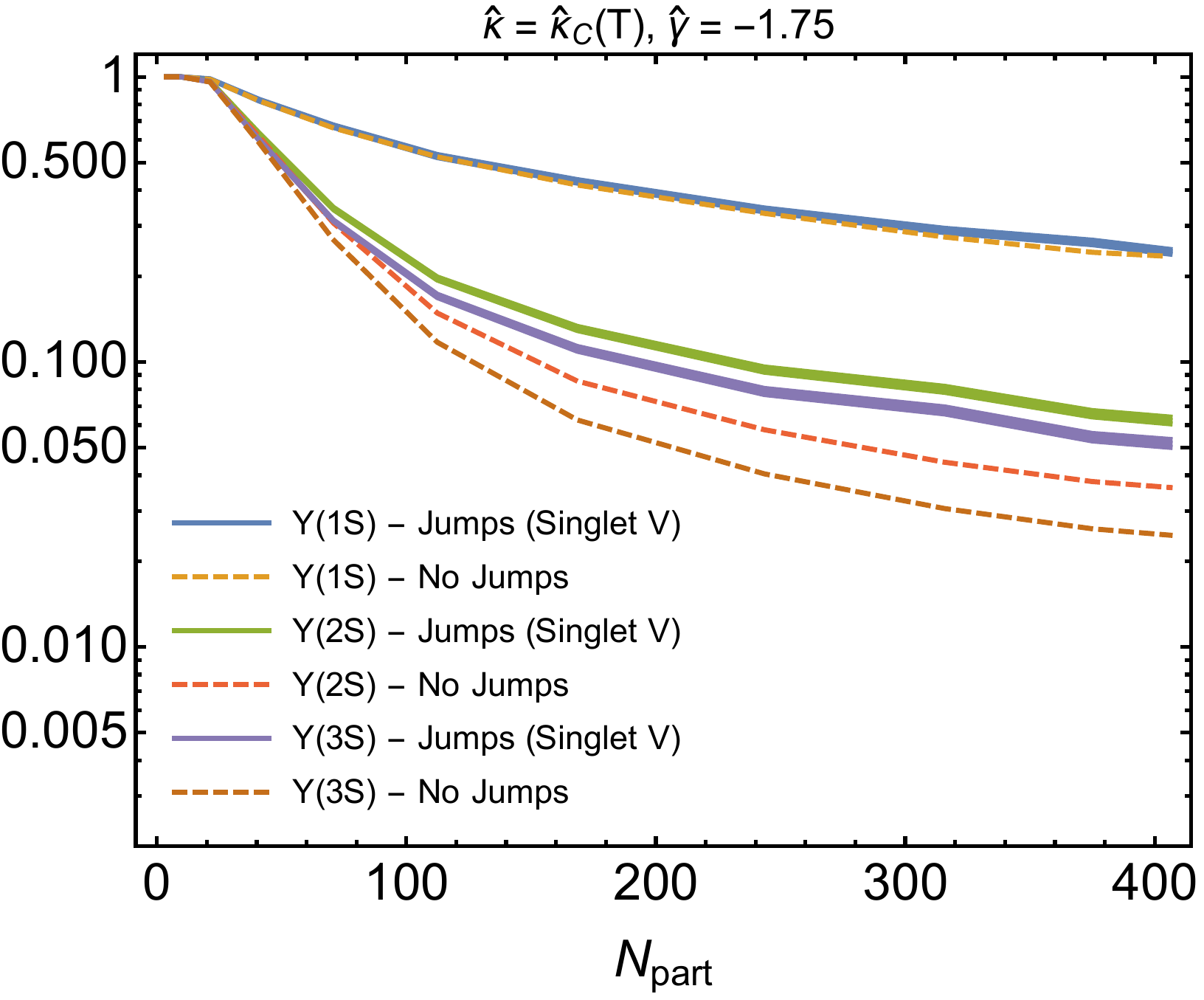}
\end{center}
\caption{Singlet $S$-wave survival probabilities versus $N_\text{part}$ obtained when replacing in the code the repulsive color octet potential with an attractive color singlet potential.
  Dashed lines show the results obtained with no quantum jumps, i.e. when the evolution is entirely described by $H_{\text{eff}}$.
  The full QTraj results are represented by continuous lines (the bands account for the statistical errors). 
  For these results we used 98304 quantum trajectories per point in $N_{\rm part}$.}
\label{fig:attractiveJumpNoJump}
\end{figure}

In order to assess what role the repulsive nature of the octet potential plays in the importance of quantum jumps, 
in figure~\ref{fig:attractiveJumpNoJump} we present QTraj results obtained when promoting in the code all color octet states to color singlet ones.
For ease of comparison, we have made the vertical scale in figure~\ref{fig:attractiveJumpNoJump} the same logarithmic scale that has been used in figure~\ref{fig:jumpNojump}.
Comparing the central panel in the top row of figure~\ref{fig:jumpNojump} with figure~\ref{fig:attractiveJumpNoJump},
we see that the effect of quantum jumps is enhanced in particular on the excited states,  
when using a framework that includes only the attractive color singlet potential.
This enhancement is due to the fact that the singlet-potential is attractive,
which causes states to have smaller average radii than when they interact through a repulsive color octet potential. 
A smaller radius leads to a larger overlap with $S$-wave states and eventually to a larger probability to jump back to a state with lower angular momentum.

\begin{figure}[ht]
\begin{center}
\includegraphics[width=0.475\linewidth]{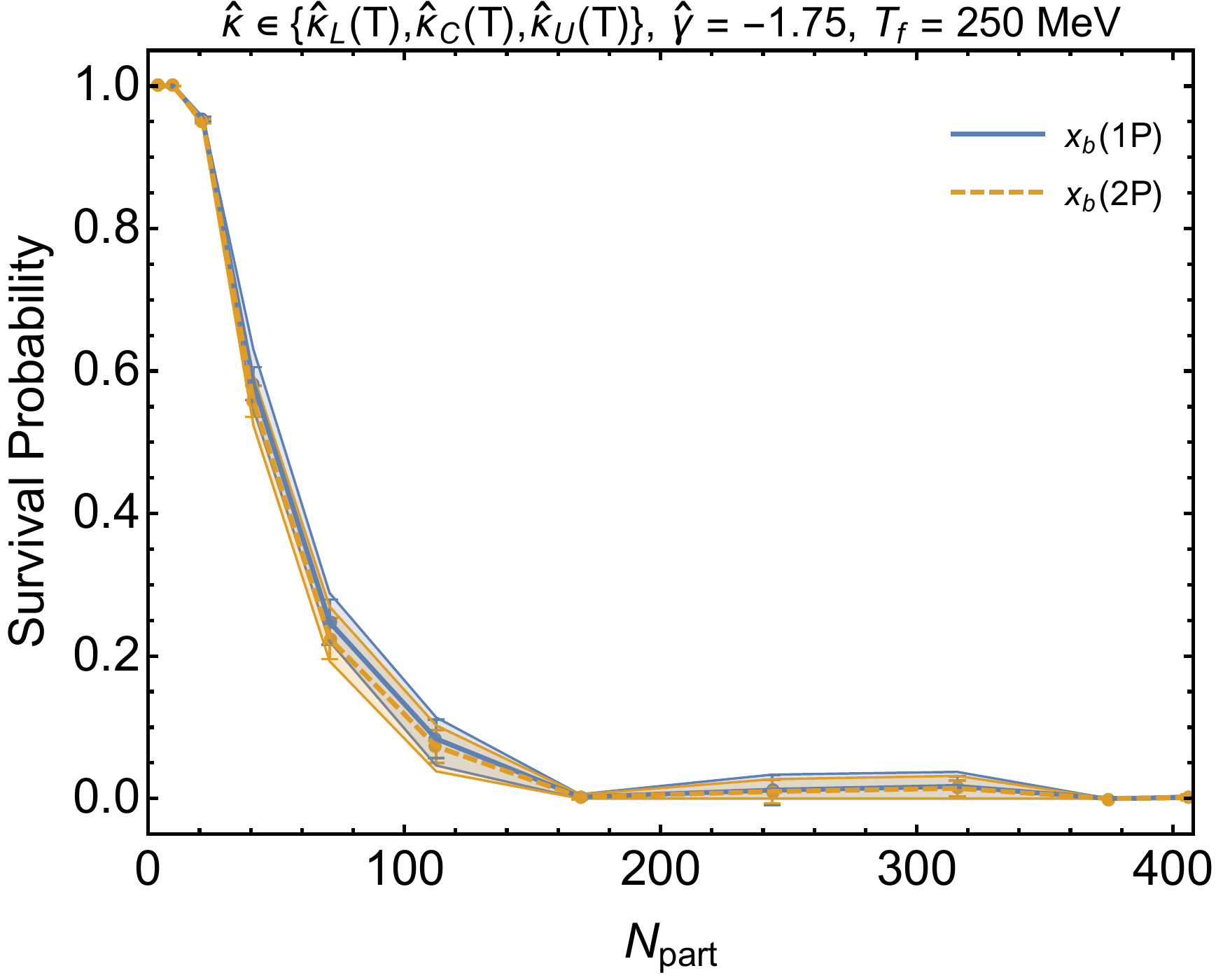}  $\;\;\;$
\includegraphics[width=0.475\linewidth]{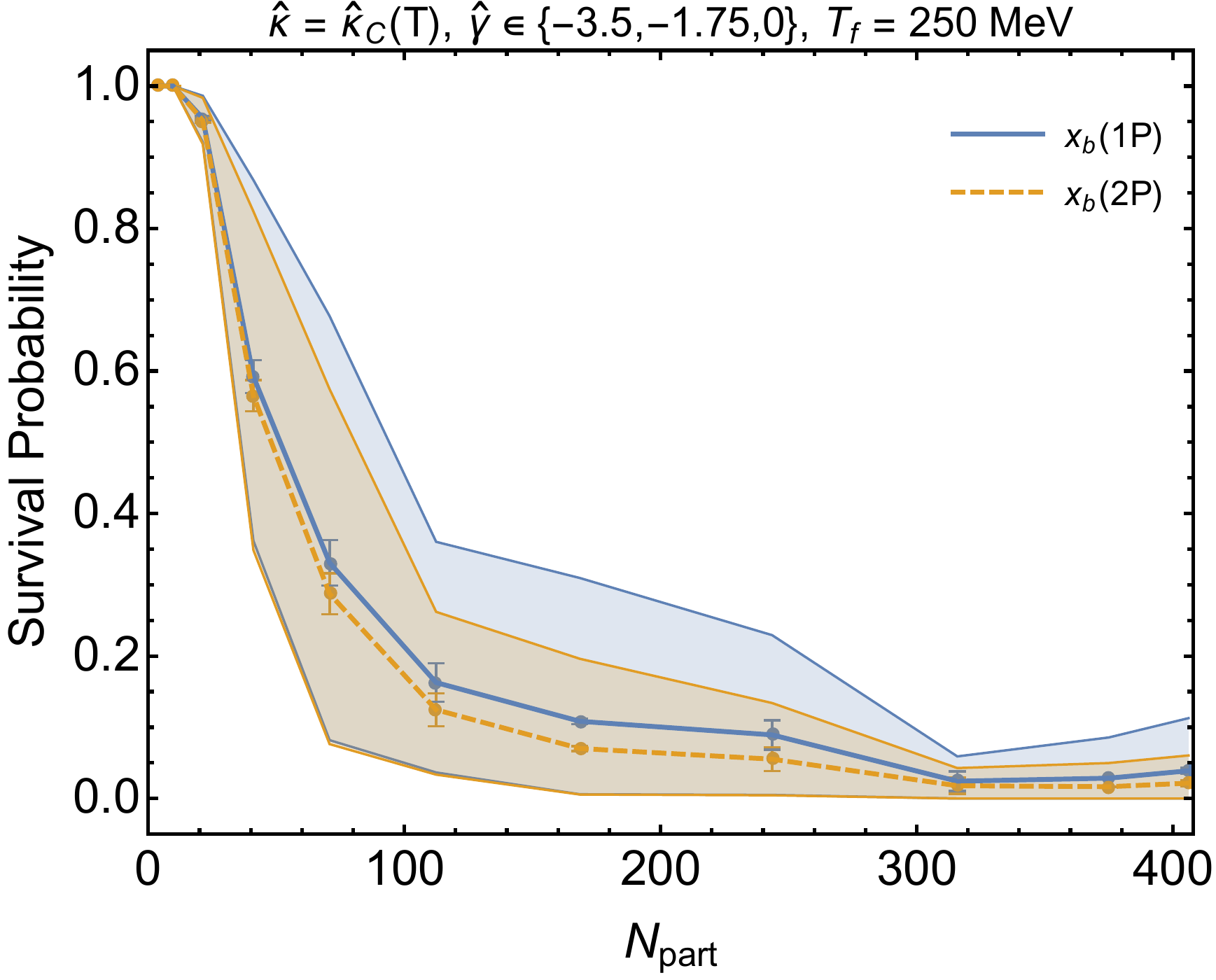} 
\end{center}
\caption{Survival probability of $P$-wave bottomonium states versus $N_{\rm part}$ using
  Gaussian initial conditions to regularize the derivative of the delta function.  
  The bands and error bars shown in the left and right panels correspond to the variations detailed in the caption of figure~\ref{fig:swaveRAA}.}
\label{fig:pwaveRAA}
\end{figure}

\subsubsection*{Singlet delta $P$-wave initial conditions}
In figure~\ref{fig:pwaveRAA}, we show the QTraj results obtained with color singlet $P$-wave ($l=1$) initial conditions.
As in figure~\ref{fig:swaveRAA}, the left and right panels correspond to varying $\hat\kappa$ and $\hat\gamma$ while holding the other coefficient fixed.
We have averaged over approximately 393216 quantum trajectories at each point in $N_\text{part}$, 
and the statistical errors are indicated by error bars on the central line for each state.
As can be seen from these figures, there is a stronger variation with $\hat\gamma$ than with $\hat\kappa$.
In the right panel, the upper bound for the $P$-wave survival probabilities has been obtained when using $\hat\gamma=0$,
with the other two values considered resulting in much stronger $P$-wave suppression.

\subsubsection*{Off-diagonal overlaps and octet initial conditions}
In the previous two subsections, we presented the singlet $S$-wave overlaps resulting from $S$-wave delta initial conditions and the singlet $P$-wave overlaps resulting from $P$-wave delta initial conditions.
Due to the fact that the full QTraj evolution can mix different angular momentum and color states, one can also consider, for example,
{\it (i)} the singlet $S$-wave overlaps resulting from $P$-wave delta initial conditions,
{\it (ii)} the singlet $P$-wave overlaps resulting from $S$-wave delta initial conditions,
and {\it (iii)} the singlet $S$-wave overlaps resulting from octet $P$-wave initial conditions.
In all three cases, we find that these contributions to the final overlaps are small and can be ignored in phenomenological applications.
We provide details concerning our findings in appendix~\ref{Ap:odo}.

\subsection{Final results including late-time feed down of excited states}
\label{sect:feeddown}
Once each quantum state is evolved using QTraj, the survival probabilities are converted into particle numbers by multiplying by
(a) the expected number of binary collisions in the centrality bin sampled and
(b) the direct production cross section for each bottomonium state.
After the number of states that survived transversal of the QGP is computed, one needs to take into account the late time feed down of excited bottomonium states.
Here, we follow refs.~\cite{Islam:2020gdv,Islam:2020bnp} and introduce a feed down matrix $F$ which collects the known information about excited bottomonium state decays available
from the Particle Data Group~\cite{pdg}.  

In the case of pp collisions, one can convert the direct production cross sections into the post feed down cross sections by multiplying a vector containing them
by the feed down matrix $\vec{\sigma}_\text{exp} = F \vec{\sigma}_\text{direct}$ with
\be
F = \left(
\begin{array}{ccccccccc}
	1 & 0.2645 & 0.0194 & 0.352 & 0.18 & 0.0657 & 0.0038 & 0.1153 & 0.077 \\
	0 & 1 & 0 & 0 & 0 & 0.106 & 0.0138 & 0.181 & 0.089 \\
	0 & 0 & 1 & 0 & 0 & 0 & 0 & 0 & 0 \\
	0 & 0 & 0 & 1 & 0 & 0 & 0 & 0.0091 & 0 \\
	0 & 0 & 0 & 0 & 1 & 0 & 0 & 0 & 0.0051 \\
	0 & 0 & 0 & 0 & 0 & 1 & 0 & 0 & 0 \\
	0 & 0 & 0 & 0 & 0 & 0 & 1 & 0 & 0 \\
	0 & 0 & 0 & 0 & 0 & 0 & 0 & 1 & 0 \\
	0 & 0 & 0 & 0 & 0 & 0 & 0 & 0 & 1 \\
\end{array}
\right) ,
\label{eq:fdm}
\ee
where the vectors $\vec{\sigma}$ collect the experimentally-observed and direct cross sections for the $\{ \Upsilon(1S), \Upsilon(2S), \chi_{b0}(1P), \chi_{b1}(1P), \chi_{b2}(1P), \Upsilon(3S), \chi_{b0}(2P), \chi_{b1}(2P), \chi_{b2}(2P) \}$ states per unit rapidity averaged over the rapidity interval $|y| \leq 2.4$.
Also note that, knowing the experimental values for the production cross-sections $\vec{\sigma}_\text{exp}$,
one can compute the direct cross sections via $\vec{\sigma}_\text{direct} = F^{-1} \vec{\sigma}_\text{exp}$.
We take the experimental cross-sections to be $\sigma_\text{exp} = \{ 57.6, 19, 3.72, 13.69, 16.1, 6.8, 3.27, 12.0, 14.15\}$\,nb.   
We note that this results in $\Upsilon(1S)$ feed down fractions of $\{0.747,0.068,0.134, 0.00776,0.0431\}$ for $1S$, $2S$, $1P$, $3S$, and $2P$ states, respectively, which is in reasonable agreement with prior analyses of feed down fractions at low transverse momentum \cite{Woeri:2015hq}.

For the $\Upsilon(1S)$,  $\Upsilon(2S)$, and $\Upsilon(3S)$ cross sections, we use the 5.02 TeV data obtained by the CMS collaboration in the rapidity interval $|y| \leq 2.4$ \cite{Sirunyan:2018nsz}.  In ref.~\cite{Sirunyan:2018nsz} the left panel of figure 3 presents $B \times d\sigma/dy$, where $B$ is the dimuon branching fraction.  
Averaging over rapidity in the interval presented in the CMS figure, we obtain $B \times d\sigma/dy\approx$ 1.44 nb, 0.37 nb, and 0.15 nb, respectively.  
Dividing by the branching fractions for $\Upsilon(1S)$,  $\Upsilon(2S)$, and $\Upsilon(3S)$ $\rightarrow \mu^+ \mu^-$, which are $\approx$ 2.5\%, 1.9\%, and 2.2\%, respectively \cite{pdg}, one obtains
\begin{equation}
	\langle d\sigma[\Upsilon(1S), \Upsilon(2S), \Upsilon(3S)]/dy \rangle_y =  \{57.6, 19, 6.8 \} \text{ nb}.
\end{equation}

For the $\chi_b$ cross sections, we make use of the measurements of ref.~\cite{Aaij:2014caa} from which, together with $\sigma[\Upsilon(1S)]$ and the ratios $\sigma[\chi_{bj}(nP)]/\sigma[\chi_{bj'}(nP)]$, all six of the necessary $\chi_{b}$ cross sections can be calculated.  We take the values of the $\mathcal{R}$ ratio from the lowest $p_{T}$ bins of tables 5 and 6 of ref.~\cite{Aaij:2014caa} (measured at $\sqrt{s}=7$ TeV and $\sqrt{s}=8$ TeV, respectively) and extrapolate to $\sqrt{s}=5$ TeV.
Assuming $\sigma[\chi_{b2}(nP)]/\sigma[\chi_{b1}(nP)] = 1.176$ \cite{HeeSok} for both the $1P$ and $2P$ states (which is consistent with available experimental data \cite{Khachatryan:2014ofa}), we use eq.~(1) of ref.~\cite{Aaij:2014caa} to extract $\sigma[\chi_{bj}(nP)]$ for $j,\, n=1,\,2$.
Less is known about the $\sigma[\chi_{b0}(nP)]$ cross-sections.  
Based on theoretical expectations \cite{HeeSok}, we take the $\chi_{b0}$ cross sections to be 1/4 of the average of the $\sigma[\chi_{b1}(nP)]$ and $\sigma[\chi_{b2}(nP)]$ cross-sections.  
This gives 
\begin{align}
	\langle d\sigma[\chi_{b0}(1P),\chi_{b1}(1P),\chi_{b2}(1P)]/dy \rangle_y &= \{ 3.72, 13.69, 16.1 \} \text{ nb,} \\
	\langle d\sigma[\chi_{b0}(2P),\chi_{b1}(2P),\chi_{b2}(2P)]/dy \rangle_y &= \{ 3.27, 12.0, 14.15 \} \text{ nb.}
\end{align}

\begin{figure}[t]
\begin{center}
\includegraphics[width=0.48\linewidth]{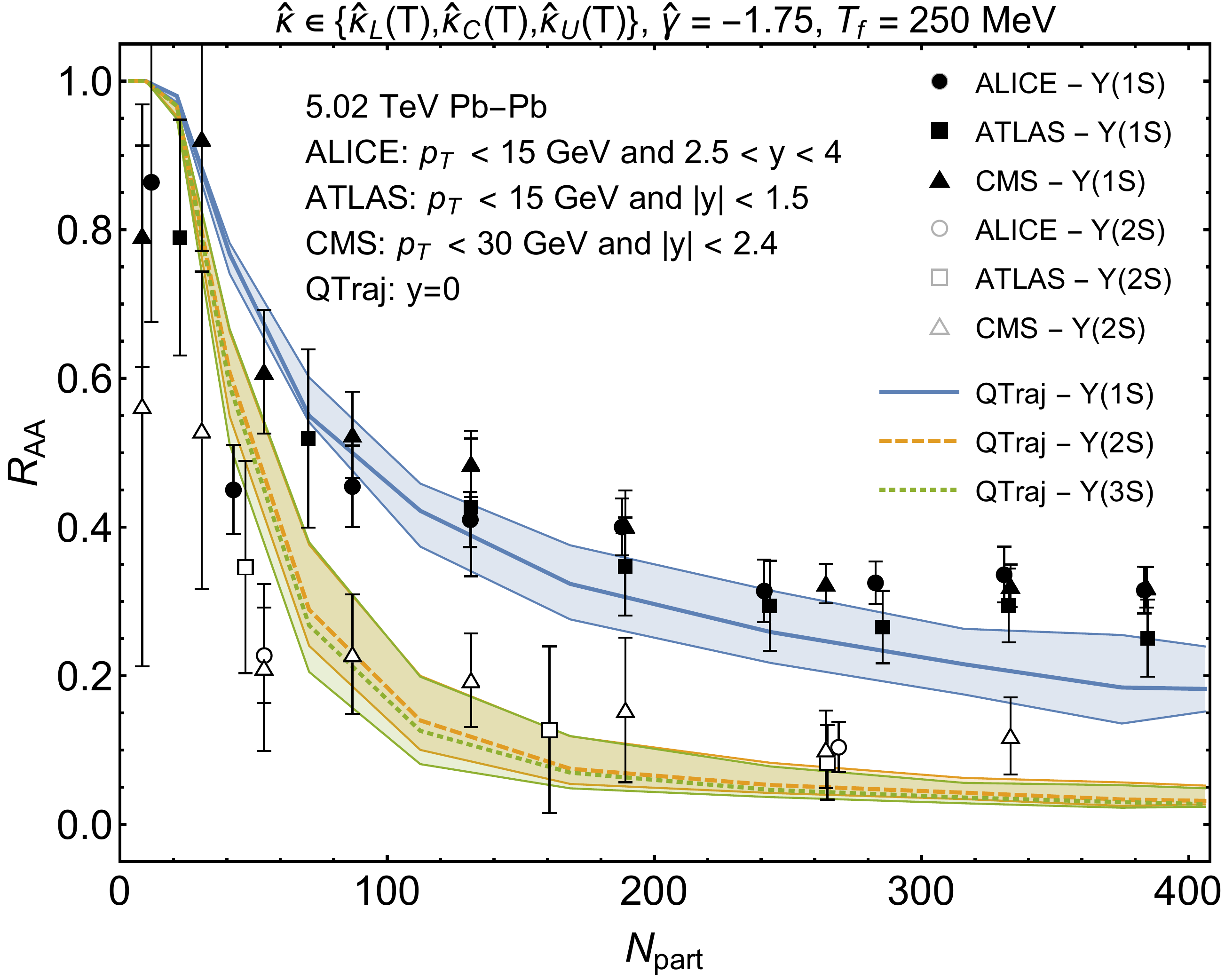}  $\;\;$
\includegraphics[width=0.48\linewidth]{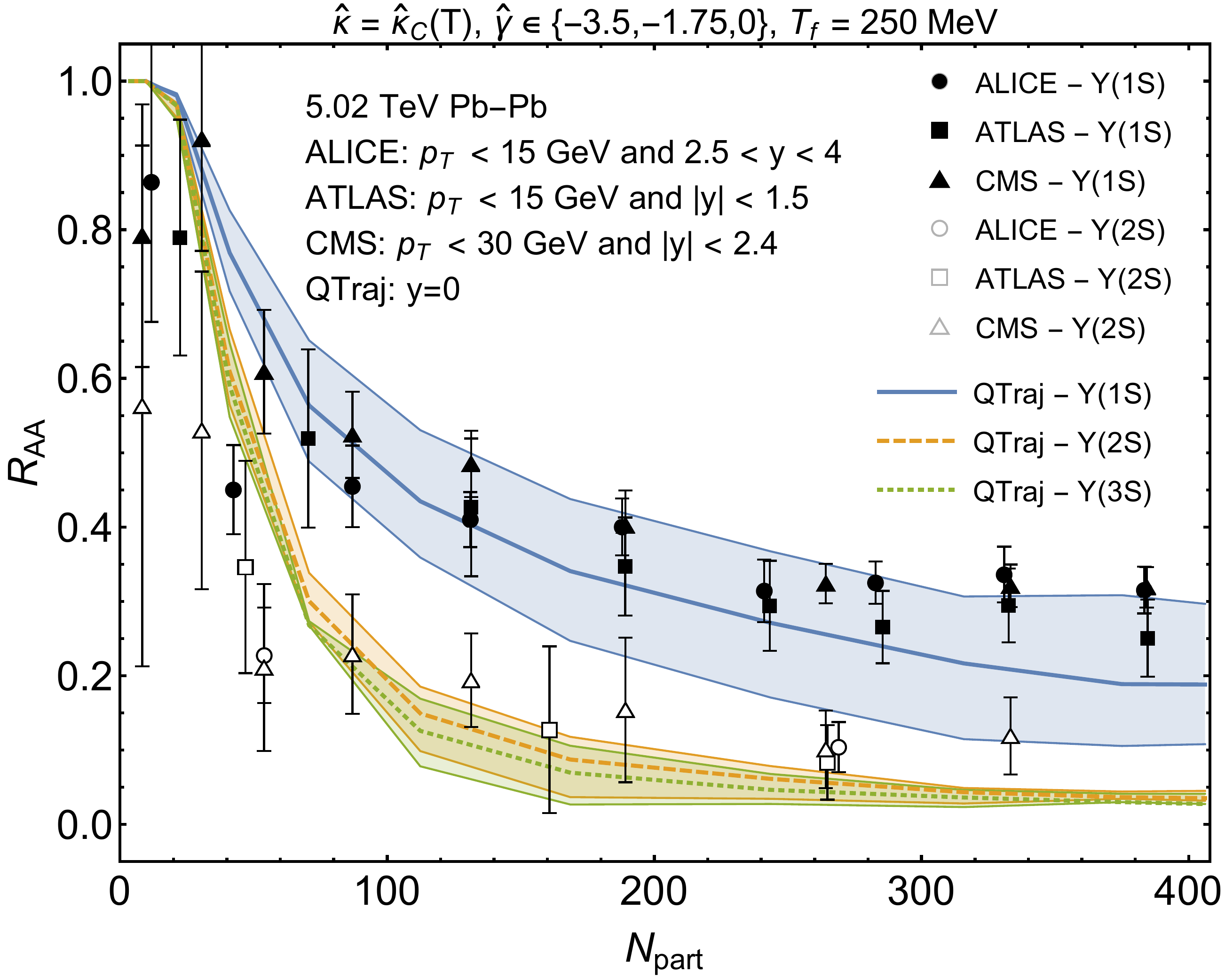}
\end{center}
\caption{$R_{AA}$ of singlet $S$-wave states versus $N_{\rm part}$ taking into account excited state feed down.
  The bands shown in the left and right panels correspond to the variations detailed in the caption of figure~\ref{fig:swaveRAA}.
  The data points are from the  ALICE \cite{Acharya:2020kls}, ATLAS \cite{ATLAS5TeV}, and CMS \cite{Sirunyan:2018nsz} collaborations.
  The experimental error bars were obtained by adding statistical and systematic uncertainties in quadrature.}
\label{fig:raaPostFeeddown}
\end{figure}

To compute the effect of final-state feed down in $AA$ collisions, we first construct a vector $\vec{N}_\text{QGP}$ containing the numbers of each state
produced at the end of each simulation (survival probability $\times \langle N_\text{bin}(b) \rangle \times  \vec{\sigma}_\text{direct}$).
We then multiply the result by the same feed down matrix used for pp feed down, i.e. $\vec{N}_\text{final} = F \vec{N}_\text{QGP}$.
The use of the same feed down matrix for both pp and $AA$ collisions is related to the fact that feed down occurs on a time scale that is much longer than the QGP lifetime.
After the feed down is complete, we compute the post feed down $R_{AA}$ for each state by dividing the final number of each state produced by the average number of binary collisions
in the sampled centrality class times the post feed down pp production cross-section for that state ($\sigma_\text{exp}^i$), giving
\be
R^{\,i}_{AA}(c) =\frac{\left( F \cdot S(c) \cdot \vec\sigma_{\rm direct} \right)^i }{\sigma^i_{\rm exp}} \, ,
\ee
where $i$ labels the state of interest, $S$ is a diagonal matrix that contains the quantum-trajectory averaged survival probabilities for each state along the diagonal, and $c$ indicates the centrality class considered.  The resulting $R^{\,i}_{AA}(c)$ contain the suppression factors for all states included in our analysis.
Since, with respect to the Hamiltonian in our QTraj simulations, states belonging to the same spin multiplet are degenerate, we take the survival probabilities of these states to be the same, i.e. $S[\chi_{bj}(nP)] = S[\chi_{b}(nP)]$ when constructing the survival probability matrix $S$.

In figure~\ref{fig:raaPostFeeddown}, we present our final results for the suppression of the $\Upsilon(1S)$, $\Upsilon(2S)$, and $\Upsilon(3S)$ as a function of $N_\text{part}$
compared to experimental data available from the ALICE \cite{Acharya:2020kls}, ATLAS \cite{ATLAS5TeV}, and CMS \cite{Sirunyan:2018nsz} collaborations.
The bands shown on the QTraj results represent the variations with respect to $\hat\kappa$ (left) and $\hat\gamma$ (right) with the systematic uncertainties propagated through feed down.
As with the pre feed down survival probability we see a large sensitivity to the choice of $\hat\gamma$ and $\hat\kappa$.
Overall, comparing the central lines with the available data we find quite reasonable agreement between the QTraj predictions and the experimental data, however, for the most central collisions, QTraj seems to show a somewhat stronger suppression than is seen in the data.

It is important to note that the minimum temperature at which we consider in-medium damping of states is $T_{\rm f} = $ 250~MeV.
The reason for this choice is that at temperatures lower than 250~MeV the hierarchy $T,m_D \gg E$ may not hold,
and one needs to solve a different set of evolution equations~\cite{Brambilla:2017zei}.
As a result, QTraj predicts that $R_{AA} = 1$ for $N_{\rm part} \lesssim 9.5$ (see table~\ref{tab:collisiondata}), when using the path-averaged temperature.
Nevertheless, damping processes do occur at low temperatures.
This calls for a future extension of the in medium evolution equations and QTraj to the low temperature region, $T_{\rm c} \lesssim T \lesssim T_{\rm f}$,
in order to describe more accurately data at very small $N_{\rm part}$.

\begin{figure}[t]
\begin{center}
\includegraphics[width=0.475\linewidth]{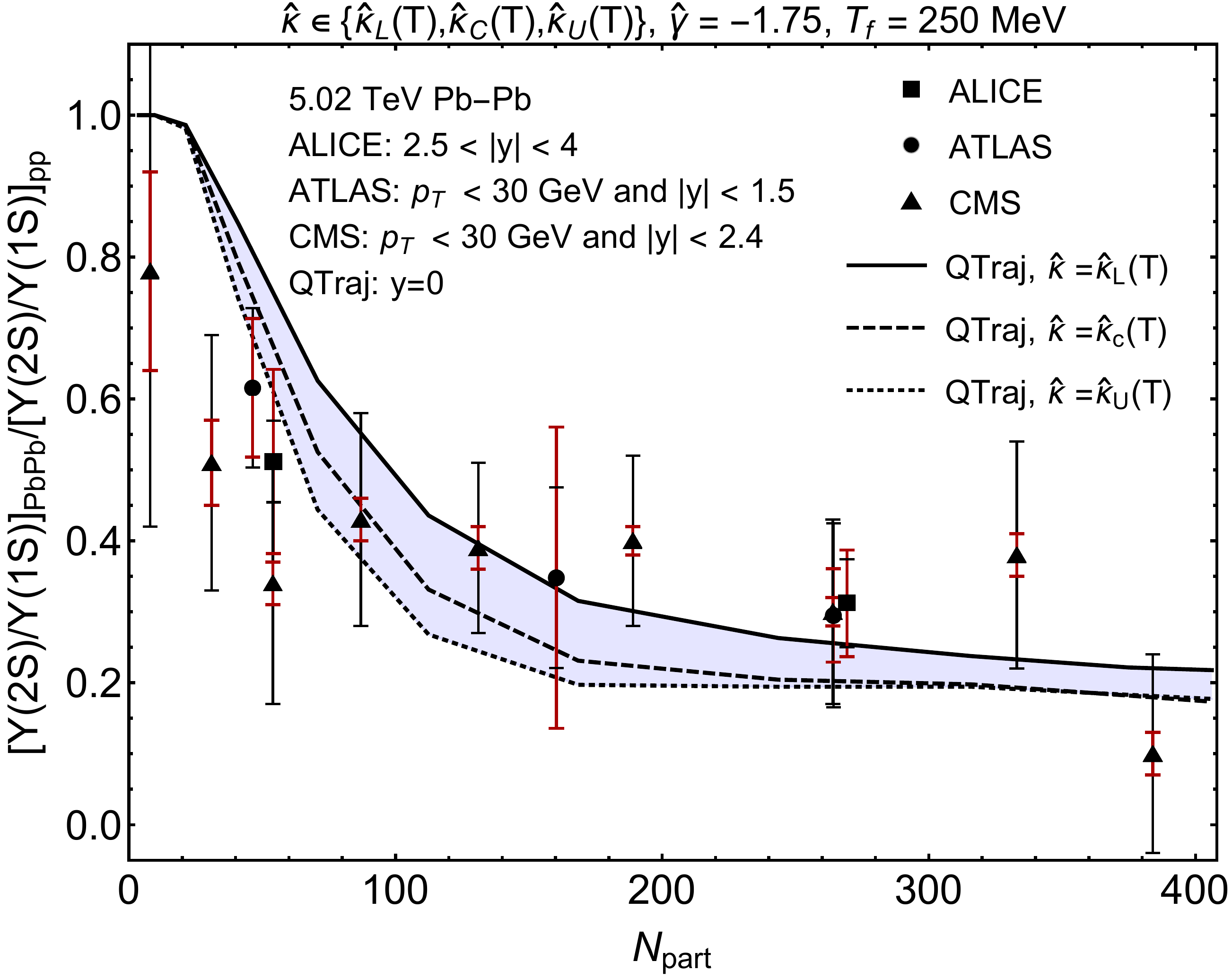}  $\;\;\;$
\includegraphics[width=0.475\linewidth]{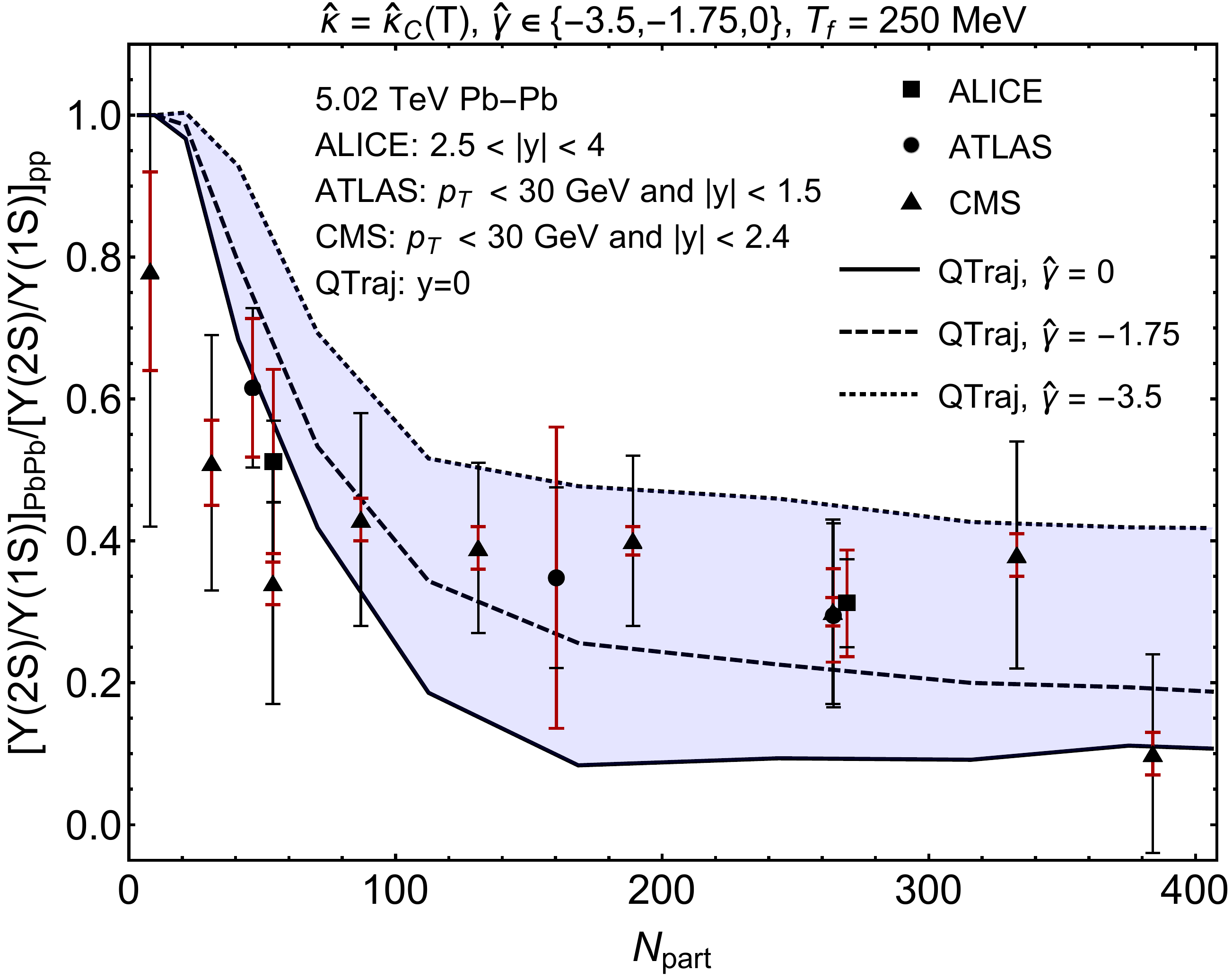}
\end{center}
\caption{Double ratio $[\Upsilon(2S)/\Upsilon(1S)]_\text{PbPb}/[\Upsilon(2S)/\Upsilon(1S)]_\text{pp}$ as a function of $N_\text{part}$.
  The bands shown in the left and right panels correspond to the variations detailed in the caption of figure~\ref{fig:swaveRAA}.
  They account for the (anti-)correlations in the survival probabilities when varying $\hat\kappa$ and $\hat\gamma$.
  The data points are from the ALICE \cite{Acharya:2020kls}, ATLAS~\cite{ATLAS5TeV}, and CMS ~\cite{Sirunyan:2017lzi} collaborations.
  Systematic and statistical experimental uncertainties are indicated by red and black error bars, respectively.}
\label{fig:doubleRatio2s}
\end{figure}

Next, we turn to the double ratios constructed from $\Upsilon(2S)$ and $\Upsilon(3S)$ to $\Upsilon(1S)$.
These are obtained by computing the ratio of $\Upsilon(nS)$ and $\Upsilon(1S)$ yields in PbPb collisions, divided by the same ratio in pp collisions;
we will indicate this double ratio with $[\Upsilon(nS)/\Upsilon(1S)]_{\text{PbPb}}/$ $[\Upsilon(nS)/\Upsilon(1S)]_{\text{pp}}$.
On the experimental side, these quantities have typically smaller systematic uncertainties, which allow for tighter constraints on theoretical models.
In figure~\ref{fig:doubleRatio2s}, we present the QTraj results for the double ratio $[\Upsilon(2S)/\Upsilon(1S)]_{\text{PbPb}}/$ $[\Upsilon(2S)/\Upsilon(1S)]_{\text{pp}}$ as a function of $N_{\rm part}$.
We compare the QTraj results with data available from the ALICE \cite{Acharya:2020kls}, ATLAS~\cite{ATLAS5TeV}, and CMS ~\cite{Sirunyan:2017lzi} experiments.
For QTraj, we show the theoretical uncertainty as a shaded blue band; for the experimental systematic and statistical uncertainties we use black and red error bars, respectively.
As can be seen from the figure, QTraj
is within 1$\sigma$ of the statistical error bars.
We see visible deviations at small $N_\text{part}$, with these again being due to the fact that we do not allow in-medium breakup at low temperatures as discussed above.
By comparing the left and right panels of figure~\ref{fig:doubleRatio2s},
we see that the QTraj results depend more on the assumed value of $\hat\gamma$ than on the value of $\hat\kappa$.
In all cases considered, QTraj predicts that the $2S$ to $1S$ double ratio depends only mildly on $N_\text{part}$ for $N_\text{part} \gtrsim 150$.

\begin{figure}[ht]
\begin{center}
\includegraphics[width=0.475\linewidth]{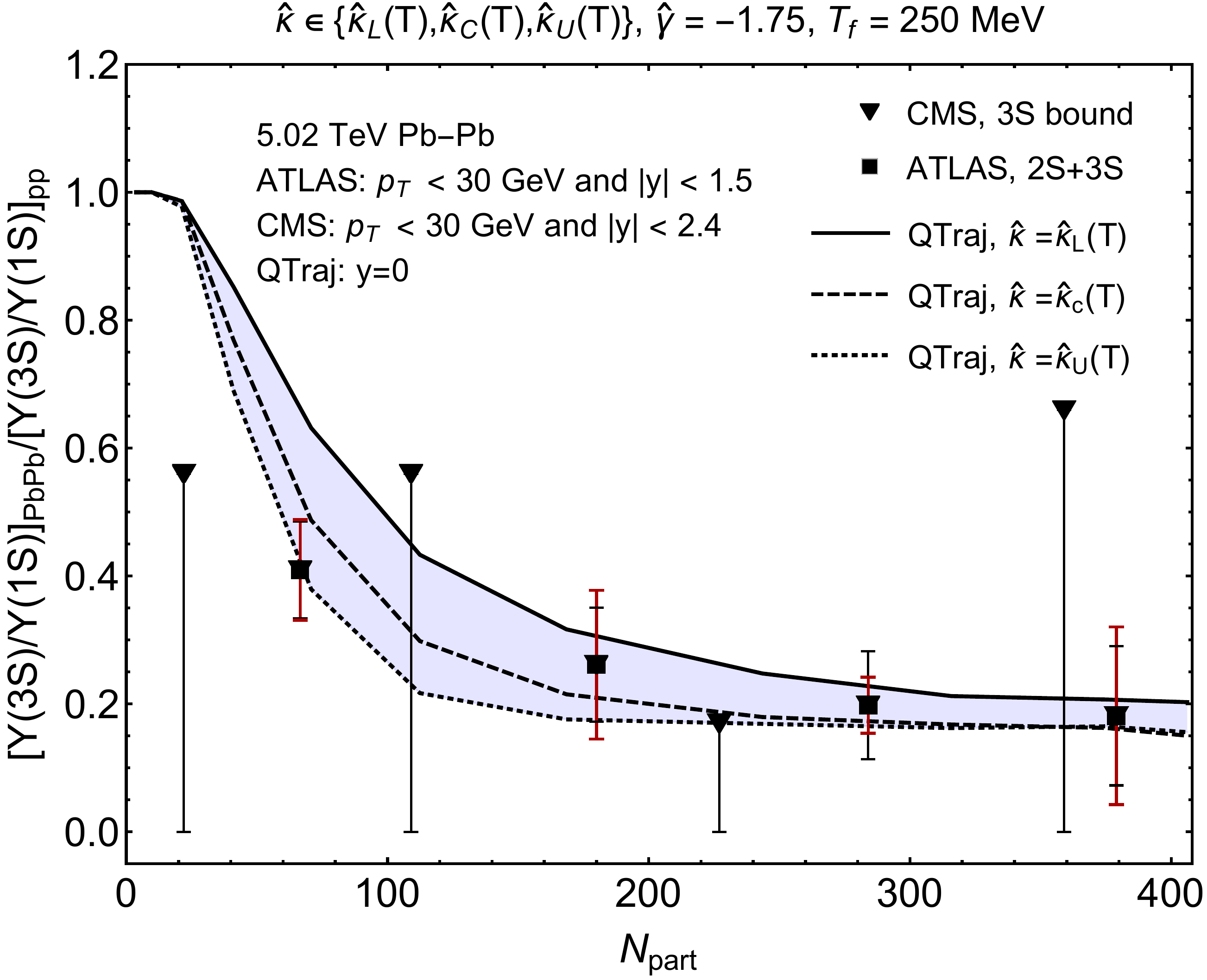}  $\;\;\;$
\includegraphics[width=0.475\linewidth]{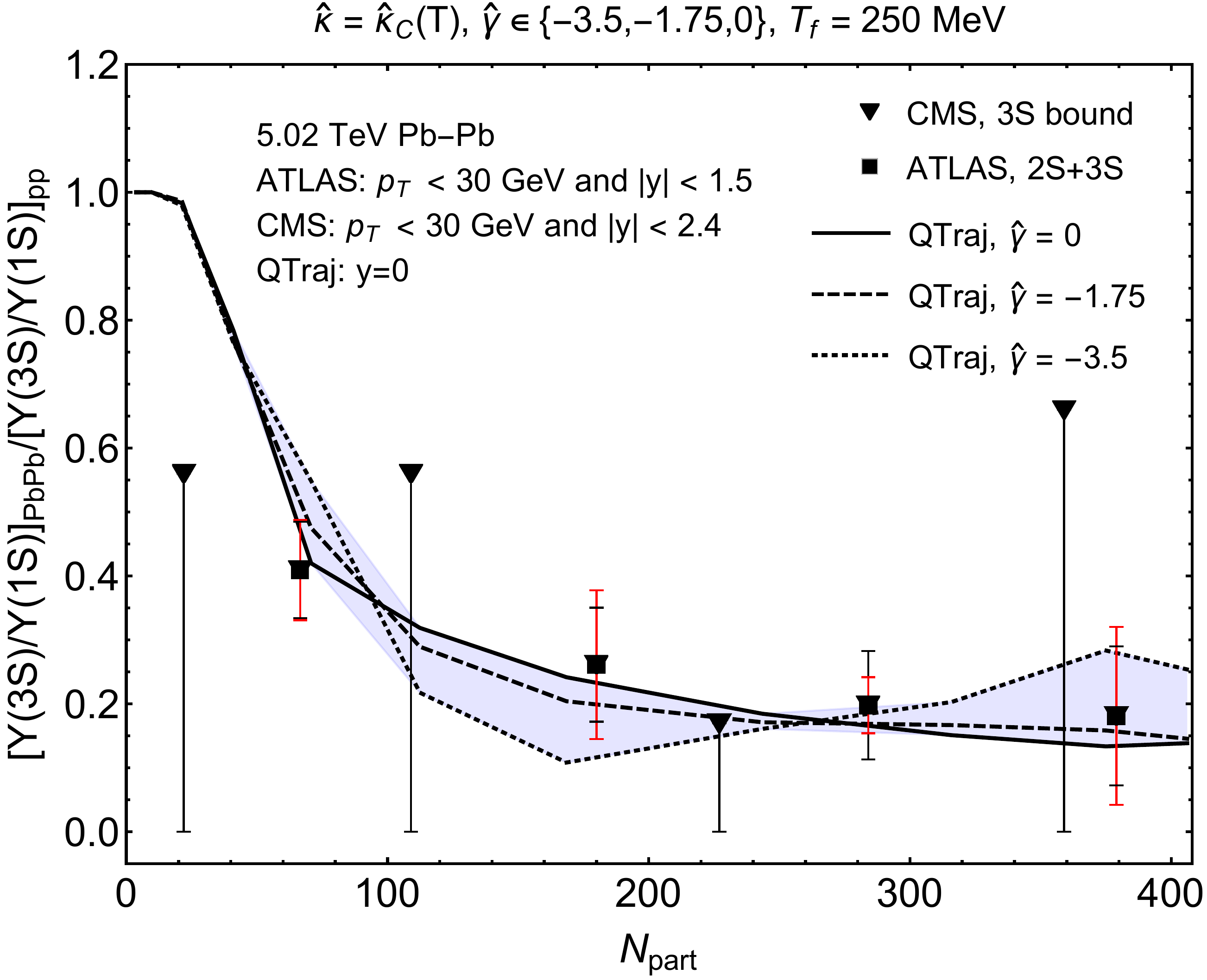}
\end{center}
\caption{Double ratio $[\Upsilon(3S)/\Upsilon(1S)]_\text{PbPb}/[\Upsilon(3S)/\Upsilon(1S)]_\text{pp}$ as a function of $N_\text{part}$.  Data and QTraj results are as detailed in the caption of figure~\ref{fig:doubleRatio2s}.}
\label{fig:doubleRatio3s}
\end{figure}

Finally, in figure~\ref{fig:doubleRatio3s} we present the QTraj results for the $3S$ to $1S$ double ratio and compare with experimental data
from the ATLAS~\cite{ATLAS5TeV} and CMS ~\cite{Sirunyan:2017lzi} experiments.  
In the case of the CMS data, the results were reported as upper bounds on the $3S$ to $1S$ double ratio.
In the case of ATLAS, the collaboration presented results only for an integrated $2S+3S$ double ratio.
As can be seen from this figure, the QTraj results for the $3S$ to $1S$ ratio depend less strongly on the assumed value of $\hat\gamma$
than in the double ratio $[\Upsilon(2S)/\Upsilon(1S)]_{\text{PbPb}}/$ $[\Upsilon(2S)/\Upsilon(1S)]_{\text{pp}}$.
Similar to the $2S$ to $1S$ double ratio, we see that QTraj predicts that the $3S$ to $1S$ ratio does not depend strongly on $N_\text{part}$ for $N_\text{part} \gtrsim 150$.
In the future, increased statistics from the ATLAS and CMS collaborations may allow for more constraints on the $3S$ to $1S$ double ratio.

\section{Conclusions}
\label{sec:conclusions}
In this paper, we have presented a new program called QTraj to solve the Lindblad equation
describing the nonequilibrium evolution of a color singlet and octet heavy quark-antiquark pair of a small radius in a strongly coupled QGP
characterized by an inverse correlation length larger than the typical quark-antiquark energy.
The evolution equations were originally derived in~\cite{Brambilla:2017zei,Brambilla:2016wgg} and solved there with the publicly available code QuTiP.
The new code QTraj presents several distinct advantages with respect to the one previously used.
It shows a sizeable reduction of the memory requirement from ${\cal O}(N^2)$ to ${\cal O}(N)$ for a system simulated with $N$ discrete points,
due to the fact that it is computing the wave-function instead of directly the 
density matrix. The need to average over a large number of trajectories is efficiently counterbalanced by running many trajectories in an embarrassingly parallel setup.

The need to average over a large number of trajectories is efficiently counterbalanced by running many trajectories in parallel.
In addition, QTraj allows to account for wave-functions of any angular momentum quantum number $l$
and uses all-points derivatives allowing for more accurate solutions of the evolution equations.
In particular, with the new code we could eliminate several of the restrictions necessary in~\cite{Brambilla:2017zei,Brambilla:2016wgg}, 
such as only including $l \leq 1$ states and using a Bjorken time evolution for the temperature profile of the plasma.

Many new investigations have become possible that have been performed in this paper for the first time.
{\it (i)} We included in the solution of the evolution equations the contributions coming from states of all angular momentum quantum numbers $l$,
    without the need of introducing a cutoff in this parameter.
{\it (ii)} We studied also the evolution the in-medium of quarkonium $P$ states.
{\it (iii)} We coupled the Lindblad equation to a realistic medium evolution based on the quasiparticle anisotropic hydrodynamics (aHydroQP) framework
for dissipative relativistic hydrodynamics.
{\it (iv)} We studied the dependence on the initial conditions.
{\it (v)} We thoroughly investigated the impact of color octet heavy quark-antiquark pairs on the evolution of the quarkonium in the medium.
{\it (vi)} We quantified the recombination contribution with respect to the effective Hamiltonian evolution, 
in this way assessing the importance of quantum jumps for heavy quarkonium evolution in the QGP.
{\it (vii)} We investigated extensively the dependence of the quarkonium nuclear modification factor $R_{AA}$
on the two parameters that characterize the quarkonium evolution in the QGP, i.e. the heavy quark momentum coefficient $\kappa$ and its dispersive counterpart $\gamma$.
In particular, we explored for the first time the impact of the temperature dependence of $\kappa/T^3$, as extracted from \cite{Brambilla:2020siz},
on the quarkonium evolution and on $R_{AA}$.

An interesting feature of our approach is the characterization of the quarkonium evolution in the QGP
in terms of only two transport coefficients $\kappa$ and $\gamma$, which are, in general, temperature dependent.
We have seen that the survival probabilities depend on the values of these parameters.
In particular, increasing or decreasing $\kappa$ results in a decreased or enhanced survival probability for all three $S$-wave bottomonium states below threshold,
while the impact of $\gamma$ is different from one state to the other.
For the $\Upsilon(1S)$, a smaller value of $\gamma$ results in a reduced survival probability, while for the $\Upsilon(2S) $ the reverse is true.
Notable is, in particular, the $\gamma$ dependence of the $P$-state survival probabilities.
Therefore, the quarkonium survival probability can act directly as a diagnostic of the QGP through these transport coefficients.

We explored and quantified the impact of the quantum jumps, i.e. the recombination effects.
We found that quantum jumps seem to only marginally affect the $\Upsilon(1S)$,
whose survival probability can be well described using just the effective Hamiltonian evolution.
For excited states, however, quantum jumps are found to give a sizeable contribution. The effect increases by increasing the value of $\kappa$.
We already commented that increasing $\kappa$ decreases the survival probabilities of the states:
quantum jumps correct and mitigate this effect with respect to the pure effective Hamiltonian evolution.

Finally, we included the late time feed down from excited states and we compared our results to the bottomonium nuclear suppression factor measured by the ALICE, ATLAS, and CMS collaborations.
It is important to emphasize that the computed bottomonium nuclear suppression factor does not depend on any free parameter as both $\kappa$ and $\gamma$
have been determined independently by lattice QCD in ref.~\cite{Brambilla:2020siz} and~\cite{Brambilla:2019tpt}, respectively.
Hence the comparison is really between a QCD prediction and data.
We obtain a good agreement with data, which is better than the one reported in \cite{Brambilla:2016wgg,Brambilla:2017zei},
confirming the importance of developing the new program in order to lift the limitations of the previous code.
The predictions of QTraj have small uncertainties that depend on the allowed values for the $\kappa$ and $\gamma$ transport coefficients.
In particular, there is a larger dependence on $\gamma$ that calls for a dedicated lattice study of this coefficient in a wide temperature range.
Our formalism, as currently implemented, does not include medium effects at temperatures below $250$~MeV.
This is an approximation that is accurate up to corrections of relative order $(a_0T)^2$, which are not included.
As already mentioned at the end of section \ref{sect:hydro}, this approximation might neglect physical effects, like thermal gluodissociation, 
  possibly more relevant for excited states, $\Upsilon(2S)$ and $\Upsilon(3S)$, than for the $\Upsilon(1S)$.
  The reason is that our assumed hierarchy of scales is more marginally realized for the former than for the latter bottomonia.
  This is visible in our results where the relative error in the survival probability coming from the variation of $\kappa$ and $\gamma$
  is indeed larger for the excited states.
Although the absolute size of the medium effects at low temperatures is presumably small (specially for the case of $\Upsilon(1S)$),
it would be desirable to include these effects in a future work to better address collisions with a very small number of participants and to improve our description of excited states.
In a forthcoming work, we will include corrections of order $E/T$ to the currently considered evolution equations, thus extending the range of validity of our method to lower temperatures \cite{nlo}.  

We considered also double ratios of $S$-wave production cross sections in pp and PbPb, which eliminated several of the systematics, both theoretically and experimentally.
We predicted that the $2S$ to $1S$ and the $3S$ to $1S$ double ratios do not depend strongly on $N_{\rm part}$ for  $N_{\rm part} \gtrsim 150$
and we made quantitative predictions for these ratios.

Using QTraj we plan to explore other interesting observables in the near future such as the $p_T$ dependence of $R_{AA}$ and $v_2$.
Moreover, we plan to relax the assumption of isotropy and solve the Lindblad equation in such cases where the $\kappa$ and $\gamma$ coefficients become tensors instead of numbers.
Longer term goals include to consider quarkonium not at rest with respect to the QGP~\cite{Escobedo:2013tca},
to extend these studies to quarkonia with larger radius
and eventually to solve the full evolution master equations \cite{Brambilla:2016wgg,Brambilla:2017zei} far from equilibrium.

\acknowledgments
We thank Hee Sok Chung for input and discussions on the $\chi_b(nP)$ production cross sections used to calculate the late-time feed down of excited states.
M.S. has been supported by the U.S. Department of Energy, Office of Science, Office of Nuclear Physics Award No.~DE-SC0013470.
M.S. also thanks the Ohio Supercomputer Center for support under the auspices of Project No.~PGS0253.  
J.H.W. has been supported by the U.S.\ Department of Energy, Office of Science, Office of Nuclear Physics and Office of Advanced Scientific Computing Research
within the framework of Scientific Discovery through Advance Computing (SciDAC) award Computing the Properties of Matter with Leadership Computing Resources.
J.H.W.'s research has been also funded by the Deutsche Forschungsgemeinschaft (DFG, German Research Foundation) - Projektnummer 417533893/GRK2575 ``Rethinking Quantum Field Theory''.
This work has received financial support from Xunta de Galicia (Centro singular de investigaci\'{o}n de Galicia accreditation 2019-2022), by European Union ERDF,
and by  the ``Mar\'{i}a  de Maeztu''  Units  of  Excellence program  MDM-2016-0692  and  the Spanish Research State Agency.
N.B., P.V. and A.V. acknowledge the support from the Bundesministerium f\"ur Bildung und Forschung project no. 05P2018
and by the DFG cluster of excellence ORIGINS funded by the Deutsche Forschungsgemeinschaft under Germany's Excellence Strategy - EXC-2094-390783311.
The QuTiP simulations have been carried out on the local theory cluster (T30 cluster) of the Physics Department of the Technische Universit\"at M\"unchen (TUM).

\appendix

\section{Change of orbital momentum during a quantum jump}
\label{Ap:om}
In order to determine the probability to jump to any given orbital angular momentum state,
we need to study the term of the Lindblad equation that goes like $\sum_n C_n\rho C_n^\dagger$.
If we consider a density matrix $\rho$ block diagonal in the quantum numbers $l$ and $m$, then the result of computing $\sum_n C_n\rho C_n^\dagger$ is also block diagonal. 
For simplicity, let us consider a density matrix that only contains states with a given orbital momentum.
Since the Lindblad equation is linear, it will be trivial to generalize to any other density matrix with spherical symmetry.
Then we can split the original term $\sum_i C_i^x\rho^l {C^x_i}^\dagger$
(where $x$ can be either $0$ or $1$, see notation in section \ref{ssec:hq}) in a sum $\sum_{l'}C^{x,l\to l'}\rho_l {C^{x,l\to l'}}^\dagger$,
where each term generates a density matrix with an orbital momentum $l'$. 

The procedure that we follow in practice is the following.
First, in the waiting time approach, we determine if a jump takes place by computing $\langle \Psi^l|\Psi^l\rangle$, 
where we explicitly write the orbital angular momentum of the wave-function to clarify the notation.
After determining whether the wave-function is affected by the collapse operator that induces singlet-octet transitions, $C_i^0$, or octet-octet transitions, $C_i^1$,
the system will jump to a state with angular momentum $l'$ with probability
\begin{equation}
\frac{\langle\Psi^l|{C^{x,l\to l'}}^\dagger C^{x,l\to l'}|\Psi^l\rangle}{\sum_{l'}\langle\Psi^l|{C^{x,l\to l'}}^\dagger C^{x,l\to l'}|\Psi^l\rangle}\,.
\end{equation} 
Since we can write $C_i^x=A^x r \left({r_i}/{r}\right)$ where $A^x$ is a matrix in color space, the previous equation is equal to
\begin{equation}
\frac{1}{2l+1}\sum_{mm'}|\langle Y_{lm}|\frac{r_i}{r}|Y_{l'm'}\rangle|^2\,,
\end{equation}
where $Y_{lm}$ are spherical harmonics and we have taken into account that the system does not have any preferred polarization.
We note that the previous quantity is only non-zero if $l'=l+1$ or $l'=l-1$.

Now we can perform the computation using the following argument. Starting from the identity
\begin{equation}
\frac{1}{2l+1}\sum_m \langle Y_{lm}|Y_{lm}\rangle=1\,,
\end{equation}
we can use the fact that the spherical harmonics can be used to expand any function on the unit sphere to deduce that
\begin{equation}
P_d^l+P_u^l=1\,,
\end{equation}
where
\begin{equation}
P_d^l=\frac{1}{2l+1}\sum_{mm'}|\langle Y_{lm}|\frac{r_i}{r}|Y_{l-1\,m'}\rangle|^2\,,
\end{equation}
and
\begin{equation}
P_u^l=\frac{1}{2l+1}\sum_{mm'}|\langle Y_{lm}|\frac{r_i}{r}|Y_{l+1\,m'}\rangle|^2\,.
\end{equation}
We can identify $P_d^l$ as the probability to jump to a state with angular momentum $l-1$ and $P_u^l$ as the probability to jump to a state with angular momentum $l+1$.
It then holds that
\begin{equation}
P_d^l=\frac{2l-1}{2l+1}P_u^{l-1}=\frac{2l-1}{2l+1}\left(1-P_d^{l-1}\right)\,.
\label{eq:angprob}
\end{equation}
From this recurrence relation we determine $P_d^l$ and $P_u^l$ knowing that $P_d^0=0$:
\begin{equation}
P_d^l=\frac{l}{2l+1}\qquad {\rm and} \qquad P_u^l=\frac{l+1}{2l+1}\,.
\label{eq:angprob2}
\end{equation}

\section{Off-diagonal overlaps}
\label{Ap:odo}
In the main body of the paper, we focused on the singlet overlaps resulting from singlet initial conditions with a fixed angular momentum $l$.
For singlet $S$-wave initial conditions we presented the resulting $S$-wave overlaps and, likewise, for singlet $P$-wave initial conditions we presented the resulting $P$-wave overlaps.
In this appendix, we demonstrate why it is consistent to ignore the off-diagonal contributions corresponding to,
e.g., singlet $P$-wave overlaps resulting from singlet $S$-wave initial conditions and singlet $S$-wave overlaps from octet $P$-wave initial conditions.
Such off-diagonal contributions are generated during the QTraj evolution due to the quantum jumps, while without quantum jumps ($H_{\rm eff}$ evolution only) such overlaps are identically zero.
We will present evidence that, when including quantum jumps, the off-diagonal contributions are small enough as to be ignored for phenomenological applications.

\begin{figure}[ht]
\begin{center}
\includegraphics[width=0.475\linewidth]{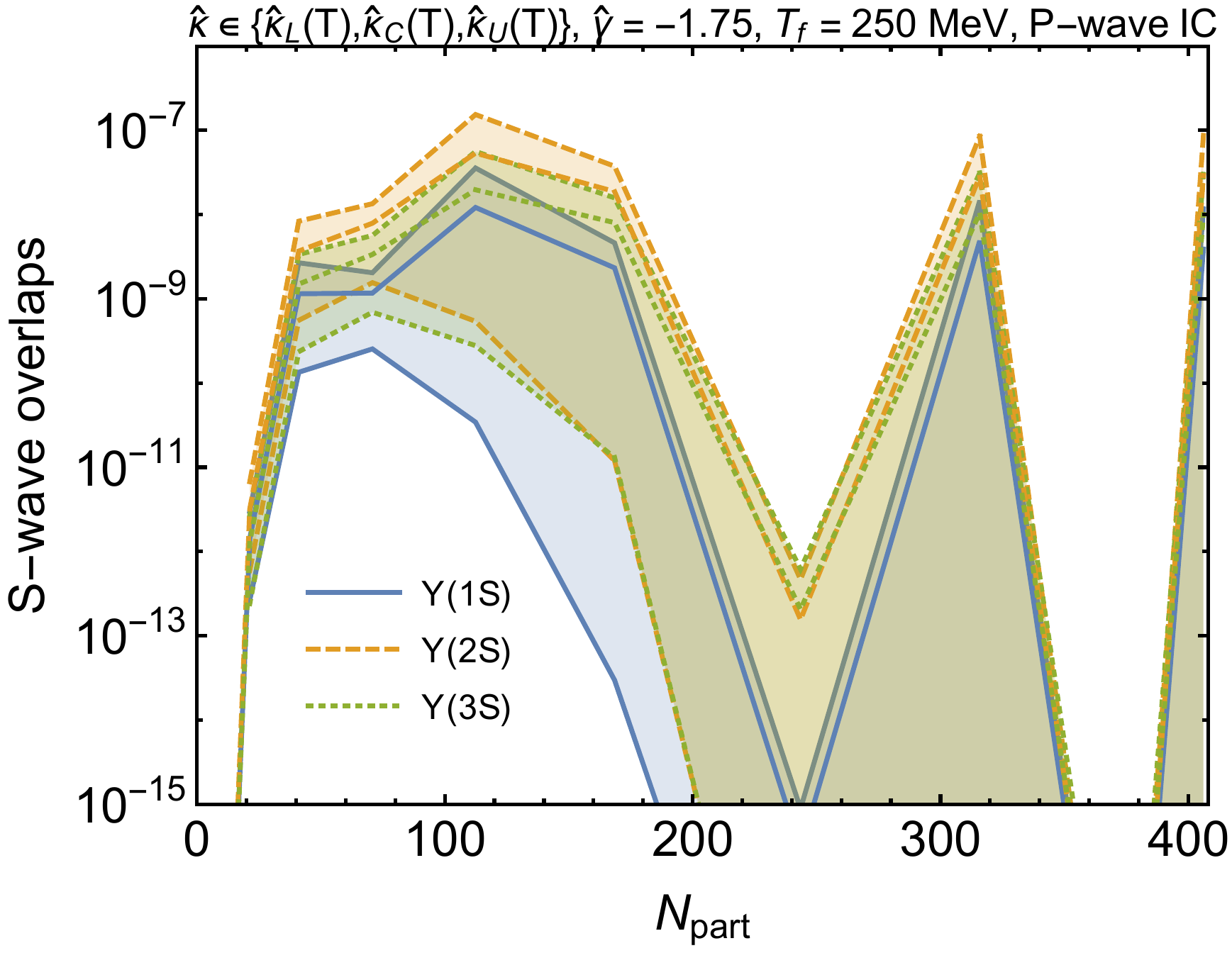} $\;\;\;$
\includegraphics[width=0.475\linewidth]{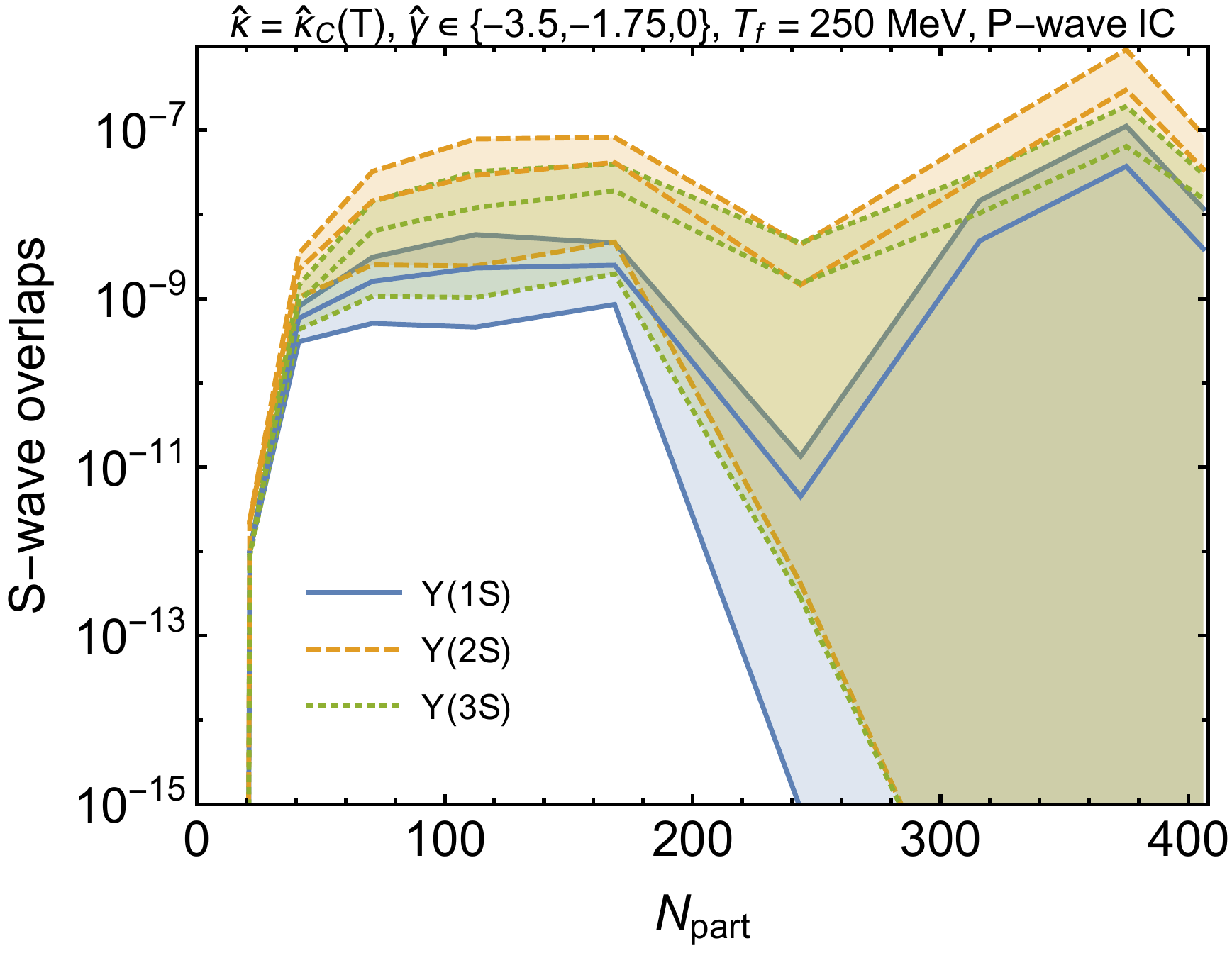}
\end{center}
\caption{Quantum-mechanical overlaps with singlet $S$-wave states obtained using singlet $P$-wave initial conditions.
The bands shown in the left and right panels correspond to the variations detailed in the caption of figure~\ref{fig:swaveRAA}.
In both panels, the central line shows the result averaged over the corresponding variation.}
\label{fig:swaveOverlaps-pwaveIC}
\end{figure}

In figure~\ref{fig:swaveOverlaps-pwaveIC}, we present the singlet $S$-wave overlaps resulting from singlet $P$-wave initial conditions as a function of $N_\text{part}$.
As in the main body of the text, the left and right panels show the variation over the assumed values of $\hat\kappa$ and $\hat\gamma$, respectively.
In order to gauge the magnitude of these numbers, we note that for a central collision the singlet $S$-wave overlaps resulting from singlet $S$-wave initial conditions
(corresponding to figure~\ref{fig:swaveRAA}) are approximately $6 \times 10^{-3}$, $1 \times 10^{-4}$, and $2 \times 10^{-5}$ for the $\Upsilon(1S)$, $\Upsilon(2S)$, and $\Upsilon(3S)$, respectively.
Comparing to the magnitude of the overlaps shown in figure~\ref{fig:swaveOverlaps-pwaveIC}, we see that the off-diagonal contribution in angular momentum is small in this case.
This contribution is additionally suppressed by the fact that the $P$-wave initial production cross sections are down by a factor of approximately four with respect to the $S$-wave production cross-section.
For this reason we ignore this contribution in our final phenomenological predictions.

\begin{figure}[ht]
\begin{center}
\includegraphics[width=0.475\linewidth]{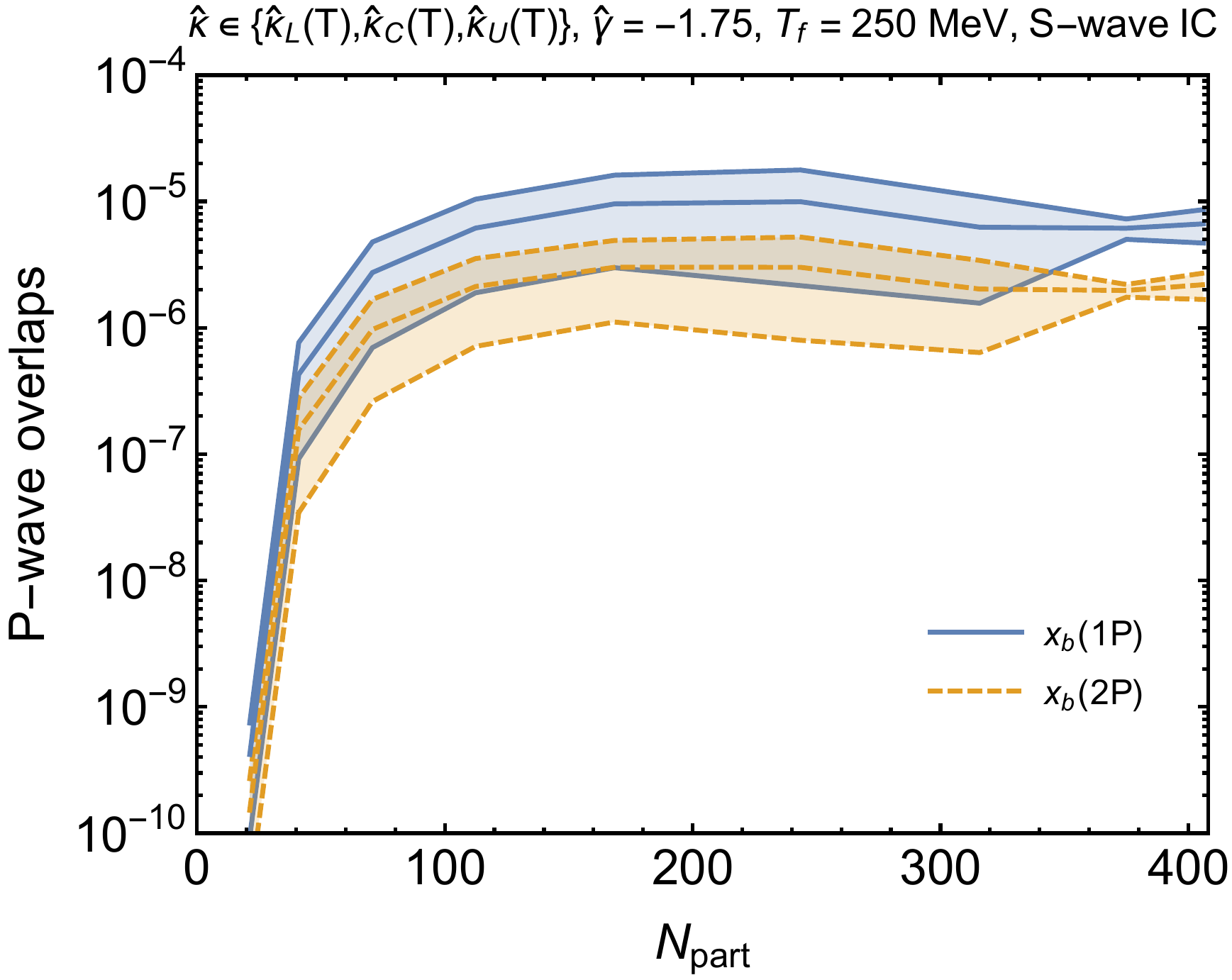} $\;\;\;$
\includegraphics[width=0.475\linewidth]{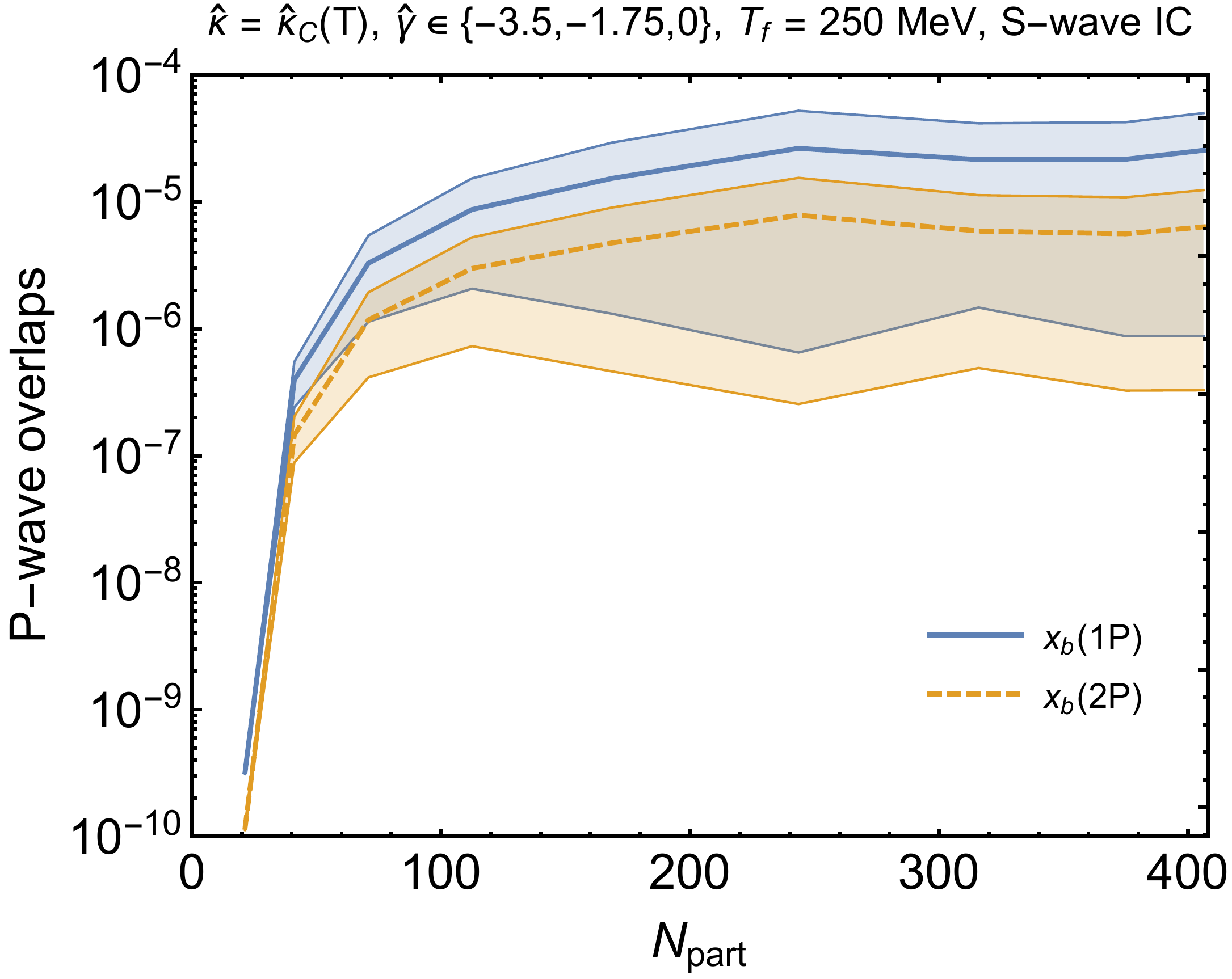}
\end{center}
\caption{Quantum-mechanical overlaps with singlet $P$-wave states obtained using singlet $S$-wave initial conditions.
The bands shown in the left and right panels correspond to the variations detailed in the caption of figure~\ref{fig:swaveRAA}.
In both panels, the central line shows the result averaged over the corresponding variation. }
\label{fig:pwaveOverlaps-swaveIC}
\end{figure}

\begin{figure}[ht]
\begin{center}
\includegraphics[width=0.475\linewidth]{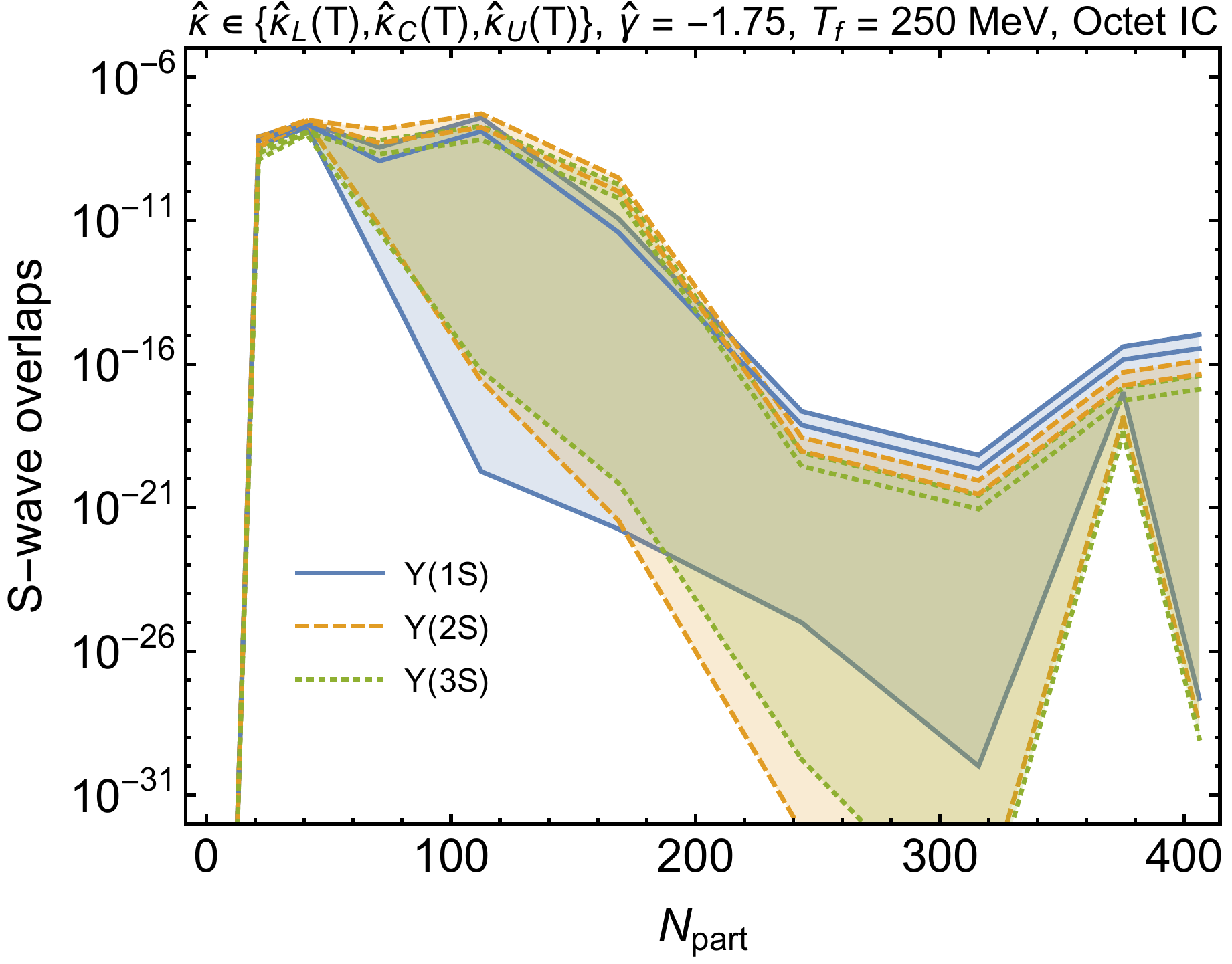} $\;\;\;$
\includegraphics[width=0.475\linewidth]{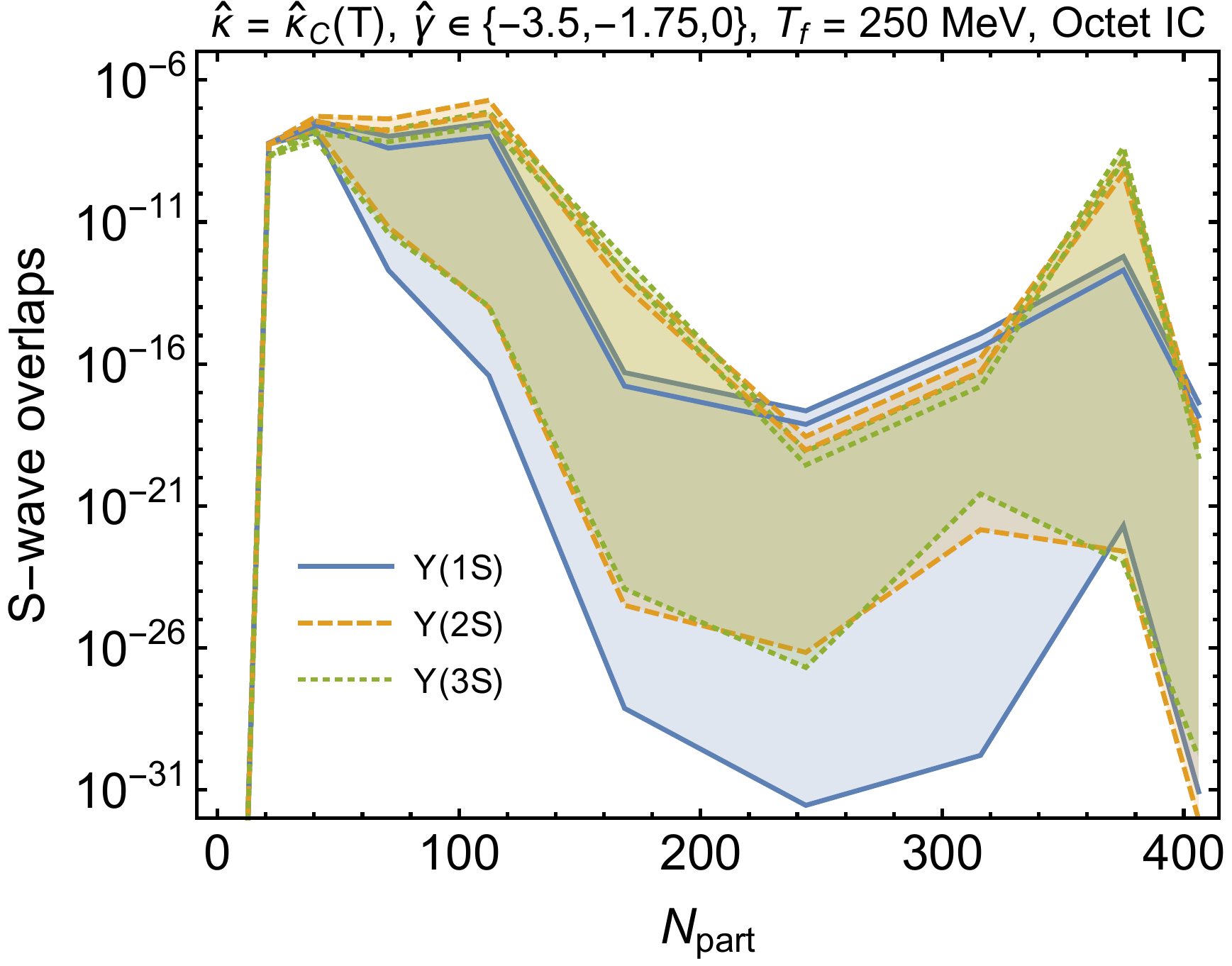}
\end{center}
\caption{Quantum-mechanical overlaps with singlet $S$-wave states obtained using octet $P$-wave initial conditions.
The bands shown in the left and right panels correspond to the variations detailed in the caption of figure~\ref{fig:swaveRAA}.
In both panels, the central line shows the result averaged over the corresponding variation. }
\label{fig:swaveOverlaps-octetIC}
\end{figure}

In figure~\ref{fig:pwaveOverlaps-swaveIC}, we present the singlet $P$-wave overlaps resulting from singlet $S$-wave initial conditions as a function of $N_\text{part}$.
In order to gauge the magnitude of these numbers, we note that for a central collision
the singlet $P$-wave overlaps resulting from singlet $P$-wave initial conditions (corresponding to figure~\ref{fig:pwaveRAA})
are approximately in between $10^{-4}$  and $10^{-7}$, and in between $10^{-5}$ and $10^{-8}$
for the $\chi_b(1P)$ and $\chi_b(2P)$, respectively, depending on the assumed values of $\hat\kappa$ and $\hat\gamma$.
These are very small numbers, whose effect falls well inside the range of variations considered when varying $\hat\kappa$ and $\hat\gamma$ (see figure~\ref{fig:pwaveRAA}).
We show that the smallness of the off-diagonal contributions is generic in appendix~\ref{sec:swavestudies},
where we present the diagonal and off-diagonal overlaps obtained when the initial wave-function is taken to be a pure $\Upsilon(1S)$ or $\Upsilon(2S)$.

In figure~\ref{fig:swaveOverlaps-octetIC}, we present the singlet $S$-wave overlaps resulting from octet $P$-wave initial conditions as a function of $N_\text{part}$.
In order to gauge the magnitude of these numbers, we note that for a central collision the singlet $S$-wave overlaps resulting from singlet $S$-wave initial conditions
(corresponding to figure~\ref{fig:swaveRAA}) are approximately $6 \times 10^{-3}$, $1 \times 10^{-4}$, and $2 \times 10^{-5}$ for the $\Upsilon(1S)$, $\Upsilon(2S)$, and $\Upsilon(3S)$, respectively.
As can be seen from figure~\ref{fig:swaveOverlaps-octetIC}, the $S$-wave overlaps resulting from octet $P$-wave initial conditions are
orders of magnitude smaller than the overlaps resulting from singlet $S$-wave initial conditions.
For this reason, we can safely ignore the off-diagonal octet-singlet contributions when considering phenomenology.

\begin{figure}[b]
	\begin{center}
		\includegraphics[width=0.46\linewidth]{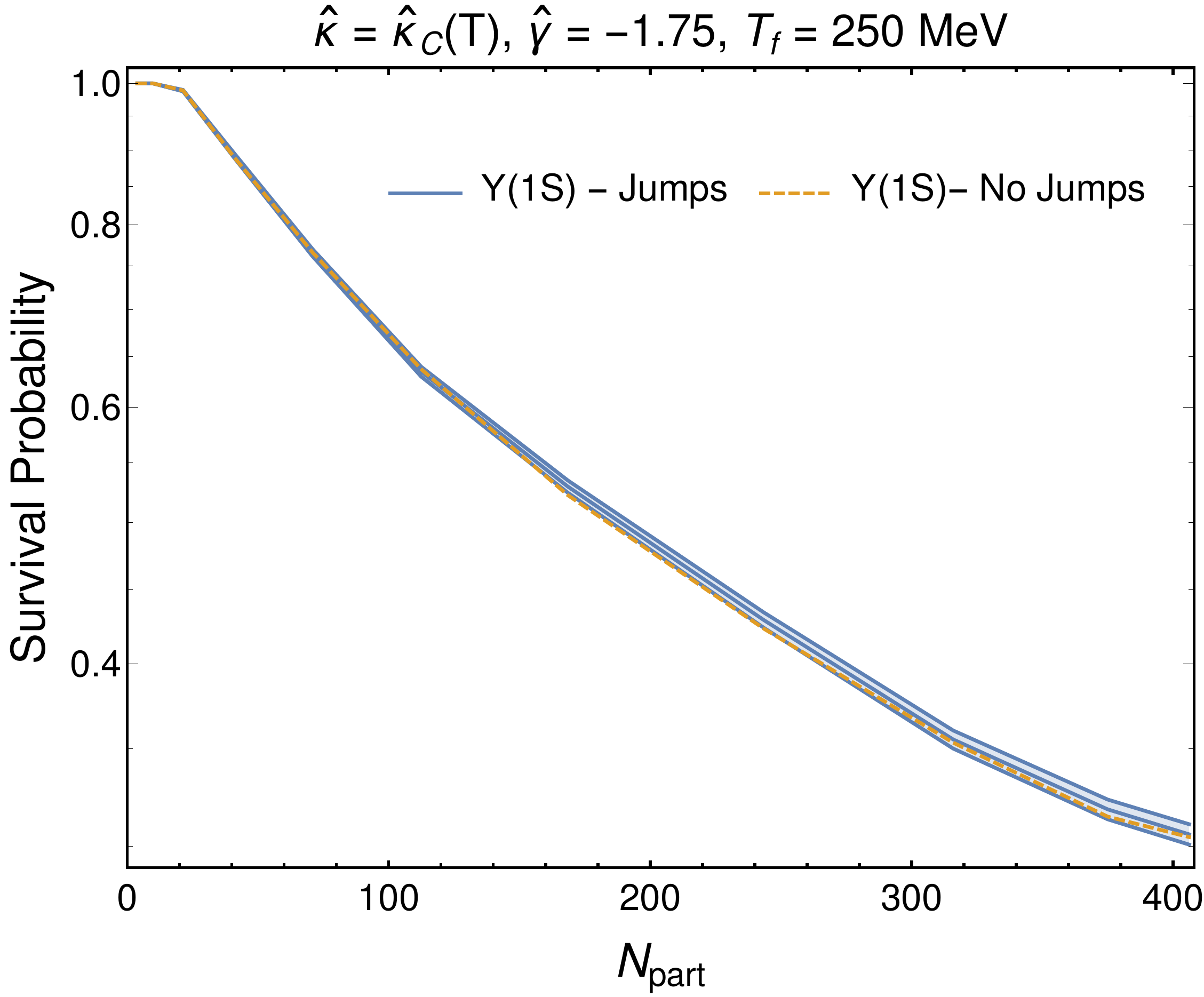} $\;\;\;$
		\includegraphics[width=0.48\linewidth]{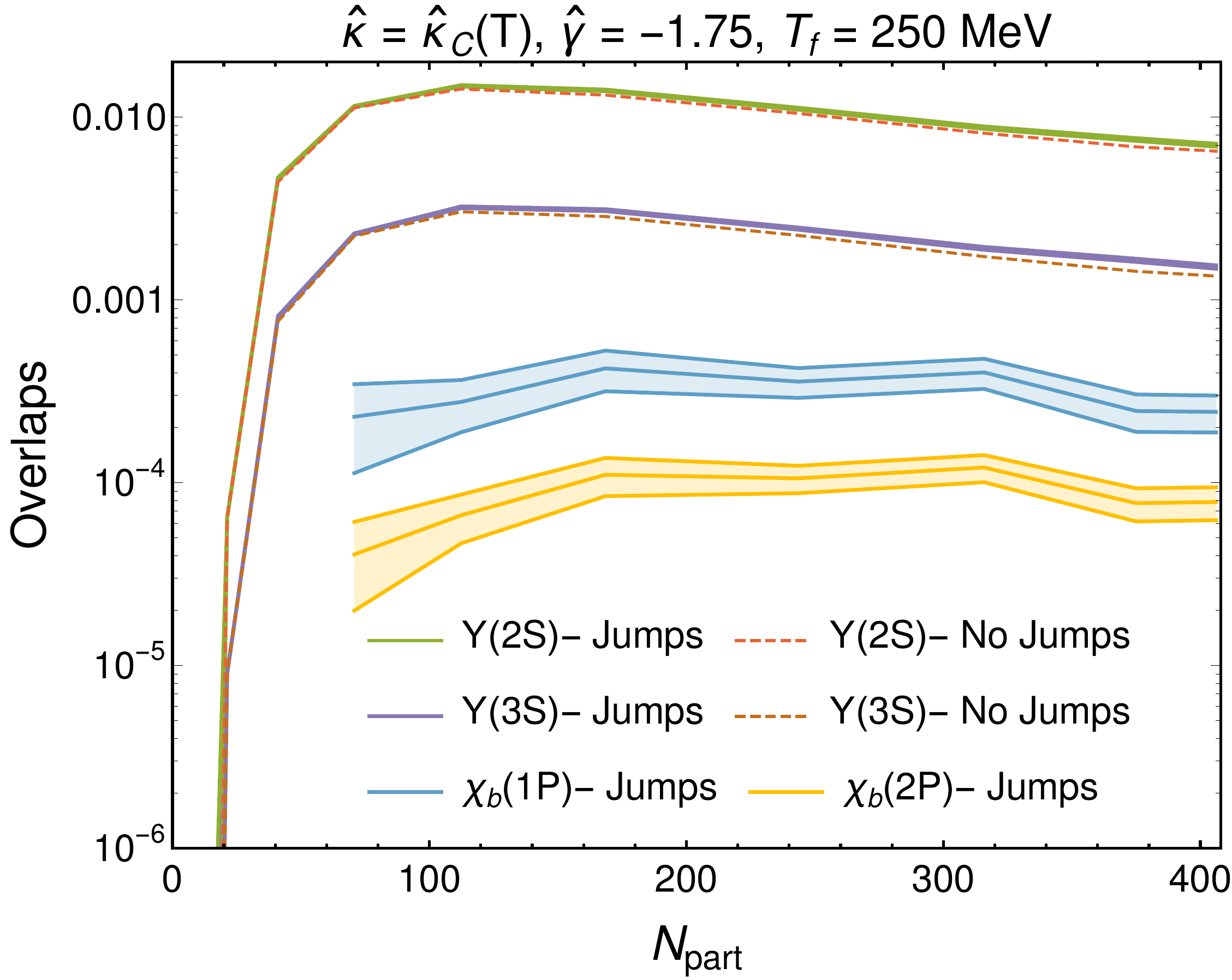}
	\end{center}
	\caption{Survival probabilities/overlaps  versus $N_{\rm part}$ with an $\Upsilon(1S)$ initial condition.
		The left panel shows the $1S$ survival probability and the right panel shows the final overlaps of the excited states.
		The continuous lines are the full QTraj evolution including jumps and the dashed lines are the cases with no jumps;
		the bands indicate the statistical uncertainty of the full QTraj result which was obtained using 8192 quantum trajectories per point in $N_{\rm part}$.}
	\label{fig:swaveOverlaps-oneSIC}
\end{figure}

\section{$S$-wave initialization studies} 
\label{sec:swavestudies}
\begin{figure}[t]
\begin{center}
\includegraphics[width=0.475\linewidth]{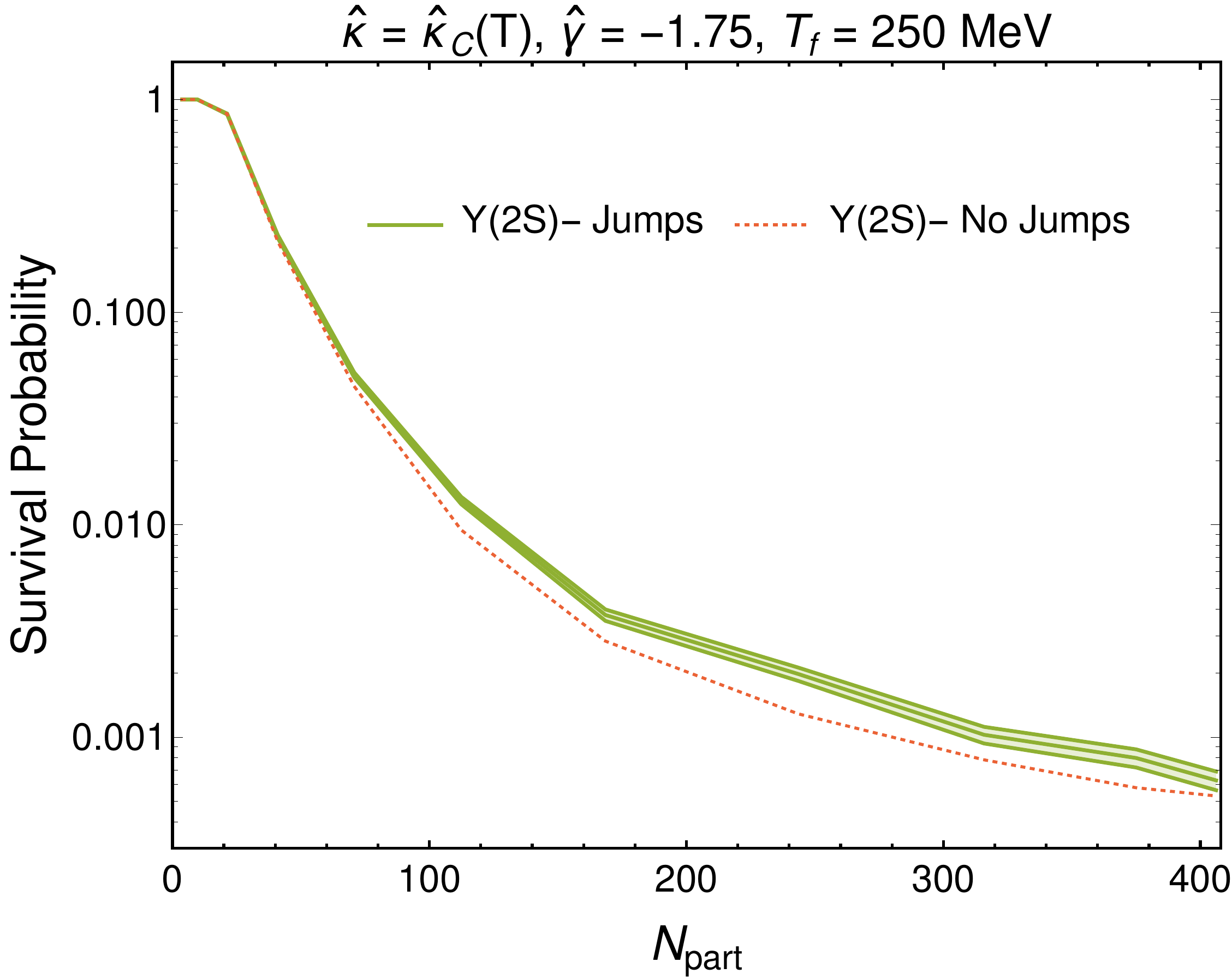} $\;\;\;$
\includegraphics[width=0.475\linewidth]{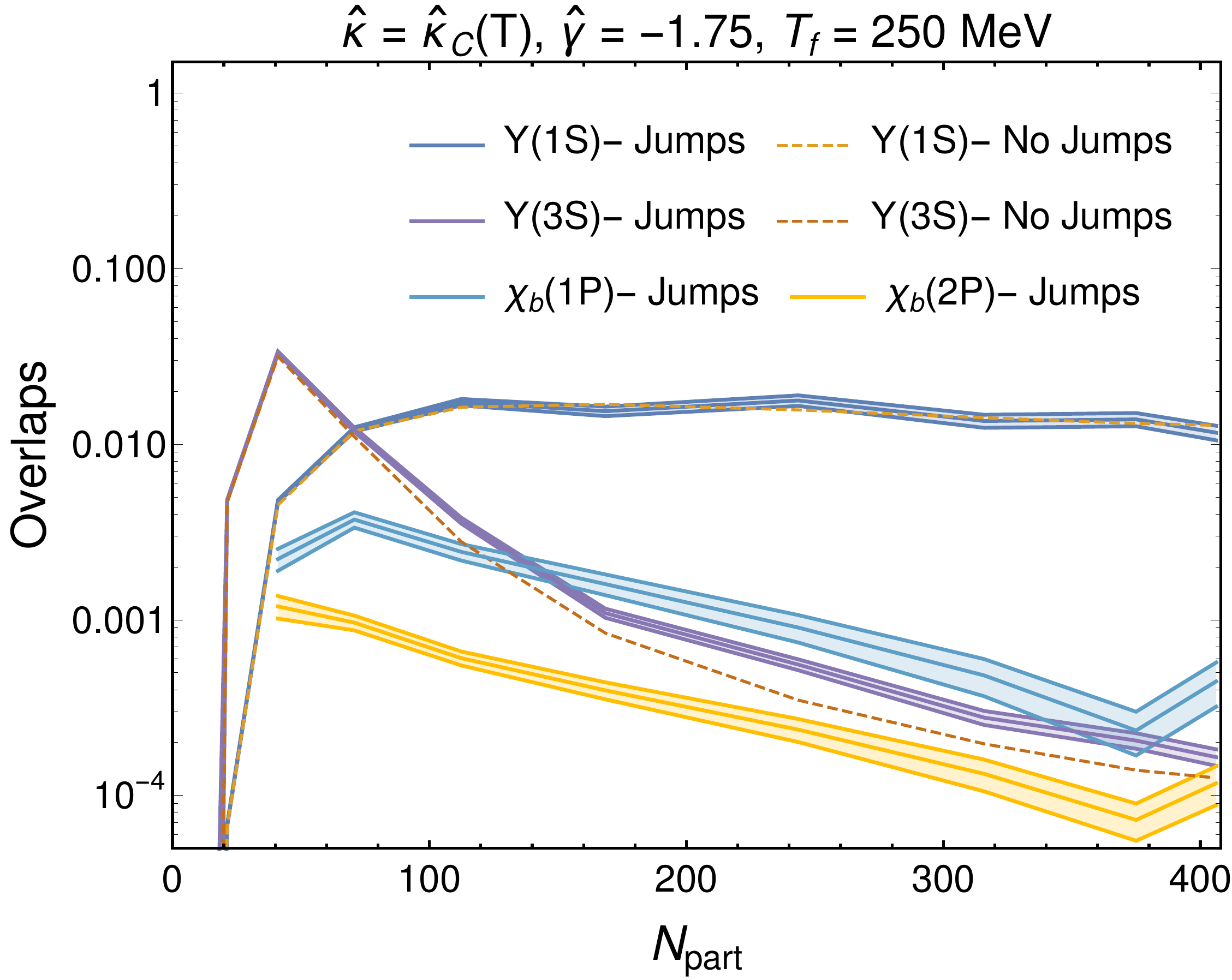}
\end{center}
\caption{Survival probabilities/overlaps versus $N_{\rm part}$ with an $\Upsilon(2S)$ initial condition.  Styling is the same as figure~\ref{fig:swaveOverlaps-oneSIC}.}
\label{fig:swaveOverlaps-twoSIC}
\end{figure}

In this appendix, we present results for the overlap factors versus $N_{\rm part}$ obtained with pure $S$-wave initial conditions. 
Figure~\ref{fig:swaveOverlaps-oneSIC} shows results for an $\Upsilon(1S)$ initial condition and figure~\ref{fig:swaveOverlaps-twoSIC} results for an $\Upsilon(2S)$ initial condition.
In both figures, we show the results for the $S$-wave overlaps with the full evolution including the jumps 
and with the evolution without the jumps using only the complex Hamiltonian $H_\text{eff}$ of eq.~\eqref{eq:Heff}. 
The overlap factor of the ground state $\Upsilon(1S)$ is marginally larger for the full evolution with jumps, and the result without jumps can serve as its lower bound.
The overlap factors for the lower excited $S$-wave states are successively suppressed by about an order of magnitude each. 
The excited $S$-waves are slightly more enhanced by the jumps, although most of the contribution is already present in the evolution without jumps.
$P$-wave states, which are generated only in the evolution with jumps, are suppressed against $S$-wave states of the same principal quantum number by another order of magnitude.
Starting from an $\Upsilon(2S)$ state, the $\Upsilon(1S)$ overtakes already for $N_{\rm part} \gtrsim 100$ and becomes the dominant state. 
$\Upsilon(3S)$ on the other hand, assumes a maximum at $N_{\rm part} \approx 50$, where it exceeds even the $\Upsilon(1S)$. 

\begin{figure}[th]
\begin{center}
\includegraphics[width=0.65\linewidth]{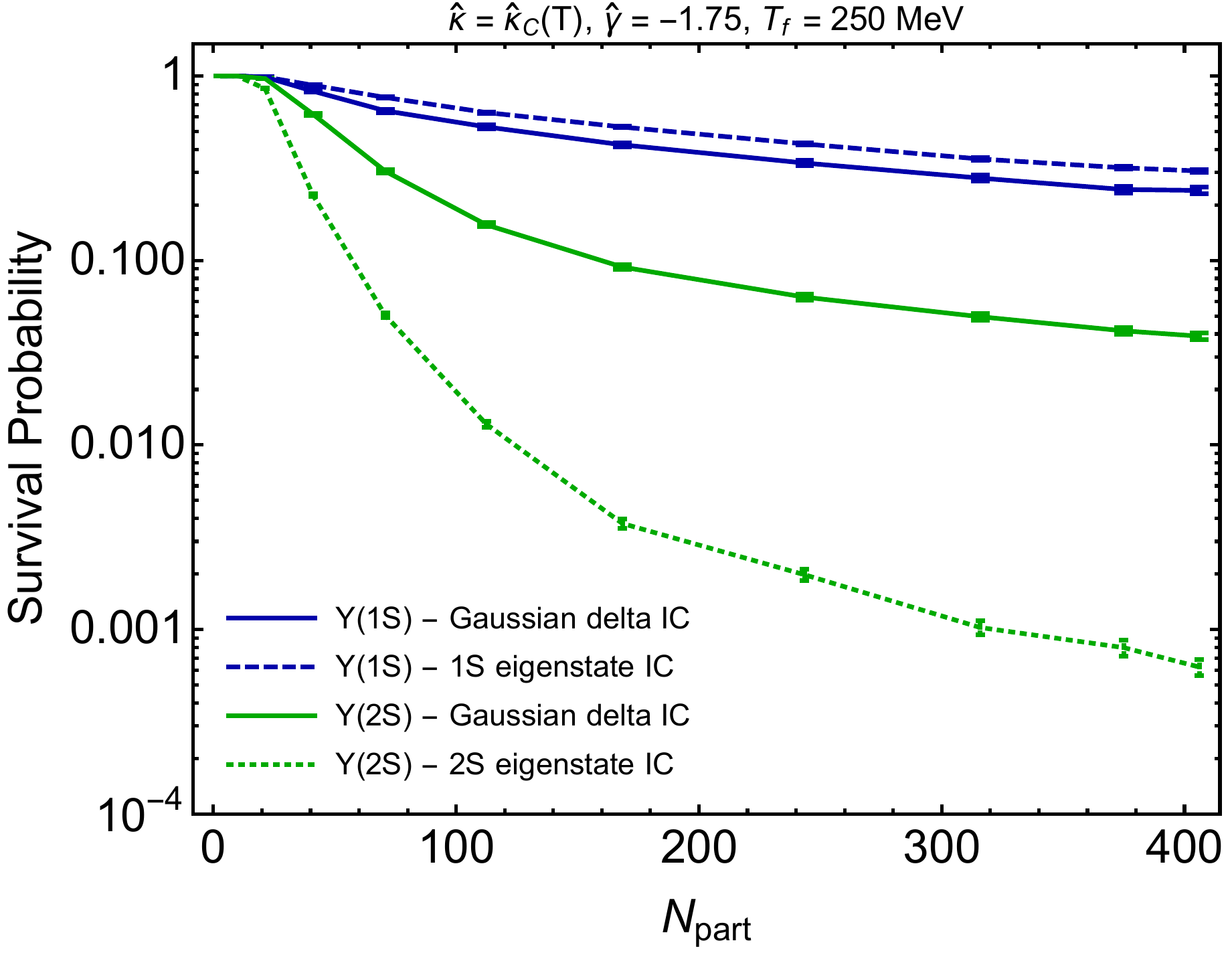}
\end{center}
\caption{$\Upsilon(1S)$ and $\Upsilon(2S)$ survival probabilities versus $N_{\rm part}$. 
Solid lines are the results obtained from the singlet $S$-wave Gaussian delta initial condition (IC) used in the main body of the paper.  
The dashed and dotted lines are the results obtained from $\Upsilon(1S)$ and $\Upsilon(2S)$ initial conditions, respectively.  
In both cases, the effect of jumps is included.
The error bars are the statistical errors associated with the average over quantum trajectories.}
\label{fig:icCompare}
\end{figure}

Finally, in figure \ref{fig:icCompare} we present a comparison of the $\Upsilon(1S)$ and $\Upsilon(2S)$ survival probabilities obtained from 
{\it (i)} $S$-wave eigenstate initial conditions and {\it (ii)} Gaussian delta initial conditions as used in the main body of the paper.  
We see less suppression of the $\Upsilon(1S)$ when using the pure $\Upsilon(1S)$ eigenstate initial condition than when using the Gaussian delta initial condition.  
For the $\Upsilon(2S)$, however, this pattern is reversed, and one sees a much larger difference between the results obtained using pure eigenstate and Gaussian delta initial conditions.  
The same pattern is observed when ignoring jumps, i.e. evolving only with the temperature-dependent $H_{\rm eff}$.
The key difference between the two types of initial conditions is that the Gaussian delta is a linear superposition of many vacuum eigenstates.  
The results indicate, therefore, that quantum state mixing due to the temperature-dependent Hamiltonian is important
and leads to a substantial reduction in the suppression of the $\Upsilon$ excited states.

\section{Effect of varying $\mathbf{T_{\rm f}}$}
\label{sec:Tf}
As discussed at the end of section~\ref{sect:hydro}, we evolve the quarkonium in the medium down to a temperature $T_{\rm f}$, which we take to be 250~MeV,
and in the vacumm from $T_{\rm f}$ to the crossover temperature $T_{\rm c}$.
The reason for the first choice is that the evolution equation set in section~\ref{ssec:hq} holds under the condition $T$, $m_D \gg E$.
Only under this condition, effects coming from the energy region $E$ may be neglected with respect to medium effects,
and medium effects can be cast in the two transport coefficients $\kappa$ and $\gamma$~\cite{Brambilla:2016wgg,Brambilla:2017zei,Brambilla:2019tpt}.
The largest Coulombic binding energy is the one of the $\Upsilon(1S)$ state; its value is $- 1/(m_ba_0^2)$, which is about 400~MeV for our choice of parameters.
The condition $T \ge T_{\rm f} = 250$~MeV guarantees reasonably well that the temperature scale, which is more properly $\pi T$,
and $m_D \approx 2 T$ are larger than 400~MeV during the quarkonium evolution in the medium. 
The reason for the second choice is that if the temperature becomes as low as the binding energy, in medium effects do not affect the potential (real and imaginary).
The potential is the vacuum one.
This is an exact statement that follows from the effective field theory description of the system~\cite{Brambilla:2008cx}.
In this situation, in medium effects, which are of relative order $(a_0T)^2$ or smaller according to the power counting of the effective field theory,
do not enter the Hamiltonian but affect the evolution equation through state dependent functions rather than transport coefficients.
They may be included systematically in the evolution equation as done in~\cite{Brambilla:2016wgg,Brambilla:2017zei}.
By neglecting order $(a_0T)^2$ corrections, which are small at low temperatures,
the quarkonium evolution for $T \le T_{\rm f}$ is therefore that one of a Coulombic bound state in the vacuum.

\begin{figure}[t!]
\begin{center}
	\includegraphics[width=1\linewidth]{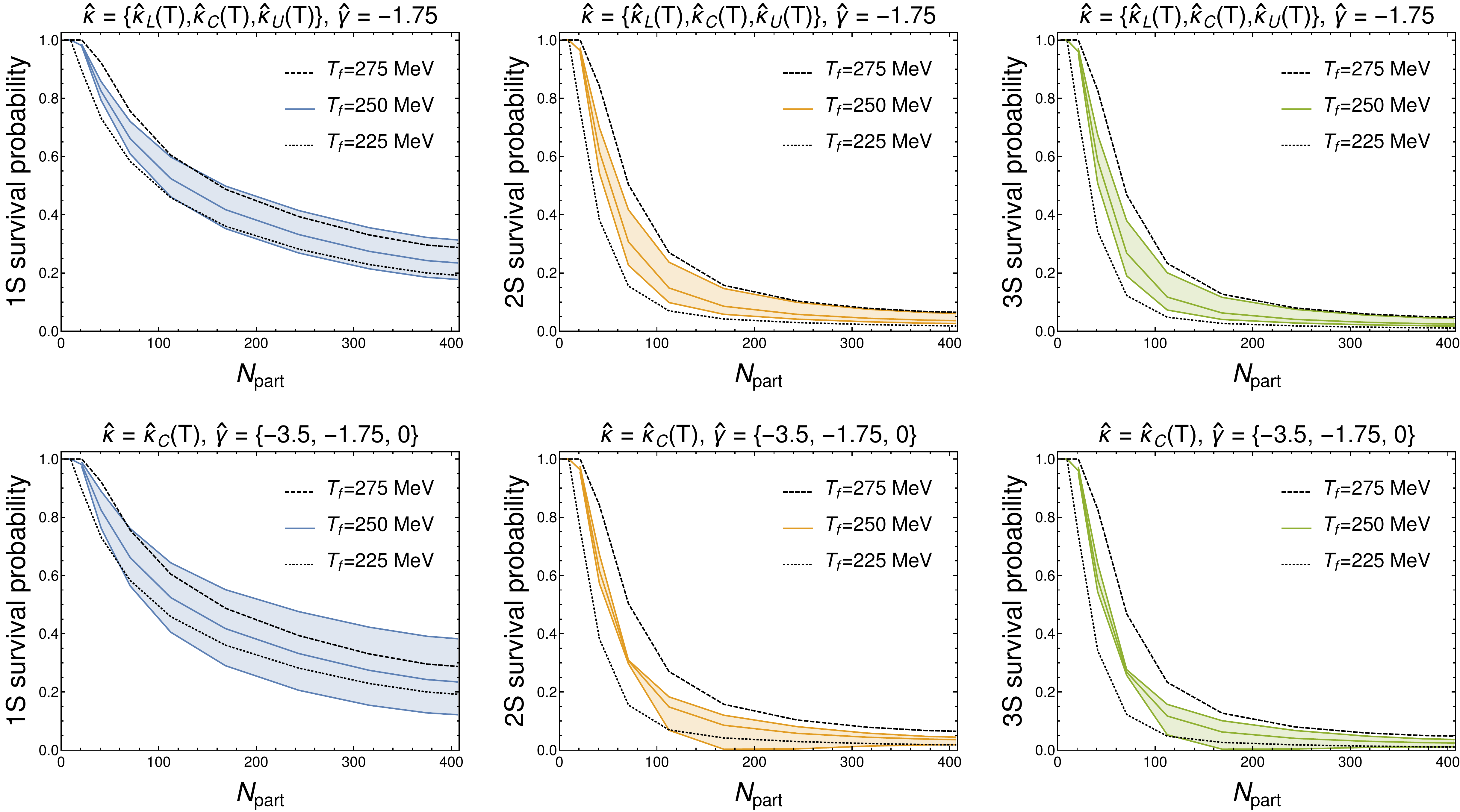}
\end{center}
\caption{
	$1S$, $2S$, and $3S$ survival probabilities of an initial Gaussian state evolved with $H_{\rm eff}$, i.e., without jumps.
	The central colored line in each panel represents the evolution down to $T_{\rm f}=250$~MeV with $\hat{\kappa}_{C}(T)$ and $\hat{\gamma}=-1.75$.
	The colored bands represent variations in $\hat{\kappa}$ (upper row) and $\hat{\gamma}$ (lower row).
	The black dashed and dotted lines represent the evolution down to $T_{\rm f}=275$~MeV and $225$~MeV, respectively, with $\hat{\kappa}_{C}(T)$ and $\hat{\gamma}=-1.75$. 
}
\label{fig:tfCompare}
\end{figure}

The value that we have chosen for $T_{\rm f}$ is somewhat arbitrary and a possible source of uncertainty in the determination of $R_{AA}$.
In figure~\ref{fig:tfCompare}, we show for $T_{\rm f} = (250 \pm 25)$~MeV
the survival probabilities of the $1S$, $2S$, and $3S$ bottomonium states computed evolving from an initial Gaussian state with $H_{\rm eff}$;
$H_{\rm eff}$ provides the bulk of the evolution as we have seen in the main body of the paper (cf. figure~\ref{fig:jumpNojump}).
The variations of the survival probabilities due to a $10\%$ uncertainty in $T_{\rm f}$ are of the same size as the ones due to the uncertainties in $\kappa$ and $\gamma$.

\section{Finite size effects}
\label{sec:finite_size}
As mentioned in section~\ref{sect:codetests}, the Gaussian initial condition used in the main body results in a rapidly spreading wave function when jumps are turned on.
As a result, one should make sure that there are not significant effects due to the finite box size ($L$) used in the simulations.
The evolution of the expectation value of $r$ varies for each quantum trajectory and in runs in which there are many jumps,
one may start to hit the edge of the box and the wave function may even reflect back from the edge of the box.
However, such situations are dominated by octet configurations with rather large $l$ and usually do not have time to reflect back to the left side of the box by the end of the run.
As a result, such cases have essentially zero overlap with the low $l$ compact bound states of interest, due to having very large $l$ and being in the octet configuration most of the time.

\begin{figure}[t!]
\begin{center}
	\includegraphics[width=0.6\linewidth]{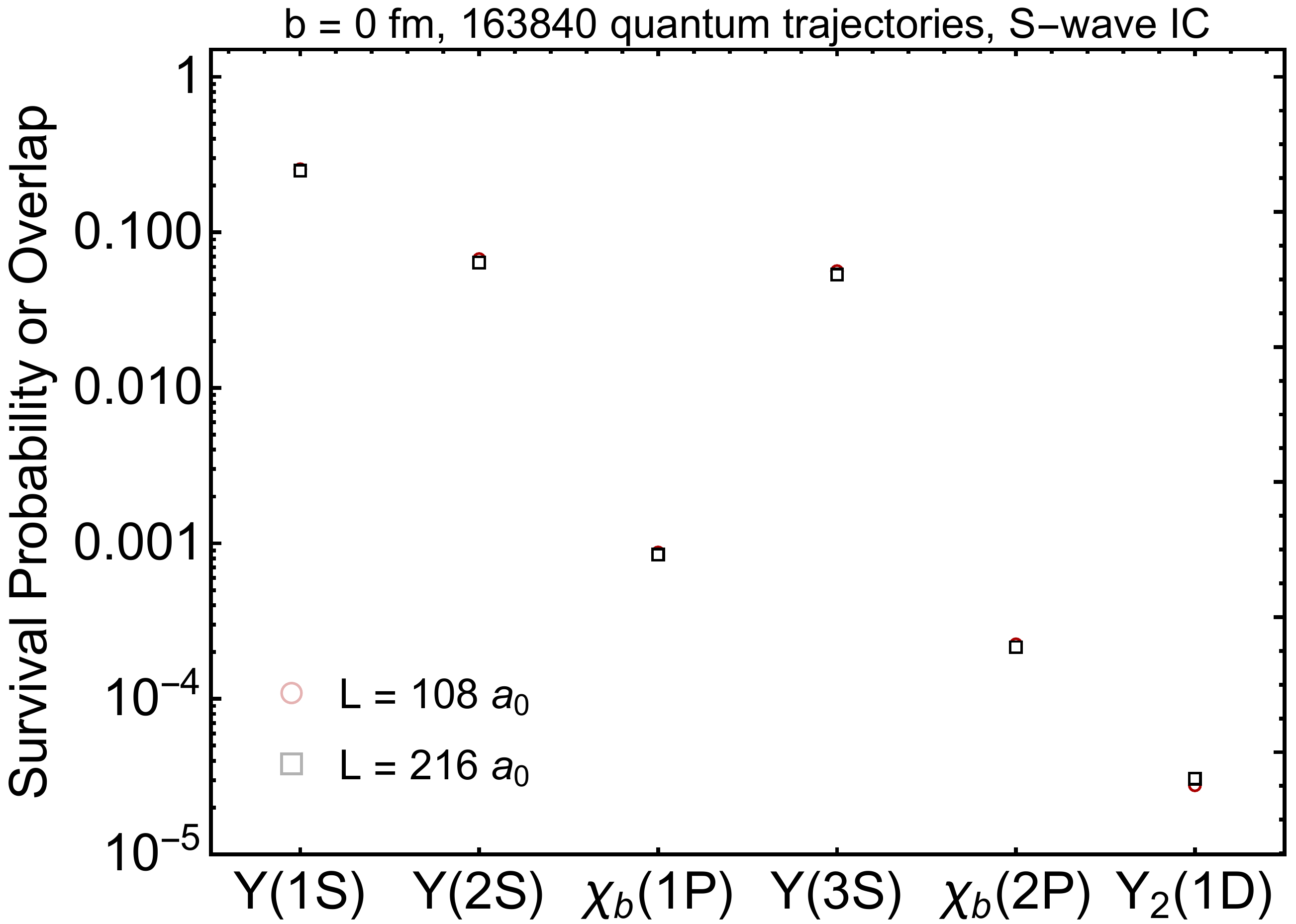}$\;$
\end{center}
\caption{
    Comparison of state survival probabilities for the $1S$, $2S$, and $3S$ states and quantum mechanical overlaps for the $1P$ and $2P$ states
    obtained using two different box sizes corresponding to $L = 108 \, a_0$ (open red circles) and $L = 216 \, a_0$ (open black squares) with fixed lattice spacing.
    In both cases, we used $S$-wave Gaussian initial conditions with zero impact parameter $b=0$ fm, $T_{\rm f}=250$ MeV,
    $\hat{\kappa}(T) = \hat{\kappa}_{C}(T)$, and $\hat{\gamma}=-1.75$.  }
\label{fig:boxsizeTest}
\end{figure}

In order to demonstrate that the results presented in the main body of the manuscript are not significantly affected by finite size effects,
in figure~\ref{fig:boxsizeTest} we present a comparison of survival probabilities for the $1S$, $2S$, and $3S$ states
and quantum mechanical overlaps for the $1P$ and $2P$ states obtained using two different box sizes
corresponding to $L = 108 \, a_0$ (open red circles) and $L = 216 \, a_0$ (open black squares) with fixed lattice spacing.
In both cases, we used $S$-wave Gaussian initial conditions with zero impact parameter $b=0$~fm, $T_{\rm f}=250$ MeV,
$\hat{\kappa}(T) = \hat{\kappa}_{C}(T)$, and $\hat{\gamma}=-1.75$.
We choose $b = 0$ fm for this test because this case has the longest evolution time and hence is the most susceptible to any finite size effects during the evolution.
As shown in figure~\ref{fig:boxsizeTest}, the $S$-wave results obtained are independent of $L$ within the statistical errors reported.  
Quantitatively, the ratio of the $L=206 \, a_0$ results to the $L=108 \, a_0$, shown in figure~\ref{fig:boxsizeTest} are $\{0.98\pm 0.04,0.95\pm 0.06,0.97\pm 0.06,0.95\pm0.07,0.96\pm 0.06,1.10\pm 0.10\}$ computed from left to right, which demonstrates that our $S$-wave results for the two box sizes are consistent with unity within statistical errors stemming from the average over quantum trajectories.  We have  performed similar tests for $P$-wave initial conditions and found that, for both box sizes, the resulting $b=0$ survival probabilities/overlaps are always extremely small.

\pagebreak

\bibliographystyle{JHEP}
\bibliography{qtraj}

\end{document}